\documentclass[12pt,a4paper]{article}%
\pdfoutput=1
\usepackage{amsmath,amssymb,amsfonts,wasysym}
\usepackage[pdftex]{graphicx}
\usepackage{color}
\usepackage{float}
\usepackage[bf,footnotesize]{caption2}
\usepackage{hyperref}%
\usepackage{amsmath}%
\usepackage{amsfonts}%
\usepackage{amssymb}
\pdfoutput=1
\numberwithin{equation}{section}
\setcaptionmargin{1cm}

\def\La{\mathcal{L}}
\def\Amp{\mathcal{A}}
\def\({\left(}
\def\){\right)}
\def\f{\frac}
\def\be{\begin{equation}}
\def\ee{\end{equation}}

\def\de{\partial}
\def\demub{\de_{\mu}}

\def\demua{\de^{\mu}}

\newcommand{\cL}{{\cal L}}

\newcommand{\lrp}[1]{\left(#1\right)}

\newcommand{\ba}{\begin{array}}
\newcommand{\ea}{\end{array}}
\newcommand{\beqa}{\begin{eqnarray}}
\newcommand{\eeqa}{\end{eqnarray}}
\newcommand{\bea}{\begin{eqnarray}}
\newcommand{\eea}{\end{eqnarray}}

\begin{document}

\begin{titlepage}
\thispagestyle{empty}
\vskip 1.0cm
\begin{center}
\hspace{3cm}\vspace{2cm}\\
{\Large\bf Scuola Normale Superiore Di Pisa}\vspace{0cm}\\%\vskip 2.0cm
{\bf Classe Di Scienze}\vspace{0cm}\\
{\bf PhD Thesis}\vspace{0cm}\\
{\bf 2010}\vspace{5cm}\\
{\Large \bf Composite Vectors and Scalars in\\ Theories of Electroweak Symmetry Breaking}\vspace{5.0cm}\\%\vskip 1.0cm
%{\large R. Barbieri$^{a,b}$, A. E. C\'arcamo Hern\'andez$^{a,b}$ and R. Torre$^{b,c}$}\\[0.7cm]{\it $^a$ Scuola Normale Superiore, Piazza dei Cavalieri 7, I-56126 Pisa, Italy}\\[5mm]{\it $^b$ INFN, Sezione di Pisa, Largo Fibonacci 3, I-56127 Pisa, Italy}\\[5mm]{\it $^c$Universit\`a di Pisa, Dipartimento di Fisica, Largo Fibonacci 3, I-56127 Pisa, Italy}
{\bf Antonio Enrique C\'arcamo Hern\'andez}\\
{\bf antonio.carcamo@usms.cl}\vspace{1.5cm}\\
{\bf Advisor: Professor Riccardo Barbieri}\vspace{1.2cm}\\
{\bf Defended at the Scuola Normale Superiore Di Pisa on 27th of January of 2011.}
\end{center}\newpage

\begin{center}
\thispagestyle{empty}
\hspace{3cm}\vspace{2cm}\\
{\Large\bf Abstract}
\end{center}
\vspace{0cm}
In the context of a strongly coupled Electroweak Symmetry Breaking, composite triplet of heavy vectors belonging to the $SU(2)_{L+R}$ adjoint representation and a composite scalar singlet under $SU(2)_{L+R}$ may arise from a new strong interaction invariant under the global $SU(2)_L\times SU(2)_R$ symmetry, which is spontaneously broken down to $SU(2)_{L+R}$. This thesis consists of two parts. The first part is devoted to the study of the heavy composite vector pair production at the LHC via Vector Boson Fusion and Drell-Yan annihilation under the assumption that the interactions among these heavy vector states and with the Standard Model gauge bosons are described by a $SU(2)_L\times SU(2)_R/SU(2)_{L+R}$ Effective Chiral Lagrangian. The expected rates of multi-lepton events from the decay of the composite vectors are also given. The second part studies the associated production at the LHC of a composite vector with a composite scalar by Vector Boson Fusion and Drell-Yan annihilation in the framework of a $SU(2) _{L}\times SU(2) _{R}/SU(2)_{L+R}$ Effective Chiral Lagrangian with massive spin one fields and one singlet light scalar. The expected rates of same sign di-lepton and tri-lepton events from the decay of the composite vector and composite scalar final state are computed. The connection of the Effective Chiral Lagrangians with suitable gauge models is elucidated. 

\newpage

\begin{flushright}
\thispagestyle{empty}
\Large{To my mother Yadira Esther, to my father Antonio Nicol\'as, to my sister Eliana Marcela, to my brother Juan David and to my girlfriend Emeline.}
\end{flushright}
\vskip 1cm
%\vspace{20cm}
%\begin{flushleft}
%\end{flushleft}

%\begin{abstract}
%\vskip 1cm \hspace{0.7cm} April 2010
\end{titlepage}

\tableofcontents
\newpage
\section{Introduction}
\subsection{Statement of the problem and purpose of the thesis}
%that the nature of the dynamics is responsible for Electroweak Symmetry Breaking (EWSB), that is, to know if it is weakly or strongly coupled
One of the most important issues to be settled by the LHC is whether the dynamics responsible for ElectroWeak Symmetry Breaking (EWSB) is weakly or strongly coupled. A weakly coupled dynamics describing the mechanism of the EWSB is provided by the Standard Model and its Supersymmetric extensions. In the Standard Model, the existence of one Higgs doublet is assumed in order to explain the generation of the masses of all the fermions and bosons. In addition to the 3 eaten up Goldstone bosons, the Higgs doublet contains one physical neutral scalar particle, called the Higgs boson, which is crucial for keeping under control unitarity in the elastic and inelastic channels of the gauge boson scattering and which allows us to extrapolate a weakly coupled model up to the Planck scale. A light Higgs boson can also successfully account for the ElectroWeak Precision Tests (EWPT).\\

The Higgs boson mass is the only unknown parameter in the symmetry breaking sector of the Standard Model. However, an upper bound on the mass of the Higgs boson can be set requiring that the quartic coupling in the Higgs self interaction potential, which grows with rising energy, should be finite at an energy scale $\Lambda $ up to which the Standard Model is assumed to be valid. If the quartic coupling in the Higgs self interaction potential becomes large, which corresponds to a heavy Higgs boson, perturbation theory in terms of this coupling breaks down. In that case, the Higgs boson becomes strongly interacting. Moreover,
the requirement of unitarity in longitudinal $WW$ scattering can be used to set an upper bound on the mass of the Higgs boson. In the Standard Model, the scattering amplitude for longitudinal $W$ bosons will violate unitarity when the mass of the Higgs boson takes values larger than about $1$ TeV \cite{Quigg}, which means that perturbation theory breaks
down and the Standard Model becomes strongly interacting for a sufficiently
heavy Higgs boson. The lower bound on the mass of the Higgs boson is determined from the requirement of vacuum stability of the scalar self interaction
potential; this lower bound depends on the mass of the top quark and on the
cutoff $\Lambda $ up to which the Standard Model
is assumed to be valid. The value of the aforementioned quartic coupling decreases when the top quark Yukawa coupling increases. For a cutoff $\Lambda =10^{16}$ GeV corresponding to the Grand Unification scale, the requirement of vacuum stability of the scalar self interaction potential, implies a lower bound of about $130$ GeV for the mass of the Higgs boson \cite{Djouadi}. The Standard Model can be self consistent up to very high energies provided that the Higgs boson is relatively light. For example, the consistency of the Standard Model up to the unification scale $\Lambda =10^{16}$ GeV sets the Higgs boson mass in the range $130$ GeV$\lesssim M_{H}$$\lesssim$ $180$ GeV \cite{Djouadi}.\\ 

%134,177
In spite of the very good agreement of the Standard Model predictions with experimental data, the Higgs boson is yet to be detected experimentally. Therefore one can say that the mechanism of EWSB responsible for the generation of the masses of all fermions and bosons remains to be explained. Moreover, the Standard Model has the hierarchy problem, which is the instability of the mass of the Higgs field against quantum corrections, which are proportional to the square of the cutoff. This means that in a quantum theory with a cutoff at the Planck scale $\Lambda\simeq 10^{19}$ GeV, the Higgs boson mass will have quantum corrections that will raise it to about the Planck scale unless an extreme fine-tuning of $34$ decimals is performed in the bare squared mass. This is the naturalness problem of the Standard Model.\\

As there is no direct experimental evidence for a Higgs particle up to date, it is natural to ask what happens if we keep all the Standard Model fields, except the Higgs boson. One can for example think of a very heavy Higgs boson and build an effective field theory below the Higgs boson mass.  The effective theory contains three of the four components of the Higgs doublet, which have become the longitudinal components of the $W^{\pm }$ and $Z$ bosons, but not the fourth component $-$ the Higgs boson. This is the starting point of the Electroweak Chiral Lagrangian (EWCL) formulation, which is inspired by the Chiral Lagrangian approach to QCD at low energies and Chiral Perturbation Theory \cite{Herrero,Manohar,Pich,Dawson:1985,Isidori}. However, as is well known, the EWCL formulation does not pass the EWPT and the unitarity considerations for $WW$ scattering (unitarity is violated at energies around $1.7$ TeV).\\
%between $1TeV$ and $1.5TeV$.\\

%These problems together with the need of having a well founded theoretical description for the Electroweak Precision Tests consistent with the experimental data, motivate the search of alternative mechanisms 
These problems can perhaps be overcome if one considers EWSB mechanisms in the framework of a strongly interacting dynamics, where the theory becomes non-perturbative above the Fermi scale and the breaking is achieved through some condensate. In the strongly interacting picture of EWSB, many models have been proposed, which predict the existence of composite particles, e.g. composite scalars \cite{Kaplan:1983, Chivukula:1993, ggpr, Zerwekh:2010, Low:2009, Contino:2009}, composite vector resonances \cite{Bagger,Pelaez:1996,SekharChivukula:2001, Csaki:2003, Barbieri:2008,Cata:2009iy,Barbieri:2010}, composite scalar and vector resonances \cite{Zerwekh:2006,Carcamo:2010} and composite fermions \cite{Barbieri:2008b}. The spin-0 and spin-1 resonances predicted by these models play a very important role in controlling unitarity in longitudinal gauge boson scattering up to the cutoff $\Lambda \simeq 4\pi v$. For appropiate couplings and masses, the exchange of the composite resonances can perhaps also account for the EWPT. Furthermore, a composite scalar does not have the hierarchy problem since quantum corrections to its mass are saturated at the compositeness scale.\\

The phenomenology of heavy vector states at high-energy colliders~\cite{He:2007ge,Accomando:2008jh,Belyaev:2008yj}, as well as their role in electroweak observables, is subject of intensive discussion. However, in most of the existing analyses specific dynamical assumptions are made such as considering  these vector states as the gauge vectors of a spontaneously broken gauge symmetry. Recent studies~\cite{Barbieri:2008,Hirn:2007} show that these assumptions may be too restrictive for generic models based on strong dynamics at the TeV scale, and only going beyond these assumptions can one successfully account for the EWPT by solely considering exchange of heavy vectors. Altogether we find it potentially useful to take a model independent approach based on an effective Lagrangian description of the new particles coming from the strong dynamics with the incorporation of the relevant symmetries, whatever they are, exact or approximate. The composite spin-0, spin-1/2 or spin-1 states arising from the unknown strong interaction, which are bound states of more fundamental constituents held together by a new strong interaction, may be the lightest non standard particles and their discovery could provide the first clue of strong EWSB at the LHC.\\
 
To understand the underlying dynamics, several measurements and observations will certainly be required. It is assumed that this new strong dynamics supposedly breaking the Electroweak Symmetry is by itself invariant under a global $SU(2)_{L}\times SU(2)_{R}$ symmetry, which is spontaneously broken to the diagonal $SU(2)_{L+R}$ subgroup. After gauging the Standard Model gauge group, the $SU(2)_L\times SU(2)_R$ global symmetry of the new strong dynamics is broken down to the $SU(2)_{L+R}$ custodial group. It is also assumed that the strong dynamics responsible for the EWSB gives rise to composite triplet of heavy vectors degenerate in mass belonging to the adjoint representation of the custodial symmetry group. These heavy vector states have a mass below the cutoff $\Lambda \simeq 4\pi v$. The study of the heavy vector pair production is crucial for distinguishing the different models since it is sensitive to many couplings and in some sense more model dependent. The heavy vector pair production at the LHC by Vector Boson Fusion and Drell-Yan annihilation is studied in the first part of this thesis under the assumption that the interactions among these heavy vector states and with the Standard Model gauge bosons are described by a $SU(2)_L\times SU(2)_R/SU(2)_{L+R}$ Effective Chiral Lagrangian. The relevant parameter space is determined by minimizing the growing energy behaviour of the scattering amplitudes for longitudinal Standard Model gauge bosons going into a pair of polarized vectors. The connection between a composite vector and a gauge vector of a spontaneously broken gauge symmetry is also investigated. The cross sections for vector pair production and the expected rates of multi-lepton events from the decay of such heavy vectors into Standard Model gauge bosons at the LHC have been computed.\\

%The expected rates of same sign di-lepton and tri-lepton events from the decay of the composite vectors into Standard Model gauge bosons are computed. 
In the second part of the thesis a light composite scalar, singlet under $SU(2)_{L+R}$ with mass $m_h\lesssim v$, is added to the $SU(2)_L\times SU(2)_R/SU(2)_{L+R}$ Effective Chiral Lagrangian. The interactions of this scalar with the Standard Model Gauge bosons and with the heavy vector pairs are introduced. The asymptotic behaviour of the elastic and inelastic channels of longitudinal SM gauge boson scattering is studied. The unitarity condition for the elastic channel of longitudinal SM gauge boson scattering is used to determine the relevant parameter space, in which the associated production of a heavy vector together with a scalar via Vector Boson Fusion and Drell-Yan annihilation at the LHC is studied. The total cross sections for the production at the LHC of a heavy vector in association with a scalar and the expected rates of same sign di-lepton and tri-lepton events from the decay of the composite vector and composite scalar final states are computed. A thorough phenomenological analysis and the evaluation of the backgrounds to such signals will be necessary to assess the visibility of composite vector pairs and composite vector-composite scalar final states at the LHC. %Finally, the conclusions are given. 

\subsection{The Electroweak Chiral Lagrangian}
Chiral Lagrangians have been extensively used to describe the phenomenon of
spontaneous symmetry breaking in strong and in weak interactions. They can be
regarded as the low energy limit of an underlying fundamental theory. The basis of this approach have been formulated by Weinberg to characterize the S matrix elements for pions interactions; after that Gasser and Leutwyler developed them building the Chiral Perturbation Theory, which describes low energy effects of strong interactions and was motivated by the fact that below the mass of the $\rho$ meson, the Hadronic spectrum contains an octet of very light pseudoscalar particles $(\pi,K,\eta)$ \cite{Ecker:1988te,Ecker:1989yg}. Inspired by the Chiral Perturbation Theory Lagrangian formalism up to $O\left(  p^{4}\right)$  developed by Ecker et al., used in the description of the low energy effects in QCD, the following EWCL can be used to formulate the EWSB without the Higgs boson:
\begin{equation}
\cL_{SB} = \frac{v^{2}}{4}\left\langle D_{\mu}U\left(  D^{\mu
}U\right)  ^{\dag}\right\rangle - \frac{v}{\sqrt{2}} \sum_{i,j} \left( \bar u_L^{(i)} d_L^{(i)} \right) 
 U \begin{pmatrix} \lambda_{ij}^u\,  u_R^{(j)} \\[0.1cm] \lambda_{ij}^d\,  d_R^{(j)} \end{pmatrix}
\label{eq:chiralLmass}+h.c~,
\end{equation}
where:
\begin{equation}%
\begin{array}
[c]{l}%
U\left(  x\right)  =e^{i\hat{\pi}\left(  x\right)  /v}\,,\qquad\hat{\pi
}\left(  x\right)  =\tau^{a}\pi^{a}=\left(
\begin{array}
[c]{cc}%
\pi^{0} & \sqrt{2}\pi^{+}\\
\sqrt{2}\pi^{-} & -\pi^{0}%
\end{array}
\right)  \,,\\
D_{\mu}U=\partial_{\mu}U-i{B}_{\mu}U+iU{W}_{\mu}\,,\qquad{W}_{\mu}=\frac
{g}{{2}}\tau^{a}W_{\mu}^{a}\,,\qquad{B}_{\mu}=\frac{g^{\prime}}{{2}}\tau
^{3}B_{\mu}^{0}\,,
\end{array}
\end{equation}
$U$ is the matrix which contains the Goldstone boson fields $\pi^a$ with $a=1,2,3$, the $\tau^{a}$ are the ordinary Pauli matrices, $\left\langle
{}\right\rangle $ denotes the trace over $SU(2)$, $\lambda_{ij}^u$ and $\lambda_{ij}^d$ are the up and down type quarks Yukawa couplings, respectively.\\

The transformation properties of the Goldstone fields under $SU(2)_{L}\times SU(2)_{R}$ are 
\begin{equation}  \label{eq5}
u \equiv \sqrt{U} \to g_{R}u h^{\dag}=hug_{L}^{\dag}\,,
\end{equation}
where $h=h\left(u,g_{L},g_{R}\right)$ is an element of $SU(2)_{L+R}$, as
defined by this very equation \cite{coleman}. The local $SU(2)_L\times U(1)_Y$ invariance is now manifest in the Lagrangian (\ref{eq:chiralLmass}) with $U$ transforming as
\begin{equation}
U \to g_L(x) \, U\, g_Y^\dagger(x)\, , \qquad g_L(x) = \exp \big(i\theta_L^a(x) \tau^a/2 \big), \quad   g_Y(x) = \exp \big(i\theta_Y(x)\tau^3/2 \big)\, .
\end{equation}
and with the $W$, $B$ and quark fields transforming in the usual way. The inclusion of the leptons is straightforward.
In the unitary gauge $\langle U \rangle = 1$, it is immediate to see that the chiral Lagrangian (\ref{eq:chiralLmass}) gives the mass terms for the $W$ and $Z$ gauge bosons with
\begin{equation}
\label{eq:rho}
\rho = \frac{M_W^2}{M_Z^2 \cos^2\theta_W} = 1 \,.
\end{equation}\\
As is well known, this relation is the consequence of the larger approximate invariance of (\ref{eq:chiralLmass}) under the $SU(2)_L\times SU(2)_R$ global transformations $U \to g_L \, U\, g_R^\dagger $, which is spontaneously broken to the diagonal custodial group $SU(2)_{C}=SU(2)_{L+R}$ by $\langle U \rangle = 1$, and explicitly broken by $g^{\prime}$ and $\lambda^u_{ij} \not = \lambda^d_{ij}$. In the limit $g^{\prime}=0$ and $\lambda^u_{ij} = \lambda^d_{ij}$, the $SU(2)_{L+R}$ custodial symmetry implies $M_W = M_Z$, which is replaced by eq.(\ref{eq:rho}) at tree level for arbitrary $g^{\prime}$. The pions transform as a triplet under the custodial symmetry group $SU(2)_{L+R}$, which plays the role of a weak isospin group when low energy pion interactions are considered.\\
 
%In the most general framework in which $g^{\prime}\ne 0$, one has additional one loop level corrections to the $\rho$ parameter proportional to $g^{\prime}$ and to the difference $\lambda^u - \lambda^d$ between the up and down type quark Yukawa couplings but these are small. \\
A term like
\begin{equation}
 c_3\, v^2 \, \left\langle{T^3U^\dagger D_\mu U \,  }\right\rangle^2 
\end{equation}
invariant under the local $SU(2)_L\times U(1)_Y$ but not under the global $SU(2)_L\times SU(2)_R$ symmetry is therefore forbidden. Its presence would undo the $\rho=1$ relation.\\

The effective Lagrangian 
%\Tr \left( D_\mu U^\dagger D_\mu U \right)
\begin{equation}
\cL_{\rm eff}  = \cL_{gauge}+ \cL_{SB},\hspace{2cm}\cL_{gauge}=-\frac{1}{2g^{2}}\left\langle W_{\mu\nu}W^{\mu\nu}\right\rangle -\frac{1}{2g^{\prime2}}\left\langle B_{\mu\nu}B^{\mu\nu}\right\rangle \
\end{equation}
provides an accurate description of particle physics, in some cases even beyond the tree level, at least up to energies below a cutoff \cite{Contino:2010t}:
\be
\Lambda = 4 \pi v \approx 3~{\rm TeV}
\ee
%(the na\"ive dimensional cut-off of $\cL_\chi^{(2)}$).
when a loop expansion ceases to be meaningful. This Lagrangian is therefore meant to describe the spontaneous breaking of the electroweak local invariance $SU(2)_L \times U(1)_{Y}\to U(1)_{Q}$ by a strong dynamics which itself breaks a global symmetry $SU(2)_L \times SU(2)_R \times U(1)_{B-L}\to SU(2)_{L+R}\times U(1)_{B-L}$. It suffers, however, of two main problems \cite{Contino:2010t}:
\begin{itemize}
 \item The violation of unitarity in $WW$ scattering, evaluated at the 
 tree-level, below the cutoff $\Lambda$.
 \item The inconsistency of the electroweak observables $S$ and $T$ when compared with the experimental data if evaluated at the one-loop level with $\Lambda$ as ultraviolet cutoff.
\end{itemize}

While the first problem requires that some action be taken, we shall not address in the following the second problem. The electroweak observables $S$ and $T$ will receive many contributions from different sources, among which cancellations may occur and which are difficult to control without an explicit model. Furthermore $S$ and $T$ will in general be sensitive to the physics at the cutoff, not controllable by the effective Lagrangians that we are using.  

%This implies that new weakly or strongly interacting dynamics with the relevant degrees of freeedom should act as a ultraviolet completion of the Lagrangian given in (\ref{eq:Luniv}) in order to restore the perturbative unitarity in $WW$ scattering. On the other hand, it is well know that a description of the electroweak observables $S$ and $T$ consistent with the experimental data is provided for a light Higgs boson. In the framework of the Electroweak Chiral Lagrangian, the Higgs boson is absent, that is, its mass is taken to be infinitely heavy so that it is decoupled from the theory. In this case new resonances below the cuttof scale $\Lambda$ will have to be introduced in order to account for a description of the electroweak observables $S$ and $T$ compatible with the experimental data.

%These problems point toward the existence of new degrees of freedom below the cut-off. This motivates the introduction of heavy vectors fields in the EW Chiral Lagrangian.
%Assuming the W/Z bosons to become strongly interacting at TeV energies, damping the rise of the elastic W/Z scattering amplitudes, is an alternative way to solve the problem of unitarity violation at high energies in the SM, without adding a relatively light Higgs boson. Naturally, the strong forces between the massive gauge bosons may be traced back to new fundamental interactions characterized by a scale of order 1 TeV

\subsection{$W_LW_L\rightarrow W_LW_L$ and $W_LW_L\rightarrow f\bar{f}$ amplitudes}
The bad high energy behaviour of the WW elastic scattering, as of the WW annihilation into a pair of fermions manifests itself when one considers longitudinally polarized vector bosons, $W_L$. In order to compute the $W_LW_L\rightarrow W_LW_L$ and $W_LW_L\rightarrow f\bar{f}$ amplitudes one takes into account the Goldstone Equivalence Theorem which states that at high energies the amplitude for the emission or absorption of longitudinally-polarized vector boson becomes equal to the amplitude for the emission or absorption of Goldstone field $\pi$ \cite{Contino:2010t}. In particular the $W^a_LW^b_L\rightarrow W^c_LW^d_L$ and $W^a_LW^b_L\rightarrow f\bar{f}$ scattering amplitudes at high energies become equal to the $\pi^a\pi^b\rightarrow \pi^c\pi^d$ and $\pi^a\pi^b\rightarrow f\bar{f}$ scattering amplitudes up to corrections of the order $O\lrp{\frac{M^2_W}{\sqrt{s}}}$ (and up to a factor of $i^N$ where $N$ is the number of Goldstone bosons):
\begin{equation}
A(W_{L}^{a}W_{L}^{b}\rightarrow W_{L}^{c}W_{L}^{d})=A\left(\pi^{a}\pi^{b}\rightarrow \pi^{c}\pi^{d}\right)\left[1+O\left(\frac{M_{W}}{\sqrt{s}}\right)  \right], \label{p5}
\end{equation}
\begin{equation}
A(W_{L}^{a}W_{L}^{b}\rightarrow f\bar{f})=-A\left(\pi^{a}\pi^{b}\rightarrow f\bar{f}\right)\left[1+O\left(\frac{M_{W}}{\sqrt{s}}\right)\right].
\label{p6}
\end{equation}\\
%so that the $\pi$'s are the degrees of freedom that are eaten in the unitary gauge to form the longitudinal polarizations of $W$ and $Z$. \\
Taking the $g^{\prime}\to 0$ limit for simplicity, isospin conservation implies that the four pions Lorentz invariant scattering amplitude can be written as:
\begin{equation}
A\left(  \pi^{a}\pi^{b}\rightarrow\pi^{c}\pi^{d}\right)  =A\left(
s,t,u\right)^{\pi\pi\to \pi\pi}\delta^{ab}\delta^{cd}+B^{\pi\pi\to \pi\pi}\left(  s,t,u\right) \delta^{ac}%
\delta^{bd}+C\left(  s,t,u\right)^{\pi\pi\to \pi\pi} \delta^{ad}\delta^{bc}.\label{xb}%
\end{equation}\\
The Bose symmetry implies that the four pions scattering amplitude should be
invariant under the exchange of pions, that is, under the exchange
$a\leftrightarrow b$ , $t\leftrightarrow u$ and $a\leftrightarrow c$ ,
$s\leftrightarrow t$. Then, the following relations are obtained:
\begin{equation}
B\left(  s,t,u\right)^{\pi\pi\to \pi\pi}=A\left(  t,s,u\right)^{\pi\pi\to \pi\pi},\hspace{1.5cm}\hspace
{1.5cm}C\left(  s,t,u\right)^{\pi\pi\to \pi\pi}=A\left(  u,t,s\right)^{\pi\pi\to \pi\pi},\label{xc}%
\end{equation}
which implies that the four pions scattering amplitude has the following
form:
\begin{equation}
A\left(  \pi^{a}\pi^{b}\rightarrow\pi^{c}\pi^{d}\right)  =A\left(
s,t,u\right)^{\pi\pi\to \pi\pi}\delta^{ab}\delta^{cd}+A\left(  t,s,u\right)^{\pi\pi\to \pi\pi}\delta^{ac}%
\delta^{bd}+A\left(  u,t,s\right)^{\pi\pi\to \pi\pi}\delta^{ad}\delta^{bc}.\label{la}%
\end{equation}\\
The function $A\left(s,t,u\right)^{\pi\pi\to \pi\pi}$ comes from the derivative interaction 
\begin{equation}
%1%
%TCIMACRO{\tciLaplace}%
%BeginExpansion
\mathcal{L}^{\pi^{4}}=\frac{1}{48v^{2}}\left\langle\left[  \pi,\partial_{\mu}\pi\right]  \left[  \pi,\partial^{\mu}\pi\right]  \right\rangle  =-\frac{1}{6v^{2}}\varepsilon^{abe}\varepsilon^{cde}\pi^{a}\pi^{c}\partial_{\mu}\pi^{b}\partial^{\mu}\pi^{d}
\label{contact0}
\end{equation}
among the four Goldstones contained in the kinetic term of $U$ in (\ref{eq:chiralLmass}) and is given by:
\begin{equation}
A\left(s,t,u\right)^{\pi\pi\to \pi\pi}=\frac{s}{v^2}.
\end{equation}\\
The growth of the $W_{L}^{a}W_{L}^{b}\rightarrow W_{L}^{c}W_{L}^{d}$ scattering amplitude with the square of the center of mass energy $\sqrt{s}$ implies a violation of perturbative unitarity.\\ 

To determine the energy at which the perturbative unitarity is violated, the $WW$ scattering amplitude is decomposed into partial waves and the unitarity condition in the $I=0$ isospin channel is applied. The fixed isospin amplitudes are given by
\cite{Barbieri:2008}:
%at energies $\sqrt{s}\approx 4\pi v\approx 3~{\rm TeV}$. 
\begin{equation}
T\left(  I=0\right)  =3A\left(  s,t,u\right)  +A\left(  t,s,u\right)
+A\left(  u,t,s\right)  =2A\left(  s,t,u\right), \label{s11a}%
\end{equation}%
\begin{equation}
T\left(  I=1\right)  =A\left(  t,s,u\right)  -A\left(  u,t,s\right),\label{s12a}%
\end{equation}
\begin{equation}
T\left(  I=2\right)  =A\left(  t,s,u\right)+A\left(  u,t,s\right)  =-A\left(  s,t,u\right) \label{s12b}%
\end{equation}
and the partial wave coefficients have the following form:
\begin{equation}
a_{l}^{I}\left(  s\right)  =\frac{1}{64\pi}\int_{-1}^{1}d\left(  \cos
\theta\right)  P_{l}\left(  \cos\theta\right)  T\left(  I\right). \label{s14}%
\end{equation}\\
Then, it follows that the partial wave coefficient $a_{0}^{0}\left(  s\right)  $ of isospin zero for the four pion scattering is given by:%

\begin{equation}
a_{0}^{0}\left(  s\right)  =\frac{1}{32\pi}\int_{-1}^{1}dyA\left(  s,t\left(
y\right)  ,u\left(  y\right)  \right)  =\frac{s}{16\pi v^{2}}.
\label{s15a}
\end{equation}\\
The strongest unitarity constraint $\left\vert a_{0}^{0}\left(  s\right)
\right\vert <1$ implies:
\begin{equation}
\sqrt{s}<1.7\hspace{0.1cm}TeV.
\end{equation}
This means that perturbative unitarity in $WW$ scattering is violated at energies $\sqrt{s}\approx 1.7~{\rm TeV}$, implying that New Physics should manifest itself at energies in the TeV range to restore unitarity in the scattering amplitudes of longitudinal gauge bosons.\\

%in this case one has a strong $W_{L}^{a}W_{L}^{b}\rightarrow W_{L}^{c}W_{L}^{d}$. This is a consequence of the non-renormalizability of the Chiral Lagrangian given in (\ref{eq:chiralLmass}).  %However the chiral implies a violation of the perturbative unitar
From $SU(2)_{L+R}$ invariance and Bose symmetry, the $\pi^{a}\pi^{b}\rightarrow f\bar{f}$ scattering amplitude is given by:
%following relation is satisfied:
\begin{equation}
A(\pi^{a}\pi^{b}\rightarrow f\bar{f})=A(s,t,u)^{\pi\pi\to f\bar{f}}\delta^{ab}
\end{equation}
where the leading contribution to this amplitude comes from the $\pi^2f\bar{f}$ contact interaction also contained in (\ref{eq:chiralLmass}) so that the function $A(s,t,u)^{\pi\pi\to f\bar{f}}$ is given by:
\begin{equation}
A(s,t,u)^{\pi\pi\to f\bar{f}}=\frac{m_f\sqrt{s}}{v^2},
\end{equation}
$m_f$ being the fermion mass. In this case, one has that the $W_{L}^{a}W_{L}^{b}\rightarrow f\bar{f}$ scattering amplitude has an asymptotic behaviour which goes as $\frac{m_f\sqrt{s}}{v^2}$ at high energies.\\
 
%This implies that at high energies the $W_{L}^{a}W_{L}^{b}\rightarrow f\bar{f}$ scattering amplitude violates perturbative unitarity and one has a strong $W_{L}^{a}W_{L}^{b}\rightarrow f\bar{f}$ scattering. In the case of non-zero fermion masses, one has that the growth of the $W_{L}^{a}W_{L}^{b}\rightarrow f\bar{f}$ scattering amplitude with $\sqrt{s}$ is a manifestation of a Strong Electroweak Symmetry Breaking.\\
%Since in the Electroweak Chiral Lagrangian formulation 
The fact that the $W_LW_L\rightarrow W_LW_L$ and $W_LW_L\rightarrow f\bar{f}$ scattering amplitudes grow at high energies as $\frac{s}{v^2}$ and $\frac{m_f\sqrt{s}}{v^2}$, respectively implies the following two possibilities:
\begin{itemize}
\item New particles should exist in order to restore unitarity well before perturbativity is lost. In this case we have a weakly coupled EWSB, possibly extrapolable to much higher energies than $4\pi v$. 
\item The $W_LW_L\rightarrow W_LW_L$ and $W_LW_L\rightarrow f\bar{f}$ scattering amplitudes grow strongly until the interaction among the four $W$'s and between two $W$'s and fermion-antifermion pair becomes non-perturbative. Nevertheless, somewhat before this to happen, some new degrees of freedom produced by the strong dynamics may emerge at the $TeV$ scale. The ultraviolet behaviour of the $W_LW_L\rightarrow W_LW_L$ and $W_LW_L\rightarrow f\bar{f}$ scattering amplitudes may be softened by the exchange of such massive composite states. In this case the appearance of new composite degrees of freedom from a strong sector could be the earliest manifestation of a strongly coupled EWSB.
\end{itemize}
It is worth to mention that the chiral formulation has the merit of isolating the problem to the sector of the Lagrangian which leads to the mass terms for the vector bosons and the fermions. Regardless of the type of dynamics ruling the EWSB mechanism, an ultraviolet completion of the EWCL given in (\ref{eq:chiralLmass}) will have to exist. The key assumption here is that the EWCL catches the main physics below the cutoff, including the properties of the new composite particles lighter than the cutoff itself.
%grows with the center of mass energy and is proportional to the fermion mass. 
%The breakdown of the perturbative unitarity for the $W_{L}^{a}W_{L}^{b}\rightarrow f\bar{f}$ scattering occurs at energies $\sqrt{s}\approx 4\pi v\approx 3~{\rm TeV}$

\subsection{Adding a composite scalar}
The simplest extension of the minimal EWCL is to add a new scalar field $h(x)$ singlet under $SU(2)_L\times SU(2)_R$. Since an elementary scalar has the hierarchy problem, a composite scalar arising from an unspecified strong dynamics is introduced so that quantum corrections to its mass are saturated at the compositeness scale. It is assumed that the Standard Model Gauge bosons are coupled to the strong sector via weak gauging: the operators involving the field strengths $W_{\mu\nu}$ and $B_{\mu\nu}$ will appear with loop suppressed coefficients, so that they can be neglected \cite{Contino:2010}. Another assumption that is made is that the Standard Model fermions are coupled to the strong sector only via the (proto)-Yukawa interactions.\\ 

Under these assumptions the most general EWSB Lagrangian has three free parameters $a$, $b$ and $c$ ~\footnote{In general $c$ will be a matrix in flavor space, but in the following it is assumed for simplicitly that it is proportional to unity in the basis in which the mass matrix is diagonal. This guarantees the absence of flavour changing neutral effects originated from the tree level exchange of $h$.} at the quadratic order in $h$ and is given by \cite{Contino:2010}:
\begin{equation}
\label{compscalarl}
\begin{split}
{\cal L}_{EWSB} =& \frac{1}{2} \left(\partial_\mu h\right)^2 - V(h) + 
\frac{v^2}{4}\left\langle D_{\mu}U\left(  D^{\mu
}U\right)  ^{\dag}\right\rangle\left( 1 + 2 a\, \frac{h}{v} + b\, \frac{h^2}{v^2}\right)\\& 
% \frac{v^2}{4} \Tr \left( D_\mu U^\dagger D_\mu U \right)\\ & 
- \frac{v}{\sqrt{2}} \sum_{i,j} \left( \bar u_L^{(i)} d_L^{(i)} \right) 
 U \left( 1+ c\, \frac{h}{v}\right)\begin{pmatrix} \lambda_{ij}^u\,  u_R^{(j)} \\[0.1cm] \lambda_{ij}^d\,  d_R^{(j)} \end{pmatrix} + h.c
%\frac{v^2}{4} \text{Tr}\left[ \left( D_\mu U \right)^\dagger \left( D_\mu \Sigma \right) \right]  \left( 1 + 2 a\, \frac{h}{v} + b\, \frac{h^2}{v^2} + \dots \right) \\ & 
%- m_\psi \, \bar\psi_L \Sigma \left( 1+ c\, \frac{h}{v} + \cdots\right) \psi_R + h.c.
%\label{compscalarl}
\end{split}
\end{equation}
where $V(h)$ is some potential, including a mass term, for $h$. As we shall see, each of these parameters controls the unitarization of a different sector of the theory. 
%It is also assumed that the scalar has a relatively low mass $m_{h}\lesssim v$ in order to successfully account for the Electroweak Precision Tests. The Lagrangian given in \ref{compscalarl} with generic parameters $a$, $b$ and $c$ gives a general parametrization of such composite Higgs theories where all other resonances have been integrated out.

\subsection{$W_LW_L\rightarrow W_LW_L$, $W_LW_L\rightarrow f\bar{f}$ and $W_LW_L\rightarrow hh$ amplitudes}
As before, the $W_{L}^{a}W_{L}^{b}\rightarrow hh$ scattering amplitude at high energies such that $\sqrt{s}>>M^2_W$ is given by:
\begin{equation}
A(W_{L}^{a}W_{L}^{b}\rightarrow hh)=-A\left(\pi^{a}\pi^{b}\rightarrow hh\right)\left[1+O\left(\frac{M_{W}}{\sqrt{s}}\right)\right].
\label{p6}
\end{equation}\\
The function $A\left(s,t,u\right)^{\pi\pi\to \pi\pi}$ receives contributions from the four pion contact interaction $\pi^4$ and from the scalar exchange $h$ and is given by:
%which comes from the derivative interaction among four Goldstones contained in the kinetic term of $U$ in (\ref{eq:chiralLmass}) is given by:
\begin{equation}
A\left(s,t,u\right)^{\pi\pi\to \pi\pi}=\lrp{1-a^2}\frac{s}{v^2}+\frac{a^2m^2_hs}{v^2\lrp{s-m^2_h}}
\end{equation}
so that the strength of the four pion scattering amplitude is controlled by the parameter $a$. For $a=1$ the exchange of the scalar unitarizes the four pions scattering amplitude and then the $W_LW_L\rightarrow W_LW_L$ scattering amplitude at high energies. In the case in which $a\ne 1$, one has a strong $W_LW_L\rightarrow W_LW_L$ scattering with violation of perturbative unitarity at energies $\sqrt{s}\approx 4\pi v/\sqrt{1-a^2}$.\\

The leading contributions to the amplitude $A\left(\pi^{a}\pi^{b}\rightarrow f\bar{f}\right)$ come from the $\pi^2f\bar{f}$ contact interaction and from the scalar exchange so that the function $A(s,t,u)^{\pi\pi\to f\bar{f}}$ is given by:
\begin{equation}
A(s,t,u)^{\pi\pi\to f\bar{f}}=\frac{m_f\lrp{1-ac}\sqrt{s}}{v^2}.
\end{equation}
Then the parameters $a$ and $c$ control the strength of the $W_LW_L\rightarrow f\bar{f}$ scattering amplitude. Perturbative unitarity is satisfied for $ac=1$.\\

On the other hand, the $\pi^{a}\pi^{b}\to hh$ scattering amplitude $\Amp\(\pi^{a}\pi^{b}\to hh \)$ receives contributions from the $\pi^{2}h^{2}$ contact interaction and from the $\pi$ and $h$ exchanges, so that the function $\Amp\(s,t,u\)^{\pi\pi\to hh}$ is given by:
\be\label{pipihh}
\Amp\(s,t,u\)^{\pi\pi\to hh}= -\f{1}{v^{2}}\(s\(b-a^{2}\)+\f{3as m_{h}^{2}}{2\(s-m_{h}^{2}\)}-2a^{2}m_{h}^{2}+\f{a^{2}m_{h}^{4}}{t}+\f{a^{2}m_{h}^{4}}{u}\)\,.
\ee \\
%\be\label{hhcontributions}
%\Amp\(\pi^{a}\pi^{b}\to hh \)= \Amp\(\pi^{a}\pi^{b}\to hh \)_{\pi^{2}h^{2}}+\Amp\(\pi^{a}\pi^{b}\to hh \)_{\pi}+ \Amp\(\pi^{a}\pi^{b}\to hh \)_{h}\,.
%\ee
This amplitude will not grow with the center of mass energy, that is, the
perturbative unitarity condition is satisfied only for $b=a^{2}$. Hence, taking all conditions at the same time, only for the choice $a=b=c=1$ the EWSB sector is weakly interacting (provided that the scalar $h$ is sufficiently light). It is not surprising that $a=b=c=1$ precisely corresponds to the Standard Model case with $h(x)$ being part, together with the $\pi$'s, of a linear Higgs doublet. Away from the unitarity point $a=b=c=1$, the scalar exchange alone will fail to fully unitarize the amplitudes for the elastic and inelastic channels of $WW$ scattering. In this case the theory will become strongly interacting at high energies. Since the Goldstone Equivalence Theorem implies that the longitudinal polarization states of $W$ and $Z$ play the role of the pions in the new strong interaction, any collider process involving the $W$ and $Z$ bosons in the initial and final states can be helpful for an experimental study of the new strong interaction. In particular discovering a Higgs-like boson and at the same time finding an excess of events in $WW\to WW$ scattering at the LHC when compared with the prediction of the Standard Model will be a signal of the growing energy behaviour of the $WW\to WW$ scattering amplitude and then an experimental manifestation of strong EWSB. Besides that, the observation of the $WW\to hh$ scattering at the LHC, which in the Standard Model has an extremely small cross section might provide an experimental evidence of composite Higgs model and strong EWSB. The advantange of the $WW\to hh$ channel with respect to the $WW\to WW$ elastic channel comes from the fact that the first is the only process providing information on the parameter $b$ and does not have pollution from transverse modes of the W \cite{Contino:2010t}.\\
%finding a  \\

\newpage
\section{Composite Vectors at the LHC}
\subsection{Chiral Lagrangian with massive spin one fields}
In this chapter we shall consider the addition to the minimal EWCL of spin-1 states, triplet under $SU(2)_{L+R}$ in analogy with the $\rho$-states of QCD. This will allows us to study the interactions of these vectors, $V_{\mu}^{a}$, with the $W$ and $Z$ in a comprehensive way. Especially in low-energy QCD studies, the heavy spin-1 states are often described by antisymmetric tensors \cite{Ecker:1988te,Ecker:1989yg}. Here we shall on the contrary make use of the more conventional Lorentz vectors, belonging to the adjoint representation of $SU(2)_{L+R}$,
\begin{equation}
V_{\mu}=\frac{1}{\sqrt{2}}\tau^{a}V_{\mu}^{a}\,,\qquad\qquad V^{\mu}\to h
{V}^{\mu}h^{\dag},
\end{equation}

%To take advantage of several studies in the QCD case\cite{}, we describe the heavy spin-1 states by antisymmetric tensors $V^{\mu\nu}$, belonging to
%the adjoint representation of $SU(2)_{L+R}$,
%\begin{equation}
%\hat{V}_{\mu\nu}=\frac{1}{\sqrt{2}}\tau^{a}V_{\mu\nu}^{a}\,,\qquad\qquad
%\hat{V}^{\mu\nu}\to h\hat{V}^{\mu\nu}h^{\dag}\,,
%\end{equation}
with $h$ defined in (\ref{eq5}).\\ 

The $SU(2)_{L}\times SU(2)_{R}$-invariant kinetic Lagrangian for the heavy
spin-1 fields is given by
\begin{equation}
\label{eq7}\mathcal{L}_{\text{kin}}^{V}=-\frac{1}{4}\left\langle \hat{V}%
^{\mu\nu}\hat{V}_{\mu\nu}\right\rangle +\frac{M_{V}^{2}}{2}\left\langle
{V}^{\mu}{V}_{\mu}\right\rangle \,.
\end{equation}
Here $\hat{V}_{\mu\nu} = \nabla_{\mu}V_{\nu}- \nabla_{\nu}V_{\mu}$ and
\begin{equation}
\label{eq8}\nabla_{\mu}{V_{\nu}}=\partial_{\mu} {V_{\nu}}+[\Gamma_{\mu
},{V_{\nu}}]\,,\qquad\Gamma_{\mu}=\frac{1}{2}\Big[u^{\dag}\left( \partial
_{\mu}-i{B}_{\mu}\right) u+u\left( \partial_{\mu}-i{W}_{\mu}\right) u^{\dag
}\Big]\,,\qquad\Gamma_{\mu}^{\dag}=-\Gamma_{\mu},%
\end{equation}
where $u$ is defined in (\ref{eq5}). Note that the covariant derivative given in the previous expression transforms homogeneously as $V_{\mu}$
itself does. The other quantity that transforms covariantly is $u_{\mu
}=u^{\dag}_{\mu}=iu^{\dag}D_{\mu}Uu^{\dag}$, which, under $SU(2)_{L+R}$, has the following transformation rule: $u_{\mu}\to hu_{\mu}h^{\dag}$.\\ 

In terms of these quantities the most general invariant terms up to a given number of vector indices is easily constructed.
%with $u=\sqrt{U}$ so that indeed $u_{\mu}\to hu_{\mu}h^{\dag}$.\\
Assuming parity invariance of the new strong interaction, the full set of
interactions up to cubic terms in the spin-1 fields is:
\begin{equation}
\mathcal{L}_{\text{int}}^{V}=\mathcal{L}_{\text{1V}}+\mathcal{L}_{\text{2V}%
}+\mathcal{L}_{\text{3V}}\,,\label{int}%
\end{equation}
where $\mathcal{L}_{\text{1V}}$, $\mathcal{L}_{\text{2V}}$ and $\mathcal{L}_{\text{3V}}$ are given by: 
%The
%$O\left(p^{2}\right) $  couplings of these heavy spin-1 fields  to the Goldstone bosons and to the Standard Model
%gauge bosons, are:%
\begin{align}
\mathcal{L}_{\text{1V}}= &  -\frac{ig_{V}}{2\sqrt{2}}\left\langle \hat{V}%
_{\mu\nu}[u^{\mu},u^{\nu}]\right\rangle -\frac{f_{V}}{2\sqrt{2}}\left\langle
\hat{V}_{\mu\nu}(u{W}^{\mu\nu}u^{\dag}+u^{\dag}{B}^{\mu\nu}u)\right\rangle
,\label{L1V}\\
& \nonumber\\
\mathcal{L}_{\text{2V}}= &  g_{1}\left\langle {V}_{\mu}{V}^{\mu}u^{\alpha
}u_{\alpha}\right\rangle +g_{2}\left\langle {V}_{\mu}u^{\alpha}{V}^{\mu
}u_{\alpha}\right\rangle +g_{3}\left\langle {V}_{\mu}{V}_{\nu}[u^{\mu},u^{\nu
}]\right\rangle +g_{4}\left\langle {V}_{\mu}{V}_{\nu}\{u^{\mu},u^{\nu
}\}\right\rangle \nonumber\\
&  +g_{5}\left\langle {V}_{\mu}\left(  u^{\mu}{V}_{\nu}u^{\nu}+u^{\nu}{V}%
_{\nu}u^{\mu}\right)  \right\rangle +ig_{6}\left\langle {V}_{\mu}{V}_{\nu
}(u{W}^{\mu\nu}u^{\dag}+u^{\dag}{B}^{\mu\nu}u)\right\rangle \,,\label{L2V}\\
& \nonumber\\
\mathcal{L}_{\text{3V}}= &  \frac{ig_{K}}{2\sqrt{2}}\left\langle \hat{V}%
_{\mu\nu}{V}^{\mu}{V}^{\nu}\right\rangle \,.\label{L3V}%
\end{align}\\
Every parameter in (\ref{int}) is dimensionless. The interactions of a single vector with the EW gauge bosons are described by $\mathcal{L}_{\text{1V}}$ and have been extensively considered in the literature. The interactions which modify the $W^2V^2$ and $WV^2$ vertex are described by $\mathcal{L}_{\text{2V}}$ and have not been considered in the literature as well as the vector self-interactions given by $\mathcal{L}_{\text{3V}}$. It can be seen that the
interactions of two pions (two longitudinal weak bosons) with the vector field $V_{\mu}$ are characterized by a coupling $g_{V}$. The interactions of the vector field $V_{\mu}$ with one longitudinal and one transverse gauge boson are characterized by the couplings $f_{V}$ and $g_{V}$. Another important fact is the mixing of the vector field $V_{\mu}$ with the Standard Model Gauge fields; this mixing is proportional to $gf_{V}$.\\ 
%This implies that the Standard Model fermions couple to the new $V$ states on% ly via the SM gauge interactions.
%
%From the total Lagrangian:
%\begin{equation}
%\mathcal{L}^{V}=\mathcal{L}_{\chi}+\mathcal{L}_{\text{kin}}^{V}+\mathcal{L}%
%_{\text{int}}^{V}\,+\mathcal{L}_{\text{contact}},\label{Ltot}%
%\end{equation}
%we leave out:

In (\ref{int}) we are not including:
\begin{itemize}
\item Operators involving 4 $V$'s, since they are not relevant to the amplitudes considered in this work.

\item Operators of dimension higher than 4, which we assume to be weighted by
inverse powers of the cutoff $\Lambda\approx3$ TeV, as suggested by naive
dimensional analysis. As such, they would contribute to the $VV$-production
amplitudes at c.o.m. energies sufficiently below $\Lambda$ by small terms
relative to the ones that we are going to compute.

\item Direct couplings between any fermion of the SM and the composite
vectors. This is plausible if the SM fermions are \textit{elementary}. The
third generation doublet could be an exception here. If this were the case,
with a large enough coupling, this would not change any of the $VV$-production
amplitudes, but might lead to a dominant decay mode of the composite vectors
into top and/or bottom quarks, rather than into $W,Z$ pairs.
\end{itemize}\hspace{1cm}\\
%The relation of $\mathcal{L}^{V}$ with the Lagrangian formulated in terms of anti-symmetric tensor fields is described in Appendix \ref{tensor}. The Lagrangian $\mathcal{L}^{V}$ corresponds to the extrapolation at highenergies of the corresponding Lagrangian used in the description of thelow-energy effects of vector meson dominance in QCD.
As we are going to see, for a consistent description of high energy WW scattering we also have to add 4-derivative terms only involving the $\pi$-fields. Their most general form is:
\begin{equation}
\mathcal{L}_{contact}=c_{1}\left\langle\left[  u^{\mu},u^{\nu}\right]  \left[  u_{\mu},u_{\nu}\right]\right\rangle+c_{2}\left\langle\left\{  u^{\mu},u^{\nu}\right\}  \left\{  u_{\mu},u_{\nu
}\right\}  \right\rangle,
\end{equation}
so that the total lagrangian will be:
\begin{equation}
\mathcal{L}^{V}=\mathcal{L}_{\chi}+\mathcal{L}_{\text{kin}}^{V}+\mathcal{L}_{\text{int}}^{V}\,+\mathcal{L}_{\text{contact}}.\label{Ltot}
\end{equation}

\subsection{Longitudinal WW scattering amplitude}

For the process $\pi^{a}\pi^{b}\rightarrow\pi^{c}\pi^{d}$ in the center of
mass frame we have:
\begin{equation}
p_{a}^{\mu}=\left(  E,0,0,p\right)  =\left(  \sqrt{\frac{s}{4}},0,0,\sqrt
{\frac{s}{4}}\right)  ,\hspace{1.5cm}\hspace{1.5cm}p_{b}^{\mu}=\left(
E,0,0,-p\right)  =\left(  \sqrt{\frac{s}{4}},0,0,-\sqrt{\frac{s}{4}}\right),
\end{equation}%
\begin{equation}
p_{c}^{\mu}=\left(  E,k\sin\theta_{CM},0,k\cos\theta_{CM}\right)  =\left(
\sqrt{\frac{s}{4}},\sqrt{\frac{s}{4}}\sin\theta_{CM},0,\sqrt{\frac{s}{4}}%
\cos\theta_{CM}\right),
\end{equation}%
\begin{equation}
p_{d}^{\mu}=\left(  E,-k\sin\theta_{CM},0,-k\cos\theta_{CM}\right)  =\left(
\sqrt{\frac{s}{4}},-\sqrt{\frac{s}{4}}\sin\theta_{CM},0,-\sqrt{\frac{s}{4}%
}\cos\theta_{CM}\right),
\end{equation}
where $\theta_{CM}$ is the scattering angle in the center of mass frame,
with:
\begin{equation}
t=-\frac{s}{2}\left(  1-\cos\theta_{CM}\right)  ,\hspace{1.5cm}\hspace
{1.5cm}u=-\frac{s}{2}\left(  1+\cos\theta_{CM}\right), \label{xa1}%
\end{equation}
and the Madelstam variables are given by:
\begin{eqnarray}
s&=&\left(  p^{a}+p^{b}\right)  ^{2}=\left(  p^{c}+p^{d}\right)  ^{2}=E_{CM}%
^{2},\hspace{2cm}t=\left(  p^{a}-p^{c}\right)  ^{2}=\left(
p^{b}-p^{d}\right)  ^{2},\notag\\
u&=&\left(  p^{a}-p^{d}\right)  ^{2}=\left(  p^{b}-p^{c}\right)  ^{2},%
\end{eqnarray}
where $E_{CM}$ is the center of mass energy.\\

The contribution due to the four point contact interaction contained in (\ref{contact0}) to the four pions scattering amplitude is:
\begin{equation}
A\left(  \pi^{a}\pi^{b}\rightarrow\pi^{c}\pi^{d}\right)  _{\pi^{4}} =\frac{s}{v^{2}}\delta^{ab}\delta^{cd}+\frac{t}{v^{2}}\delta^{ac}\delta
^{bd}+\frac{u}{v^{2}}\delta^{ad}\delta^{bc}.\label{s4}%
\end{equation}\\
On the other hand, expanding the contact interaction in $\mathcal{L}_{\text{contact}}$ to 4-th order in the $\pi$-fields, we obtain:
\begin{equation}%
%TCIMACRO{\tciLaplace}%
%BeginExpansion
\mathcal{L}%
%EndExpansion
_{1}^{\pi^{4}}   =-\frac{8c_{1}}{v^{4}}\varepsilon^{abe}\varepsilon
^{cde}g^{\mu\lambda}g^{\nu\rho}\partial_{\mu}\pi^{a}\partial_{\nu}\pi
^{b}\partial_{\lambda}\pi^{c}\partial_{\rho}\pi^{d},
\end{equation}
\begin{equation}%
%TCIMACRO{\tciLaplace}%
%BeginExpansion
\mathcal{L}%
%EndExpansion
_{2}^{\pi^{4}}  =\frac{8c_{2}}{v^{4}}\delta^{ab}\delta^{cd}g^{\mu\lambda
}g^{\nu\rho}\partial_{\mu}\pi^{a}\partial_{\nu}\pi^{b}\partial_{\lambda}%
\pi^{c}\partial_{\rho}\pi^{d}.
\end{equation}\\
From the previous expressions, the following contributions due to $%
%TCIMACRO{\tciLaplace}%
%BeginExpansion
\mathcal{L}%
%EndExpansion
_{1}^{\pi^{4}}$ and $%
%TCIMACRO{\tciLaplace}%
%BeginExpansion
\mathcal{L}%
%EndExpansion
_{2}^{\pi^{4}}$ to the function $A\left(  s,t,u\right)  $ of the expression
(\ref{la}) are obtained:
\begin{equation}
A_{\pi^{4}}^{\left(  1\right)  }\left(  s,t,u\right)  = -\frac{8c_{1}}{v^{4}}\left(  s^{2}+2ut\right),\hspace{3cm}A_{\pi^{4}}^{\left(  2\right)  }\left(  s,t,u\right)  =\frac{8c_{2}}{v^{4}}\left(
t^{2}+u^{2}\right). \label{s8}%
\end{equation}\\
The Lagrangian which describes the $\pi^{2}V$ interaction is given by
\begin{equation}%
%TCIMACRO{\tciLaplace}%
%BeginExpansion
\mathcal{L}%
%EndExpansion
^{\pi^{2}V}=\frac{g_{V}}{v^{2}}\varepsilon^{abe}\left(  g^{\mu\kappa}%
g^{\nu\eta}-g^{\mu\eta}g^{\nu\kappa}\right)  \partial_{\mu}\pi^{a}%
\partial_{\nu}\pi^{b}\partial_{\kappa}V_{\eta}^{e},\label{pa}%
\end{equation}
which implies that the contribution due to $%
%TCIMACRO{\tciLaplace}%
%BeginExpansion
\mathcal{L}%
%EndExpansion
^{\pi^{2}V}$ to the function $A\left(  s,t,u\right)  $ of the expression
(\ref{la}) is:
\begin{equation}
A_{V}\left(  s,t,u\right)  =\frac{g_{V}^{2}}{v^{4}}\left[
s^{2}+2ut+M_{V}^{2}\left(  \frac{t\left(  u-s\right)  }{t-M_{V}^{2}}%
-\frac{u\left(  s-t\right)  }{u-M_{V}^{2}}\right)  \right].
\end{equation}\\
Therefore, the function $A\left(  s,t,u\right)  $ which describes the four
pions scattering amplitude is given by:
\begin{equation}
A\left(  s,t,u\right)  =\frac{s}{v^{2}}-\frac{8c_{1}}{v^{4}}\left(
s^{2}+2ut\right)  +\frac{8c_{2}}{v^{4}}\left(  t^{2}+u^{2}\right)
+\frac{g_{V}^{2}}{v^{4}}\left[  s^{2}+2ut+M_{V}^{2}\left(  \frac{t\left(
u-s\right)  }{t-M_{V}^{2}}-\frac{u\left(  s-t\right)  }{u-M_{V}^{2}}\right)
\right]
.
\end{equation}
The cancellation of the terms which go as $\frac{s^{2}}{v^{4}}$ in the four
pions scattering amplitude is guaranteed only when:
\begin{equation}
c_{2}=0,\hspace{2cm}\hspace{2cm}c_{1}=\frac{g_{V}^{2}}{8},\label{o1}%
\end{equation}
which we shall adopt from now on. We shall come back to these relations in the following.\\

In this case, the function $A\left(  s,t,u\right)  $ takes the following
form:
\begin{equation}
A\left(  s,t,u\right) =\frac{s}{v^{2}}-\frac{G_{V}^{2}}{v^{4}}\left[  3s+M_{V}^{2}\left(
\frac{s-u}{t-M_{V}^{2}}+\frac{s-t}{u-M_{V}^{2}}\right)  \right] \label{o2}%
\end{equation}
where we have set $g_VM_V=G_V$.
\subsection{Unitarity condition on WW elastic scattering}
Imposing the condition
\begin{equation}
G_V=\frac{v}{\sqrt{3}}
\end{equation}
gives a good high energy behaviour of the WW scattering amplitude. This may be however a too strong condition. We shall be content by requiring no violation of unitarity for $\sqrt{s}$ below the cutoff $\Lambda$.\\

The partial wave coefficient $a_{0}^{0}\left(  s\right)  $ of isospin zero for the four pion scattering is given by:
\begin{equation}
a_{0}^{0}\left(  s\right)  =\frac{1}{32\pi}\int_{-1}^{1}dyA\left(  s,t\left(
y\right)  ,u\left(  y\right)  \right)  =\frac{M_{V}^{2}}{16\pi v^{2}}\left\{
x\left(  1-\frac{3G_{V}^{2}}{v^{2}}\right)  +\frac{2G_{V}^{2}}{v^{2}}\left[
\left(  2+x^{-1}\right)  \ln\left(  x+1\right)  -1\right]  \right\}
\label{s15a}%
\end{equation}
where
\begin{equation}
x=\frac{s}{M_{V}^{2}},\hspace{1.5cm}\hspace{1.5cm}y=\cos\theta_{CM}.%
\label{s16a}%
\end{equation}\\
The strongest unitarity constraint $\left\vert a_{0}^{0}\left(  s\right)
\right\vert <1$ for any energy up to $\sqrt{s}=\Lambda$ implies:
\begin{equation}
\left\vert a_{0}^{0}\left(  s=\Lambda^{2}\right)  \right\vert =\frac{M_{V}%
^{2}}{16\pi v^{2}}\left\vert \left\{  \frac{\Lambda^{2}}{M_{V}^{2}}\left(
1-\frac{3G_{V}^{2}}{v^{2}}\right)  +\frac{2G_{V}^{2}}{v^{2}}\left[  \left(
2+\frac{M_{V}^{2}}{\Lambda^{2}}\right)  \ln\left(  \frac{\Lambda^{2}}%
{M_{V}^{2}}+1\right)  -1\right]  \right\}  \right\vert <1.\label{s17a}%
\end{equation}\\
Imposing the strongest unitarity constraint up to $\Lambda=4\pi v\simeq 3$ TeV, the allowed
region in the $\left(  M_{V},G_{V}\right)  $ plane is obtained and shown in
Figure 1. 

\includegraphics[width=15cm,height=15cm,angle=0]{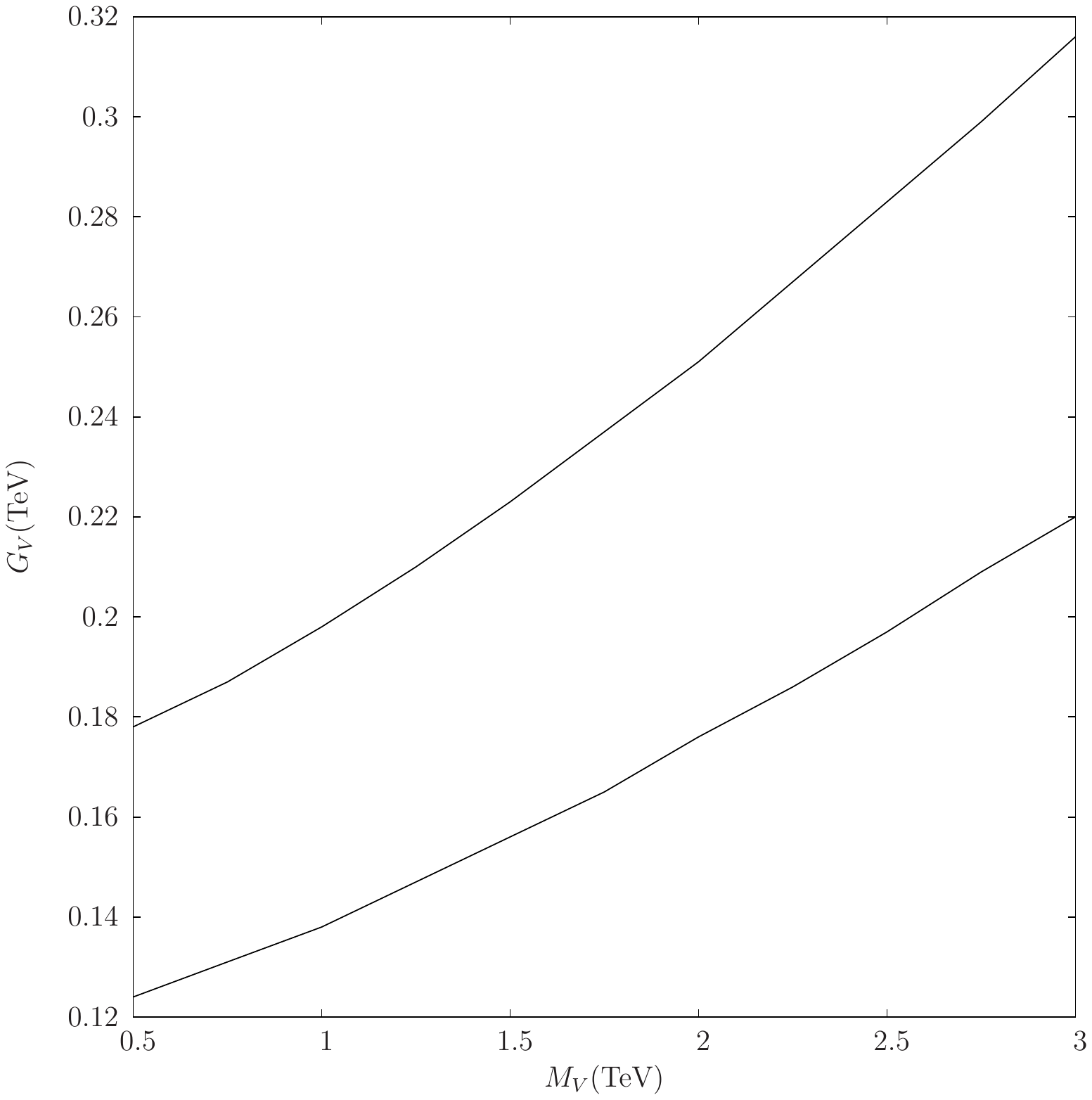}\vspace{-6cm}
\begin{center}
{\footnotesize \textbf{Figure 1: Strongest unitarity constraint in the $%
(M_{V},G_{V})$ plane for the process $\pi ^{a}\pi ^{b}\rightarrow \pi
^{c}\pi ^{d}$ at\\ $\sqrt{s}=3$ TeV.}}
\end{center}

\subsection{$W_{L} W_{L} \rightarrow V_{\lambda} V_{\lambda^{\prime}}$
helicity amplitudes}

\label{helicityamp}
\label{helicityamp}

In this Section we calculate the scattering amplitudes for two longitudinal $%
W$-bosons into a pair of heavy vectors of any helicity $\lambda,
\lambda^{\prime}= L, +, -$. To simplify the explicit formulae, we take full
advantage of $SU(2)_{L+R}$ invariance by considering the $g^{\prime}=0$
limit, so that $Z \approx W^{3}$. We also work at high energy, such that 
\begin{equation}
\sqrt{s},~\sqrt{-t},~\sqrt{-u},~M_{V} >> M_W\,,
\end{equation}
%so that all the amplitudes have the generic form
%\begin{equation}
%\mathcal{A} \approx\frac{s}{v^{2}} f(\frac{t}{s}, \frac{M_{V}^{2}}{s},
%\frac{G_{V}^{2}}{v^{2}}, \frac{K_{V} G_{V}}{M_{V}^{2}}, g_{i})
%\end{equation}
which allows us to make use of the equivalence theorem, i.e. 
\begin{equation}
\mathcal{A}(W_{L}^{a} W_{L}^{b} \rightarrow V_{\lambda}^{c} V_{\lambda
^{\prime}}^{d}) \approx -\mathcal{A}(\pi^{a} \pi^{b} \rightarrow
V_{\lambda}^{c} V_{\lambda^{\prime}}^{d})\,.
\end{equation}
This restriction will be dropped in Sections \ref{crosssec} and \ref{crosssecqq}, where we shall present numerical results, although the
limitations of the effective Lagrangian approach will remain.

There are in fact four such independent amplitudes: 
\begin{align}
& \mathcal{A}(W_{L}^{a} W_{L}^{b} \rightarrow V_{L}^{c} V_{L}^{d})\,, \\
& \mathcal{A}(W_{L}^{a} W_{L}^{b} \rightarrow V_{+}^{c} V_{-}^{d})\,, \\
& \mathcal{A}(W_{L}^{a} W_{L}^{b} \rightarrow V_{+}^{c} V_{+}^{d}) = 
\mathcal{A}(W_{L}^{a} W_{L}^{b} \rightarrow V_{-}^{c} V_{-}^{d})
\end{align}
and 
\begin{equation}
\quad\mathcal{A}(W_{L}^{a} W_{L}^{b} \rightarrow V_{L}^{c} V_{+}^{d}) = -%
\mathcal{A}(W_{L}^{a} W_{L}^{b} \rightarrow V_{L}^{c} V_{-}^{d})\,.
\end{equation}\\
By $SU(2)_{L+R}$ invariance the general form of these amplitudes is 
\begin{equation}
\mathcal{A}(W_{L}^{a} W_{L}^{b} \rightarrow V_{\lambda}^{c} V_{\lambda
^{\prime}}^{d}) = \mathcal{A}_{\lambda\lambda^{\prime}}(s, t, u) \delta
^{ab}\delta^{cd} + \mathcal{B}_{\lambda\lambda^{\prime}}(s, t, u) \delta
^{ac}\delta^{bd} + \mathcal{C}_{\lambda\lambda^{\prime}}(s, t, u) \delta
^{ad}\delta^{bc}\,,  \label{invampl}
\end{equation}
where, by Bose symmetry, it is simple to prove that 
\begin{equation}
\mathcal{A}_{\lambda\lambda^{\prime}}(s, t, u)= \mathcal{A}_{\lambda
\lambda^{\prime}}(s, u, t)~ {\text{and}}~ \mathcal{C}_{\lambda\lambda^{%
\prime}}(s, t, u)= \mathcal{B}_{\lambda\lambda^{\prime}}(s, u, t)~{\text{for}%
}~ \lambda\lambda ^{\prime}= LL, + -, ++\,,
\end{equation}
whereas 
\begin{equation}
\mathcal{A}_{L+}(s, t, u)= - \mathcal{A}_{L+}(s, u, t)~ {\text{and}}~ 
\mathcal{C}_{L+}(s, t, u)= -\mathcal{B}_{L+}(s, u, t)\,.
\end{equation}\\
These amplitudes receive contributions from:

i) contact interactions, $\pi^2 V^2$, contained in $\mathcal{L}_{\text{kin}%
}^V$ and proportional to unity (with an overall $1/v^2$ factored out) or
contained in $\mathcal{L}_{2V}$ and proportional to $g_i, i=1,\dots, 5$;

ii) one-$\pi$ exchange, proportional to $g_V^2$, contained in $\mathcal{L}%
_{1V}$;

iii) one-$V$ exchange, proportional to $g_V g_K$, with $g_V$ contained in $%
\mathcal{L}_{1V}$ and $g_K$ in $\mathcal{L}_{3V}$.\\

For ease of the reading, we keep first only the contributions with $\mathcal{%
L}_{2V}$ and $\mathcal{L}_{3V}$ set to zero, so that\footnote{%
In all these functions the variables are in the order $(s, t, u)$ and are
left understood.}:
\begin{itemize}
\item For $\lambda\lambda^{\prime}=LL$ 
\begin{align}
& \mathcal{A}_{LL}^{1V}=-\frac{G_{V}^{2}s}{v^{4}\left( s-4M_{V}^{2}\right) }%
\left[ \frac{\left( t+M_{V}^{2}\right) ^{2}}{t}+\frac{\left(
u+M_{V}^{2}\right) ^{2}}{u}\right]\,, \\
& \mathcal{B}_{LL}^{1V}=\frac{u-t}{2v^{2}}+\frac{G_{V}^{2}s\left(
u+M_{V}^{2}\right) ^{2}}{v^{4}u\left( s-4M_{V}^{2}\right) }\,.
\end{align}

\item For $\lambda\lambda^{\prime}=+-$ 
\begin{align}
& \mathcal{A}_{+-}^{1V}=\frac{2G_{V}^{2}M_{V}^{2}\left( t+u\right) \left(
tu-M_{V}^{4}\right) }{v^{4}tu\left( s-4M_{V}^{2}\right) }\,, \\
& \mathcal{B}_{+-}^{1V}=\frac{2G_{V}^{2}M_{V}^{2}\left( M_{V}^{4}-tu\right) 
}{uv^{4}\left( s-4M_{V}^{2}\right) }\,.
\end{align}

\item For $\lambda\lambda^{\prime}=++$ 
\begin{align}
& \mathcal{A}_{++}^{1V}=\frac{2G_{V}^{2}M_{V}^{2}\left( t+u\right) \left(
M_{V}^{4}-tu\right) }{v^{4}tu\left( s-4M_{V}^{2}\right) }\,, \\
& \mathcal{B}_{++}^{1V}=\frac{\left( t-u\right) }{2v^{2}}-\frac {%
2G_{V}^{2}M_{V}^{2}\left( M_{V}^{4}-tu\right) }{uv^{4}\left(
s-4M_{V}^{2}\right) }\,.
\end{align}

\item For $\lambda\lambda^{\prime}=L+$ 
\begin{align}
& \mathcal{A}_{L+}^{1V}=\frac{\sqrt{2}G_{V}^{2}M_{V}^{3}\left( t-u\right) 
\sqrt{s\left( tu-M_{V}^{4}\right) }}{v^{4}tu\left( s-4M_{V}^{2}\right)}\,, \\
& \mathcal{B}_{L+}^{1V}=-\frac{\sqrt{s\left( tu-M_{V}^{4}\right) }\left\{
v^{2}su+4M_{V}^{2}\left[ G_{V}^{2}\left( M_{V}^{2}+u\right) -v^{2}u\right]
\right\} }{2\sqrt{2}uv^{4}M_{V}\left( s-4M_{V}^{2}\right)}\,.
\end{align}
\end{itemize}

Here and in the following, we set 
\begin{equation}
G_V \equiv g_V M_V\,,~ F_V \equiv f_V M_V\,,
\end{equation}
adopting a notation familiar in the description of spin-1 states by
anti-symmetric Lorenz tensor fields.\\ 
%As discussed in Appendix \ref{tensor} these same amplitudes would indeed be obtained using anti-symmetric tensors instead of Lorentz vectors to describe the spin-1 states.

Switching on $\mathcal{L}_{2V}$ and $\mathcal{L}_{3V}$ gives an extra
contribution to the various amplitudes:

\begin{itemize}
\item For $\lambda \lambda ^{\prime }=LL$ 
\begin{align}
& \Delta \mathcal{A}_{LL}=\left( g_{1}-g_{2}\right) \frac{s\left(
s-2M_{V}^{2}\right) }{v^{2}M_{V}^{2}}+\left( g_{4}-g_{5}\right) \frac{s\left[
2M_{V}^{2}\left( 3M_{V}^{2}-s\right) +t^{2}+u^{2}\right] }{%
v^{2}M_{V}^{2}\left( s-4M_{V}^{2}\right) }\,, \\
& \Delta \mathcal{B}_{LL}=g_{2}\frac{s\left( s-2M_{V}^{2}\right) }{%
v^{2}M_{V}^{2}}+\frac{s\left( t-u\right) }{v^{2}M_{V}^{2}}\left( g_{3}+\frac{%
g_{K}g_{V}}{4}\frac{s+2M_{V}^{2}}{s-M_{V}^{2}}\right) +g_{5}\frac{s\left[
2M_{V}^{2}\left( 3M_{V}^{2}-s\right) +t^{2}+u^{2}\right] }{%
v^{2}M_{V}^{2}\left( s-4M_{V}^{2}\right) }.
\end{align}

\item For $\lambda \lambda ^{\prime }=+-$ 
\begin{align}
& \Delta \mathcal{A}_{+-}=4\left( g_{4}-g_{5}\right) \frac{\left(
M_{V}^{4}-tu\right) }{v^{2}\left( s-4M_{V}^{2}\right) }\,, \\
& \Delta \mathcal{B}_{+-}=4g_{5}\frac{\left( M_{V}^{4}-tu\right) }{%
v^{2}\left( s-4M_{V}^{2}\right) }\,.
\end{align}

\item For $\lambda \lambda ^{\prime }=++$ 
\begin{align}
& \Delta \mathcal{A}_{++}=2\left( g_{1}-g_{2}\right) \frac{s}{v^{2}}+4\left(
g_{4}-g_{5}\right) \frac{\left( tu-M_{V}^{4}\right) }{v^{2}\left(
s-4M_{V}^{2}\right) }\,, \\
& \Delta \mathcal{B}_{++}=2g_{2}\frac{s}{v^{2}}+\frac{4g_{5}\left(
tu-M_{V}^{4}\right) }{v^{2}\left( s-4M_{V}^{2}\right) }-\frac{%
g_{K}g_{V}s(t-u)}{2v^{2}\left( s-M_{V}^{2}\right) }\,.
\end{align}

\item For $\lambda \lambda ^{\prime }=L+$ 
\begin{align}
& \Delta \mathcal{A}_{L+}=\left( g_{4}-g_{5}\right) \frac{\left( t-u\right) 
\sqrt{2s\left( tu-M_{V}^{4}\right) }}{v^{2}M_{V}\left( s-4M_{V}^{2}\right) }%
\,, \\
& \Delta \mathcal{B}_{L+}=\frac{\sqrt{2s\left( tu-M_{V}^{4}\right) }}{%
v^{2}M_{V}}\left[ g_{5}\frac{t-u}{s-4M_{V}^{2}}+\left( g_{3}+\frac{g_{K}g_{V}%
}{2}\frac{s}{s-M_{V}^{2}}\right) \right] \,.
\end{align}
\end{itemize}

\subsection{Asymptotic behaviour of the $W_{L} W_{L} \rightarrow V_{\protect%
\lambda} V_{\protect\lambda^{\prime}}$ amplitudes}

\label{asymptot}

For arbitrary values of the parameters all these amplitudes grow at least as 
$s/v^2$ and some as $s^2/(v^2M_V^2)$ or as $s^{3/2}/(v^2 M_V)$. As readily
seen from these equations, there is on the other hand a unique choice of the
various parameters that makes all these amplitudes growing at most like $%
s/v^2$, i.e. 
\begin{equation}
g_V g_K= 1,~~g_3=-\frac{1}{4},~~ g_1 = g_2 =g_4 = g_5 = 0,  \label{special}
\end{equation}
whereas $f_V$ and $g_6$ are irrelevant. With this choice of parameters the
various helicity amplitudes simplify to

\begin{itemize}
\item For $\lambda\lambda^{\prime}=LL$%
\begin{align}
& \mathcal{A}_{LL}^{\text{gauge}}=-\frac{G_{V}^{2}s}{v^{4}\left(
s-4M_{V}^{2}\right) }\left[ \frac{\left( t+M_{V}^{2}\right) ^{2}}{t}+\frac{%
\left( u+M_{V}^{2}\right) ^{2}}{u}\right]\,,  \label{p1a} \\
& \mathcal{B}_{LL}^{\text{gauge}}=\frac{u-t}{2v^{2}}+\frac{G_{V}^{2}s\left(
u+M_{V}^{2}\right) ^{2}}{v^{4}u\left( s-4M_{V}^{2}\right) }-\frac{3s(u-t)}{%
4v^{2}\left( s-M_{V}^{2}\right) }\,.
\end{align}

\item For $\lambda\lambda^{\prime}=+-$%
\begin{align}
& \mathcal{A}_{+-}^{\text{gauge}}=\frac{2G_{V}^{2}M_{V}^{2}\left( t+u\right)
\left( tu-M_{V}^{4}\right) }{v^{4}tu\left( s-4M_{V}^{2}\right) }\,, \\
& \mathcal{B}_{+-}^{\text{gauge}}=\frac{2G_{V}^{2}M_{V}^{2}\left(
M_{V}^{4}-tu\right) }{uv^{4}\left( s-4M_{V}^{2}\right) }\,.  \label{p6}
\end{align}

\item For $\lambda\lambda^{\prime}=++$ 
\begin{align}
& \mathcal{A}_{++}^{\text{gauge}}=\frac{2G_{V}^{2}M_{V}^{2}\left( t+u\right)
\left( M_{V}^{4}-tu\right) }{v^{4}tu\left( s-4M_{V}^{2}\right) }\,,
\label{p3} \\
& \mathcal{B}_{++}^{\text{gauge}}=-\frac{M_{V}^{2}\left( t-u\right) }{%
2v^{2}\left( s-M_{V}^{2}\right) }-\frac{2G_{V}^{2}M_{V}^{2}\left(
M_{V}^{4}-tu\right) }{uv^{4}\left( s-4M_{V}^{2}\right) }\,.  \label{p4}
\end{align}

\item For $\lambda\lambda^{\prime}=L+$ 
\begin{align}
& \mathcal{A}_{L+}^{\text{gauge}}=\frac{\sqrt{2}G_{V}^{2}M_{V}^{3}\left(
t-u\right) \sqrt{s\left( tu-M_{V}^{4}\right) }}{v^{4}tu\left(
s-4M_{V}^{2}\right) }\,,  \label{p7} \\
& \mathcal{B}_{L+}^{\text{gauge}}=-\frac{\sqrt{2}G_{V}^{2}M_{V}\left(
M_{V}^{2}+u\right) \sqrt{s\left( tu-M_{V}^{4}\right) }}{uv^{4}\left(
s-4M_{V}^{2}\right) }+\frac{M_{V}\sqrt{s\left( tu-M_{V}^{4}\right) }}{\sqrt{2%
}v^{2}\left( s-M_{V}^{2}\right) }\,.  \label{p8}
\end{align}
\end{itemize}
\hspace{2cm}\\
We show in Section \ref{hidden-gauge} that the relations (\ref{special}),
and so the special form of the $W_{L} W_{L} \rightarrow V_{\lambda}
V_{\lambda^{\prime}}$ helicity amplitudes, arise in a minimal {gauge} model
for the vector $V_\mu$. In the generic framework considered here, some
deviations from (\ref{special}) may occur. In such a case the asymptotic
behaviour of the various amplitudes will have to be improved, e.g., by the
occurrence of heavier composite states, vectors and/or scalars, with
appropriate couplings. Note in any event that, even sticking to the
relations (\ref{special}), the amplitudes for longitudinally-polarized
vectors grow as $s/v^2$ for any value of $G_V^2$.

\subsection{Drell--Yan production amplitudes}

\label{DYampl}

At the parton level there are four Drell--Yan production amplitudes, related
to each other by $SU(2)$- invariance (in the $g^\prime$ limit, as usual): 
\begin{equation}
|\mathcal{A}(u\bar{d}\rightarrow V^+ V^0)| = |\mathcal{A}(d\bar{u}%
\rightarrow V^- V^0)|= \sqrt{2} |\mathcal{A}(u\bar{u}\rightarrow V^+ V^-)|= 
\sqrt{2}|\mathcal{A}(d\bar{d}\rightarrow V^+ V^-)|.
\end{equation}\\
They receive contributions from: i) $W (Z)$-exchange diagrams, with the $W
(Z)$ coupled to a pair of composite vectors either through their covariant
kinetic term, $\mathcal{L}^V_{\text{kin}}$, or via $g_6$ in $\mathcal{L}%
_{2V} $; ii) light-heavy vector mixing diagrams proportional to $f_V g_K$
with these couplings contained in $\mathcal{L}_{1V}$ and $\mathcal{L}_{3V}$.
Their modulus squared, summed over the polarizations of the final-state
vectors and averaged over colour and polarization of the initial fermions,
can be written as 
\begin{equation}
<|\mathcal{A}(u\bar{d}\rightarrow V^+ V^0)|^2> = \frac{g^4}{1536 M_V^6 s^2
(s-M_V^2)^2} F(s, t-u, M_V^2),  \label{DYsquared}
\end{equation}
with $F$ organized in different powers of $s$: 
\begin{equation}
F(s, t-u, M_V^2) = F^{(6)}(s, t-u, M_V^2) + F^{(5)}(s, t-u, M_V^2) +
F^{(\leq 4)}(s, t-u, M_V^2)\,,
\end{equation}
where 
\begin{align}
& F^{(6)}= (g_K f_V - 4 g_6)^2 M_V^2 s^4[s^2-(t-u)^2], \\
& F^{\left( 5\right) } = 4M_{V}^{4}s^{3}\left\{ \left(
g_{K}f_{V}-4g_{6}\right) {}^{2}\left[ 2s^{2}+(t-u)^{2}\right] \right.  \notag
\\
&\quad +\left. \left( g_{K}f_{V}-4g_{6}\right) {}\left[ 2\left(
7g_{6}-3\right) s^{2}+2\left( g_{6}-1\right) (t-u)^{2}\right] +2\left(
1-2g_{6}\right) ^{2}\left[ s^{2}+(t-u)^{2}\right] \right\},  \label{z7a} \\
& F^{(\leq 4)} = 4M_{V}^{6}\left\{ -3s^{2}f_{V}^{2}g_{K}^{2} \left[
3s^{2}+(t-u)^{2}+4M_{V}^{2}s\right] -4M_{V}^{4}\left[ \left( 8g_{6}\left(
g_{6}+2\right) -25\right) s^{2}+3(t-u)^{2}\right] \right.  \notag \\
&\quad+\left. 2f_{V}g_{K}s\left[ s\left\{ \left( 26g_{6}+9\right)
s^{2}+\left( 2g_{6}+7\right) (t-u)^{2}\right\} -6M_{V}^{2}\left[ \left(
4g_{6}-3\right) s^{2}+(t-u)^{2}\right] -24sM_{V}^{4}\right] \right.  \notag
\\
&\quad+\left. 2M_{V}^{2}s\left[ \left( 28g_{6}^{2}+9\left( 8g_{6}-3\right)
\right) s^{2}+\left( 4g_{6}^{2}+13\right) (t-u)^{2}\right] \right.  \notag \\
&\quad-\left. 4s^{2}\left[ 3g_{6}\left( g_{6}+8\right) s^{2}+\left(
5g_{6}^{2}+4\right) (t-u)^{2}\right] -48M_{V}^{6}s\right\}.  \label{z7b}
\end{align}
$F^{(5)}$ is written in such a way as to make evident what controls its
high-energy behaviour after the dominant $F^{(6)}$ is set to zero by taking $%
g_K f_V = 4 g_6$. In general, these amplitudes squared grow at high energy
as $(s/M_V^2)^2$, which is turned to a constant behaviour for 
\begin{equation}
g_K f_V= 2, ~~ g_6 = \frac{1}{2}.
\end{equation}\\
In this special case the function $F$ in eq. (\ref{DYsquared}) acquires the
form 
\begin{equation}
F^{\text{gauge}} =4M_{V}^{6}\left\{ s^{2}\left[ s^{2}-\left( t-u\right) ^{2}%
\right] +4M_{V}^{2}s\left[ 2s^{2}+(t-u)^{2}\right] -12M_{V}^{4}\left[
3s^{2}+(t-u)^{2}\right] -48M_{V}^{6}s\right\}\,.  \label{z7}
\end{equation}

\subsection{\textit{Composite} versus \textit{gauge} models}

\label{hidden-gauge}

Before studying the physical consequences for the LHC of the amplitudes
calculated in the previous Sections, we consider the connection between a 
\textit{composite} vector, as discussed so far, and a gauge vector of a
spontaneously broken symmetry \cite{Barbieri:2008b, Ecker:1989yg}. For
concreteness we take a gauge theory based on $G=SU(2)_{L}\times
SU(2)_{R}\times SU(2)^{N}$ broken to the diagonal subgroup $%
H=SU(2)_{L+R+\ldots}$ by a generic non-linear $\sigma$-model of the form 
\begin{equation}
\mathcal{L}_{\chi}=\sum_{I,J}v_{IJ}^{2}\langle
D_{\mu}\Sigma_{IJ}(D^{\mu}\Sigma_{IJ})^{\dagger}\rangle~,\qquad\Sigma_{IJ}%
\rightarrow g_{I}\Sigma _{IJ}g_{J}^{\dagger}~,  \label{eq:one}
\end{equation}
where $g_{I,J}$ are elements of the various $SU(2)$ and $D_{\mu}$ are
covariant derivatives of $G$. Both the gauge couplings of the various $SU(2)$
groups and $\mathcal{L}_{\chi}$ are assumed to conserve parity. This \textit{%
gauge} model includes as special cases or approximates via deconstruction
many of the models in the literature~\cite%
{Csaki:2003,Casalbuoni:1985,Nomura,Barbieri:2003pr,Foadi:2003xa,Georgi:2004iy}. The connection between a gauge model and a
composite model for the spin-1 fields is best seen at the Lagrangian level
by a suitable field redefinition, as we now show.\\

For the clarity of exposition let us first consider the simplest $N=1$ case,
based on $SU(2)_L\times SU(2)_C \times SU(2)_R$, i.e. on the Lagrangian 
\begin{equation}
\mathcal{L}^{\text{gauge}}_V= \mathcal{L}_{\chi}^{\text{gauge}} -\frac{1}{2
g_C^2}\left\langle v_{\mu\nu} v^{\mu\nu}\right\rangle -\frac{1}{2 g^2}%
\left\langle W_{\mu\nu} W^{\mu\nu}\right\rangle -\frac{1}{2 g^{\prime 2}}
\left\langle B_{\mu\nu} B^{\mu\nu}\right\rangle \,,  \label{gaugeL}
\end{equation}
where 
\begin{equation}
v_\mu = \frac{g_C}{2} v_\mu^a \tau^a
\end{equation}
is the $SU(2)_C$-gauge vector and the symmetry-breaking Lagrangian is
described by 
\begin{equation}
\mathcal{L}_\chi^{\text{gauge}}=\frac{v^{2}}{2} \left\langle
D_{\mu}\Sigma_{RC}\left(D^{\mu}\Sigma_{RC}\right)^{\dag}\right\rangle +\frac{%
v^{2}}{2} \left\langle
D_{\mu}\Sigma_{CL}\left(D^{\mu}\Sigma_{CL}\right)^{\dag}\right\rangle.
\label{symm-break}
\end{equation}\\
Denoting collectively the three gauge vectors by 
\begin{equation}
v_\mu^I = (W_\mu, v_\mu, B_\mu),~~ I = (L, C, R),
\end{equation}
one has for the two bi-fundamental scalars $\Sigma_{IJ}$ 
\begin{equation}
D_\mu \Sigma_{IJ} = \partial_\mu \Sigma_{IJ} -i v_\mu^I \Sigma_{IJ} +i
\Sigma_{IJ} v_\mu^J.
\end{equation}
The $\Sigma_{IJ}$ can be put in the form $\Sigma_{IJ} = \sigma_I
\sigma_J^\dag$, where $\sigma_I$ are the elements of $SU(2)_I/H$,
transforming under the full $SU(2)_L\times SU(2)_C \times SU(2)_R$ as $%
\sigma_I \rightarrow g_I \sigma_I h^\dag$.\\

As the result of a gauge transformation 
\begin{equation}
v_\mu^I \rightarrow \sigma_I^{\dag} v_\mu^I \sigma_I  +i \sigma_I ^\dag
\partial_\mu \sigma_I \equiv \Omega_\mu^I,~~ \Sigma_{IJ} \rightarrow
\sigma_I^\dag \Sigma_{IJ} \sigma_J =1,
\end{equation}
the symmetry-breaking Lagrangian reduces to 
\begin{equation}
\mathcal{L}_\chi^{\text{gauge}}=\frac{v^{2}}{2} \left\langle (\Omega_\mu^R
-\Omega_\mu^C)^2 \right\rangle +\frac{v^{2}}{2} \left\langle (\Omega_\mu^L
-\Omega_\mu^C)^2 \right\rangle,  \label{new-symm-break}
\end{equation}
or, after the gauge fixing $\sigma_R =\sigma_L^+ \equiv u$ and $\sigma_C=1$,
to 
\begin{equation}
\mathcal{L}_\chi^{\text{gauge}}={v^{2}} \left\langle (v_\mu -i\Gamma_\mu)^2
\right\rangle +\frac{v^{2}}{4} \left\langle u_\mu^2 \right\rangle,
\label{new-new-symm-break}
\end{equation}
where 
\begin{equation}
u_\mu = \Omega_\mu^R - \Omega_\mu^L,~~~ \Gamma_\mu = \frac{1}{2i}
(\Omega_\mu^R + \Omega_\mu^L)
\end{equation}
coincide with the same vectors defined in Section 2.\\

We can finally make contact with the Lagrangian (\ref{Ltot}) by setting 
\begin{equation}
v_\mu = V_\mu + i \Gamma_\mu
\end{equation}
and by use of the identity \cite{Ecker:1989yg} 
\begin{equation}
v_{\mu\nu} = \hat{V}_{\mu\nu} -i [V_\mu, V_\nu] +\frac{i}{4} [u_\mu, u_\nu]
+ \frac{1}{2}(u{W}_{\mu\nu}u^{\dag}+u^{\dag}{B}_{\mu\nu}u).
\end{equation}\\
With the further replacement $V_\mu \rightarrow g_C/\sqrt{2} V_\mu$, $%
\mathcal{L}_V^{\text{gauge}}$ coincides as anticipated with $\mathcal{L}^V$
in (\ref{Ltot}) for 
\begin{equation}
g_C = \frac{1}{2 g_V}
\end{equation}
in the special case of (\ref{special}) and $g_6=1/2, f_V = 2 g_V, M_V = g_K
v/2$ (or $G_V = v/2$).

\subsubsection{More than a single gauge vector}

To discuss the case of more than one vector, i.e. $N>1$, one decomposes the
vectors associated to $SU(2)^N$ with respect to parity as 
\begin{equation}
\Omega_i^\mu = v_i^\mu + a_i^\mu,~~\Omega_{P(i)}^\mu = v_i^\mu - a_i^\mu,
~~i=1,\dots,N,
\end{equation}
so that under $SU(2)_L\times SU(2)_R$ 
\begin{equation}
v_{i}^{\mu} \rightarrow hv_{i}^{\mu}h^{\dagger}+ih\partial^{\mu}
h^{\dagger},\quad a_{i}^{\mu}\rightarrow ha_{i}^{\mu}h^{\dagger}.
\label{gaugeinv}
\end{equation}\\
In terms of these fields the gauge Lagrangian becomes 
\begin{equation}
\mathcal{L}_{\mathrm{\text{gauge}}}=\mathcal{L}_{\text{gauge,SM}}-\sum_{i}%
\frac{1}{2g_{i}^{2}}\left[ \langle\left( v_{i}^{\mu\nu}-i[a_{i}^{\mu
},a_{i}^{\nu}]\right) ^{2}\rangle+\langle\left(
D_{V}^{\mu}a_{i}^{\nu}-D_{V}^{\nu}a_{i}^{\mu}\right) ^{2}\rangle\right] ~,
\label{eq:Lgaugef}
\end{equation}
where $v_{i}^{\mu\nu}$ are the usual field strengths and 
\begin{equation}
D_{V}^{\mu}a_i^{\nu}=\partial^{\mu}a_i^{\nu}-i[v_i^{\mu},a_i^{\nu}].
\end{equation}\\
At the same time, as a generalization of eq. (\ref{new-new-symm-break}) in
the $N=1$ case, the symmetry-breaking Lagrangian will be the sum of two
separated quadratic forms in the parity-even and parity-odd fields of the
type 
\begin{equation}
\mathcal{L}_\chi^{\text{gauge}}=\mathcal{L}_m^V(v_i^\mu -i\Gamma^\mu)+ 
\mathcal{L}_m^A(u^\mu, a_i^\mu) \,.  \label{multi-symm-break}
\end{equation}
The dependence of $\mathcal{L}_m^V$ on the variables $v_i^\mu -i\Gamma^\mu$
follows from (\ref{gaugeinv}).\\

Concentrating on the parity-even fields only, by setting 
\begin{equation}
v_i^\mu = V_i^\mu + i \Gamma^{\mu}
\end{equation}
and by the replacements $V_i^\mu \rightarrow g_i/\sqrt{2} V_i^\mu$, the
Lagrangian of the $SU(2)_{L}\times SU(2)_{R}\times SU(2)^{N}$ model,
restricted to the parity-even vectors, becomes a diagonal sum of $\mathcal{L}%
^{V_i}$, each with $g_1=g_2=g_4=g_5=0, g_3=- 1/4, g_6=1/2$ and $%
g_{V_i}=f_{V_i}/2=1/g_{K_i}$, except that the $V_i^\mu$ are not mass
eigenstates. Going to the mass-eigenstate basis maintains all the couplings
quadratic in the $V_i^\mu$ unaltered as well as the relation $f_V=2 g_V$ for
the individual mass-eingenstate vectors. On the other hand, the trilinear
couplings $g_{K_i}$ get spread among the mass eingenstates (still called $%
V_i^\mu$), so that 
\begin{equation}
\mathcal{L}_{\text{3V}}= \frac{i\hat{g}_K^{lmn}}{2\sqrt{2}} \left\langle 
\hat{V}_{\mu\nu}^l {V}^{\mu}_m {V}^{\nu}_n\right\rangle.  \label{L3Vgen}
\end{equation}\\
Picking up the lightest vector only, $i=1$, this implies $\hat{g}_K^{111} 
\hat{g}_{V_1}\neq 1$, where the \textit{hat} denotes the couplings of the
physical mass eigenstates. By the orthogonality of the rotation matrix that
brings to the mass basis, it is easy to prove, however, the following sum
rule over the full set of vectors\footnote{For related sum rules, see \cite{SekharChivukula:2008mj}}
\begin{equation}
\Sigma_i \hat{g}_{V_i} \hat{g}_K^{inn} = \frac{1}{2} \Sigma_i \hat{f}_{V_i} 
\hat{g}_K^{inn} = 1  \label{sumrule}
\end{equation}
for any fixed $n$. This ensures that the asymptotic behaviour of the
amplitudes studied above would not be worse than in the case of a single
gauge vector, but only at $s > M_{V_i}^2$ for any $i$.

\subsection{Pair production cross sections by vector boson fusion}

\label{crosssec}

\begin{figure}[ptb]
\begin{minipage}[b]{8.2cm}
\centering
\includegraphics[width=8cm]{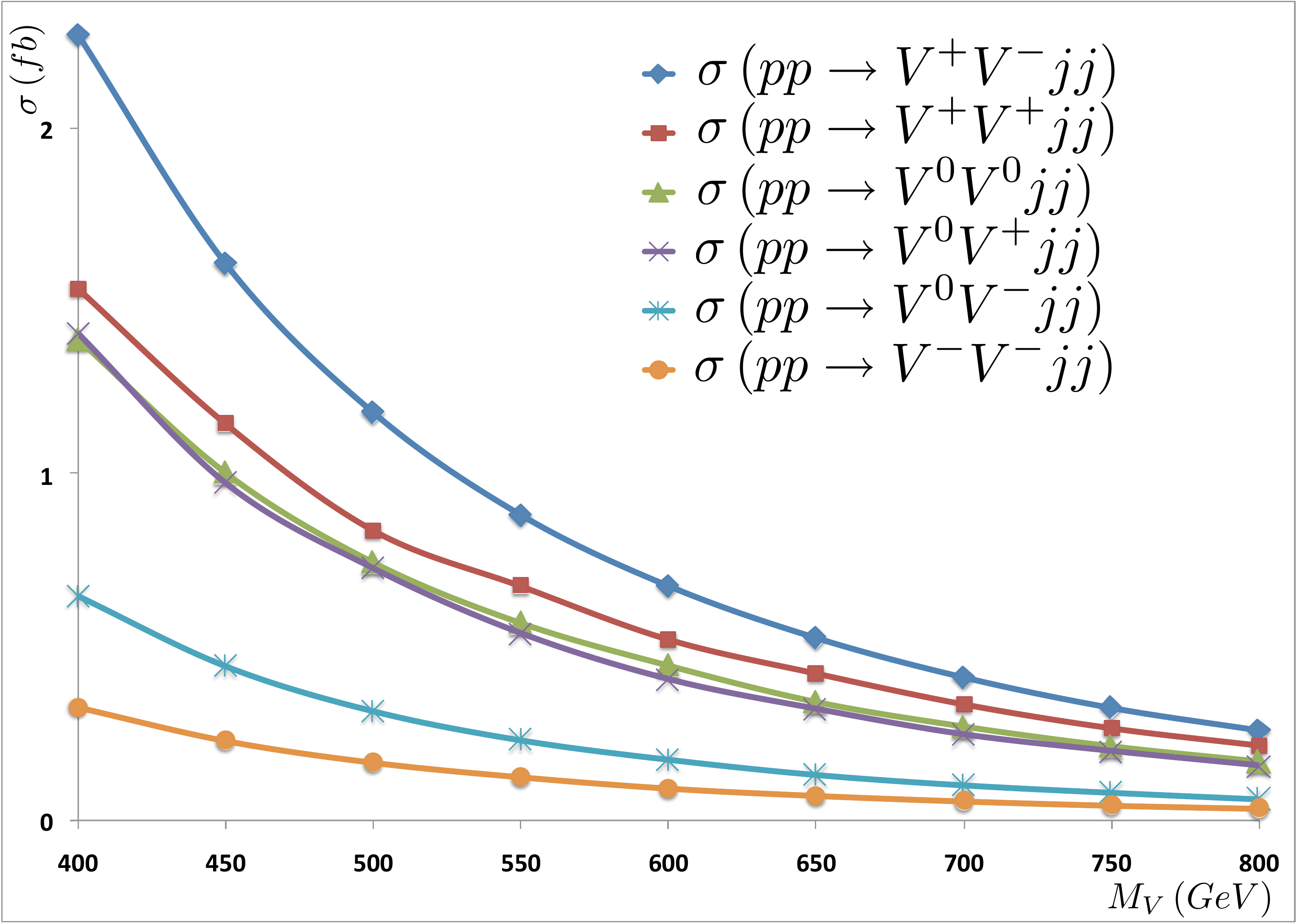}
	\\\vspace{4mm} \footnotesize{(\ref*{Fig_VBF}.a)}
\end{minipage}
\ \hspace{2mm} \hspace{3mm} \ \begin{minipage}[b]{8.5cm}
\centering
\includegraphics[width=8cm]{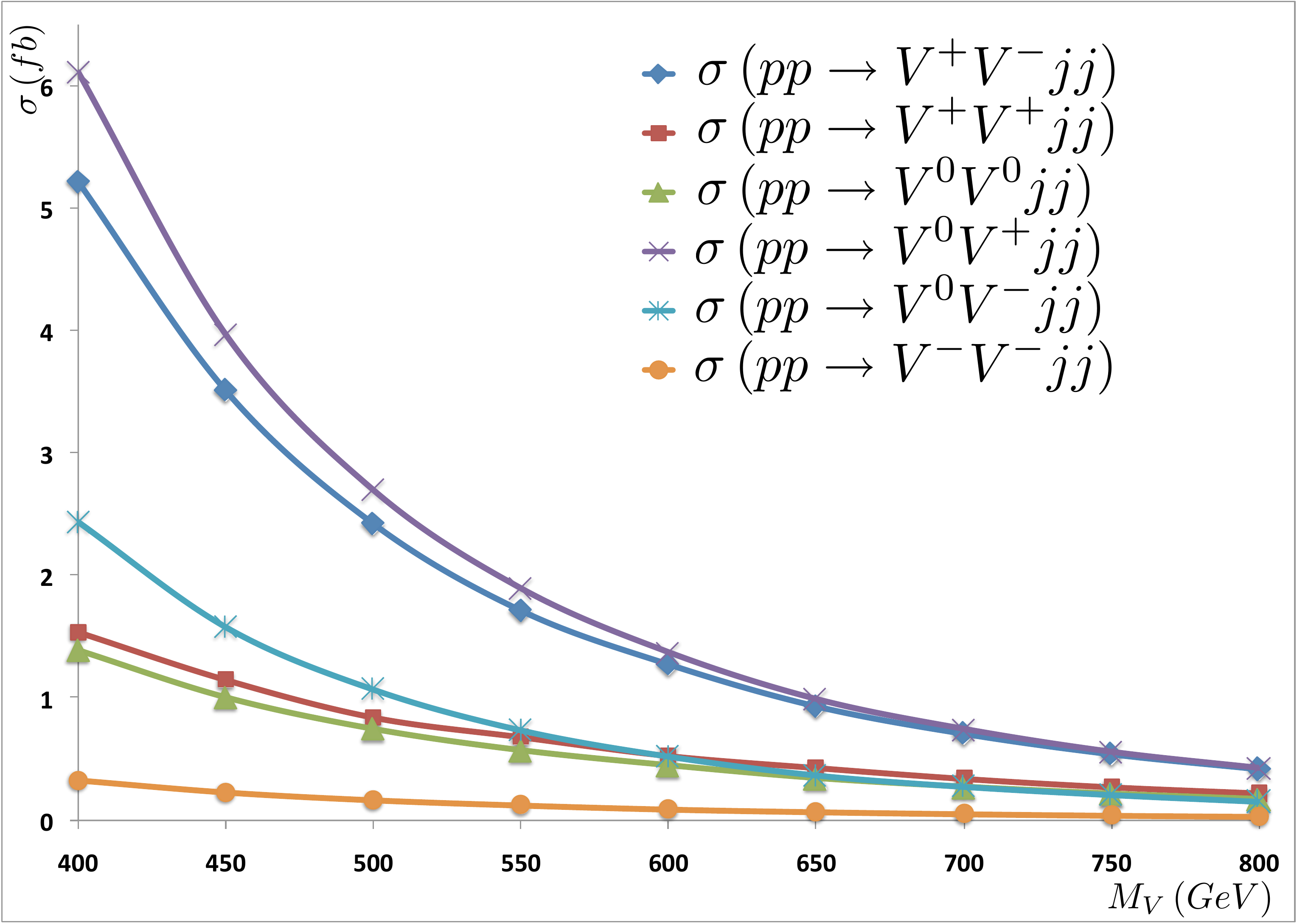}
	\\\vspace{4mm} \footnotesize{(\ref*{Fig_VBF}.b)}
\end{minipage}
\caption{Total cross sections for pair production of heavy vectors via vector
boson fusion in a gauge model (\ref*{Fig_VBF}.a) and a composite model
(\ref*{Fig_VBF}.b) as functions of the heavy vectors masses. See text for the
choice of parameters and acceptance cuts.}%
\label{Fig_VBF}%
\end{figure}

In this Section we compute the LHC production cross section at $\sqrt{S}= 14$
TeV from VBF of two heavy vectors in the different charge configurations
\begin{align}
&  pp \rightarrow W^{+}W^{-}, ZZ, \gamma\gamma, \gamma Z + qq \rightarrow
V^{+}V^{-} + qq ~(\rightarrow W^{+} Z~W^{-}Z + qq),\\
&  pp \rightarrow W^{+}W^{-}, ZZ + qq \rightarrow V^{0}V^{0} + qq
~(\rightarrow W^{+}W^{-}W^{+}W^{-} + qq),\\
&  pp \rightarrow W^{\pm}W^{\pm}+ qq \rightarrow V^{\pm}V^{\pm}+ qq
~(\rightarrow W^{\pm}Z~W^{\pm}Z + qq),\\
&  pp \rightarrow W^{\pm}Z, W^{\pm}\gamma+ qq \rightarrow V^{\pm}V^{0} + qq
~(\rightarrow W^{\pm}Z~W^{+} W^{-} + qq).
\end{align}\\
In the last step of these equations we have indicated the final state due to
the largely dominant decay modes of the heavy vectors into $WW$ or $WZ$ (See
e.g. \cite{Barbieri:2008b}). The cross sections are summed over all the
polarizations of the heavy spin-1 fields. In the calculation of the cross
sections we reintroduce the hypercharge coupling $g^{\prime}\neq0$ and we make
standard acceptance cuts for the forward quark jets,
\begin{equation}
p_{T} > 30~\text{GeV},~~|\eta|< 5.
\end{equation}\\
These cross sections depend in general on a number of parameters. Fig.
\ref{Fig_VBF}.a shows the total cross sections for the different charge
channels with all the parameters fixed as in the minimal gauge model, eq.
(\ref{special}), and $G_{V} =g_{V} M_{V} = 200$ GeV. A value of $G_{V}$
between 150 and 200 GeV keeps the elastic $W_{L} W_{L}$-scattering amplitude
from saturating the unitarity bound below $\Lambda$, almost independently from
$M_{V} < 1.5$ TeV \cite{Bagger,Barbieri:2008b}. $M_{V}$ is taken to range from 400 to 800 GeV. A value of $M_{V}$ above 800 GeV would lead to a threshold for the
vector-boson-fusion subprocess dangerously close to the cut-off scale of the
effective Lagrangian. We have checked that the typical centre-of-mass energy
of $WW\to VV$ is on average well below 2.5 TeV, even for the highest $M_{V}$
that we consider.\\

%These cross sections, which involve all polarizations of the intermediate light bosons, have also a weak dependence on $F_V$, which is set to zero.

As discussed in Sections \ref{helicityamp}-\ref{hidden-gauge}, the parameters
of the minimal gauge model damp the high energy behaviour of the different
amplitudes. Not surprisingly, therefore, any deviation from them leads to
significantly larger cross sections, as it may be the case already in a gauge
model with more than one vector. As an example, this is shown in Fig.
\ref{Fig_VBF}.b, where all the parameters are kept as in Fig. \ref{Fig_VBF}.a,
except for $g_{K} g_{V} = 1/\sqrt{2}$ rather than 1, having in mind a
compensation of the growing amplitudes by the occurrence of (a) significantly
heavier vector(s) (See eq. \ref{sumrule}). Furthermore, both in the VBF case
and in the DY case, to be discussed below, it must be stressed that the
deviations from the minimal gauge model are quite dependent on the choice of
the parameters, with cross sections that can be even higher than those in Fig.
\ref{Fig_VBF}. In turn, these cross sections have to be considered as
indicative, given the limitations of the effective Lagrangian approach.\\

To calculate the cross sections, we have used the matrix-element generator
CalcHEP \cite{calchep}, which allows one to obtain the exact amplitude for a
process such as $q_{1}q_{2}\to VVq_{3}q_{4}$ via intermediate off-shell vector
bosons. As a check, the results so obtained have been compared with the same
 cross sections in the Effective Vector Boson Approximation, using the analytic
 amplitudes in Sect. \ref{helicityamp}, for $g^{\prime}=0$ and without
 acceptance cuts. While being a factor of $1.5\div2$ systematically lower, the
 exact results are confirmed in their $M_{V}$-dependence and in the relative
 size of the different charge channels.

\subsection{Drell--Yan pair production cross sections}

\label{crosssecqq}

The DY process is an additional source of $V$-pair production at the LHC.
From the elementary parton-level amplitudes $q\bar q\to V^+V^-$ and $q_i\bar
q_j\to V^\pm V^0$ of Section \ref{DYampl}, the physical cross sections for
the different charge channels 
\begin{align}
& pp \rightarrow V^+V^- , \\
& pp \rightarrow V^\pm V^0
\end{align}
are readily computed. In general, the cross sections depend in this case on
3 parameters other than $M_V$: $f_V, g_K$ and $g_6$.\\

\begin{figure}[tbp]
\begin{minipage}[b]{8.2cm}
   \centering
   \includegraphics[width=8cm]{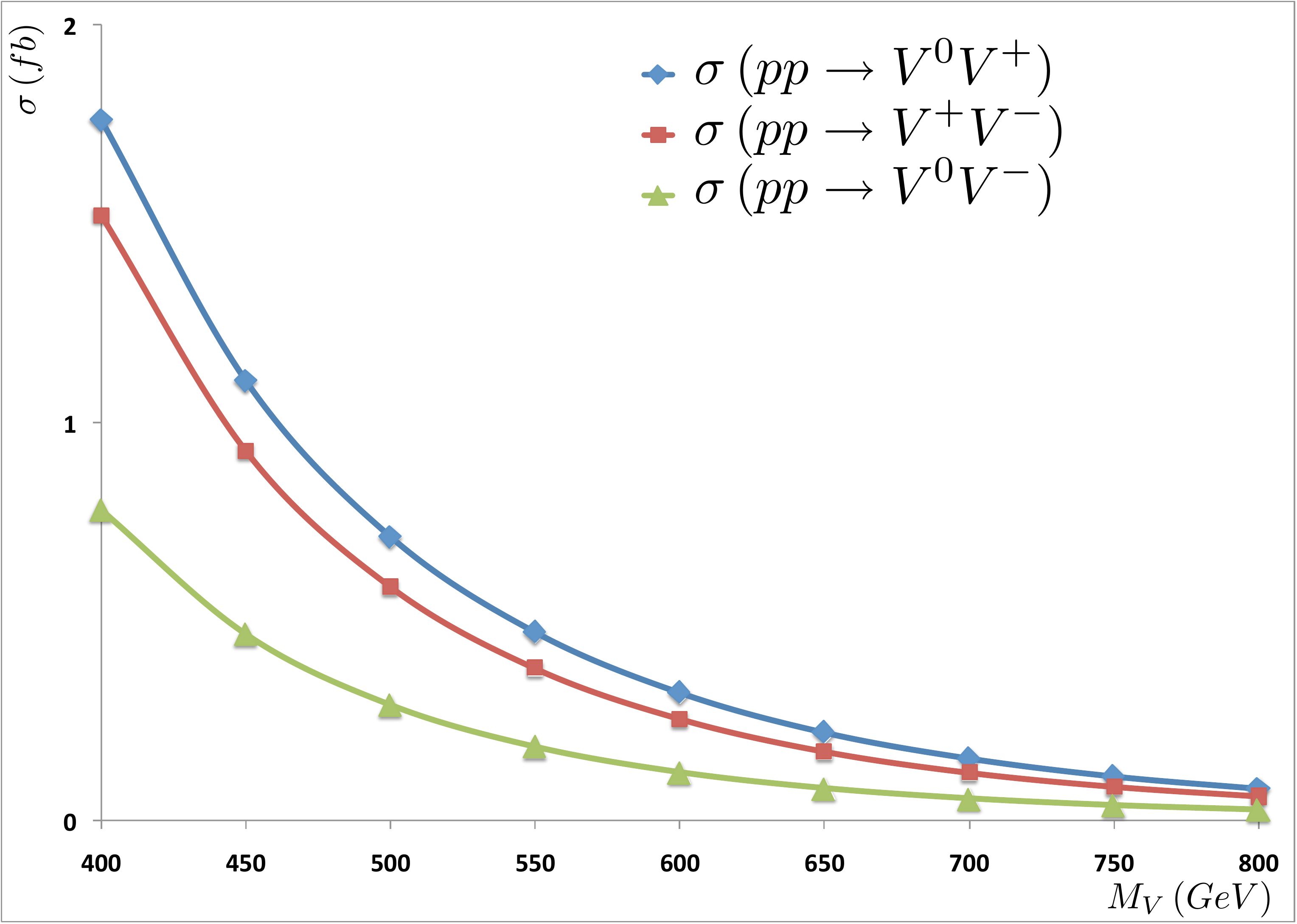}
	\\\vspace{4mm} \footnotesize{(\ref*{Fig_DY}.a)}
 \end{minipage}
\ \hspace{2mm} \hspace{3mm} \ 
\begin{minipage}[b]{8.5cm}
  \centering
   \includegraphics[width=8cm]{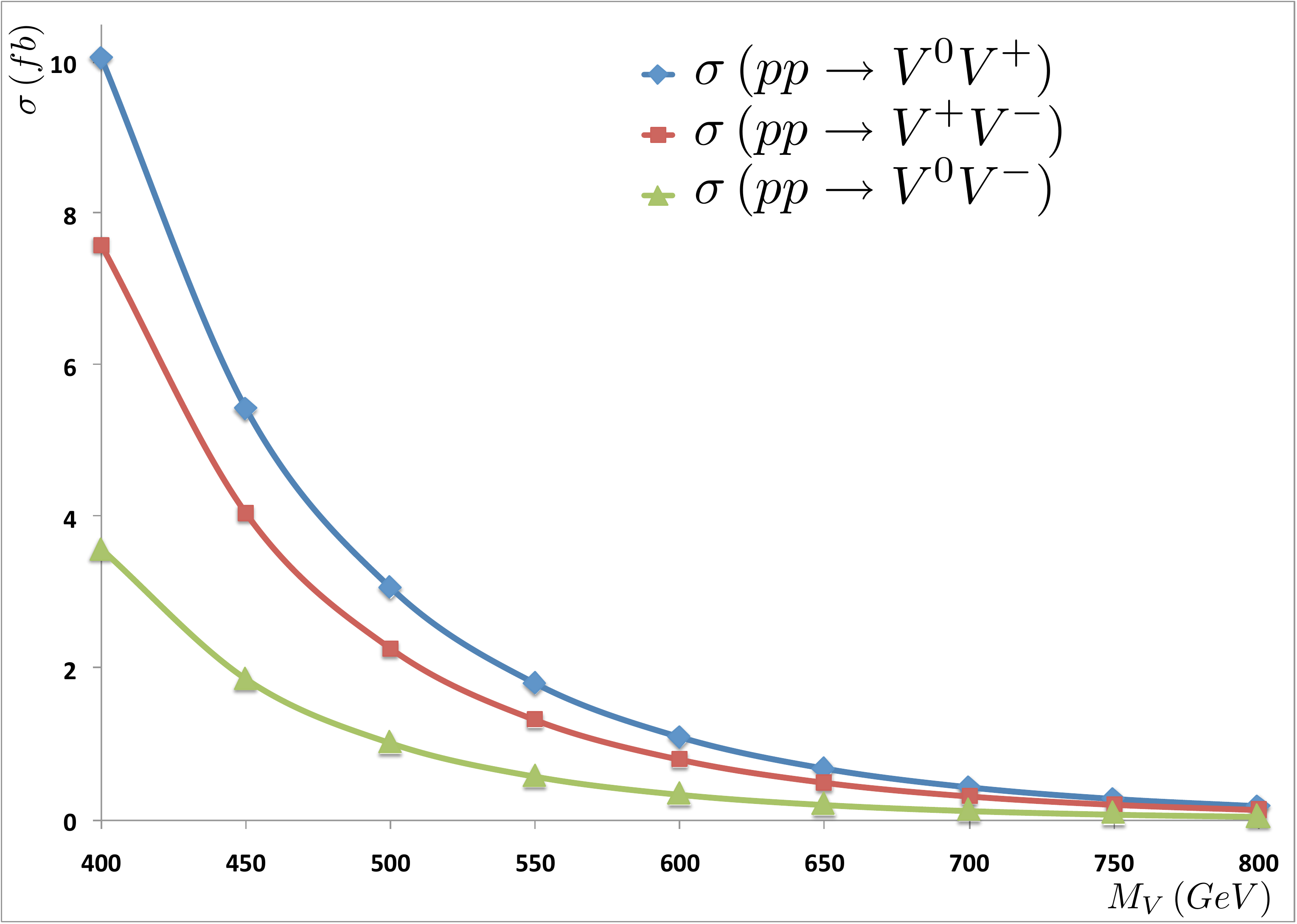}
	\\\vspace{4mm} \footnotesize{(\ref*{Fig_DY}.b)}
 \end{minipage}
\caption{Total cross sections for pair production of heavy vectors via
Drell--Yan $q\bar{q}$ annihilation in a gauge model (\protect\ref*{Fig_DY}%
.a) and a composite model (\protect\ref*{Fig_DY}.b) as functions of the
heavy vectors masses. See text for the choice of parameters.}
\label{Fig_DY}
\end{figure}
As for the vector boson fusion, we show in Fig. \ref{Fig_DY}.a the three
cross sections for the values taken by the parameters in the minimal gauge
model, $f_V g_K =2, g_6=1/2$, and for $F_V = f_V M_V = 400$ GeV
(corresponding to $f_V = 2 g_V$ and $G_V = g_V M_V = 200$ GeV as in Fig. \ref%
{Fig_VBF}.a). On the other hand, similarly to Fig. \ref{Fig_VBF}.b, we show
in Fig. \ref{Fig_DY}.b the cross sections for $f_V g_K =\sqrt{2}, g_6=1/2$
and still $F_V = f_V M_V = 400$ GeV.

\subsection{Same-sign di-lepton and tri-lepton events}

\label{signals}

After decay of the composite vectors,
\begin{equation}
V^{\pm}\rightarrow W^{\pm}Z, ~~V^{0} \rightarrow W^{+} W^{-},
\end{equation}
each $VV$-production channel, either from VBF or from DY, leads to final
states containing 2 $W$'s and 2 $Z$'s, from $V^{+} V^{-}$ and $V^{\pm}V^{\pm}%
$, 3 $W $'s and 1 $Z$, from $V^{+} V^{0}$, or 4 $W$'s from $V^{0} V^{0}$
%\footnote{%Or in fact multi-top events, see Section \ref{basicL}.}.
In fact, all final states, except for $V^{+} V^{-}$, contain at least a pair
of equal sign $W$'s, i.e., after $W \rightarrow e \nu, \mu\nu$, a pair of
same-sign leptons. In most cases there are at least 3 $W$'s, i.e. also 3 leptons.

\begin{table}[th]
\begin{center}
{\small
\begin{tabular}
[c]{|c|c|c|}\hline
& di-leptons & tri-leptons\\\hline
VBF (MGM) & 16 & 3\\\hline
DY (MGM) & 5 & 1\\\hline
VBF (comp) & 28 & 6\\\hline
DY (comp) & 18 & 4\\\hline
\end{tabular}
}
\end{center}
\caption{Number of events with at least two same-sign leptons or three leptons
($e$ or $\mu$ from $W$ decays) from vector boson fusion (VBF) or Drell--Yan
(DY) at LHC for $\sqrt{S}=14$ TeV and $\int\mathcal{L} dt = 100$~fb$^{-1}$ in
the minimal gauge model (MGM) or in a composite model (comp) with the
parameters as in Figs. \ref{Fig_VBF}-\ref{Fig_DY} and $M_{V}= 500$ GeV.}%
\label{tabcut}%
\end{table}

\begin{table}[th]
\begin{center}
{\small
\begin{tabular}
[c]{|c|c|c|}\hline
& di-leptons($\%$) & tri-leptons($\%$)\\\hline
$V^{0} V^{0} $ & 8.9 & 3.2\\\hline
$V^{\pm}V^{\pm}$ & 4.5 & -\\\hline
$V^{\pm}V^{0}$ & 4.5 & 1.0\\\hline
\end{tabular}
}
\end{center}
\caption{Cumulative branching ratios for at least two same-sign leptons or
three leptons ($e$ or $\mu$) in the $W$-decays from two vectors in the given
charge configuration.}%
\label{BRs}%
\end{table}

At the LHC with an integrated luminosity of 100 inverse femtobarns and
$\sqrt{S}$ = 14 TeV, putting together all the different charge configurations,
one obtains from $W \rightarrow e \nu, \mu\nu$ decays the number of same-sign
di-leptons and tri-lepton events given in Table \ref{tabcut} for $M_{V}$ = 500
GeV. The other parameters are fixed as in the Minimal Gauge Model (and
labelled MGM) or as in Figs. \ref{Fig_VBF}.b-\ref{Fig_DY}.b for VBF and for DY
in the previous two Sections (and labelled \textit{comp}). These numbers of
events are based on the cross sections in Figs. \ref{Fig_VBF}-\ref{Fig_DY} and
on the branching ratios for the various charge channels listed in Table
\ref{BRs}. The numbers of events for different values of $M_{V}$ are also
easily obtained. As already noticed, depending on the parameters, the number
of events in the \textit{composite} case could also be significantly higher.
No attempt is made, at this stage, to compare the signal with the background
from SM sources. To see if a signal can be observed a careful analysis will be
required, with a high cut on the scalar sum, $H_{t}$, of all the transverse
momenta and of the missing energy in each event probably playing a crucial
role. The use of the leptonic decays of the $Z$ might also be important.

\newpage

\section{A ``composite'' scalar-vector system at the LHC}

In this chapter we are interested to the case in which both a vector and a light scalar are relevant with a mass below the cutoff $\Lambda\approx 3$ TeV. In this case the role of unitarization of the different scattering channels is played both by the scalar and the vector (an example of this phenomenon is discussed for Technicolor models in \cite{Foadi:2008}). In particular, the unitarity in the elastic longitudinal gauge boson scattering does not completely constrain the couplings of the scalar and the vector to the gauge bosons, but implies a relation among them. Therefore in this case there is a wider region in the parameter space that is reasonable from the point of view of unitarity, at least in the elastic channel. In this framework we are interested to study the phenomenology of the associated scalar-vector production, that is peculiar to the present case\footnote{We shall not impose the constraints coming from the EWPT since further effects can be present, e.g. due to new fermionic degrees of freedom, that obscure their interpretation and/or a strong sensitivity to the physics at the cutoff may be involved which we do not pretend to control.}.

\subsection{The basic Lagrangian}

\label{sec2} We are interested to study a scalar-vector system in the
framework of Strongly Interacting EWSB by adopting an approach as model
independent as possible. Nevertheless, for our approach to make sense,
we have to make some assumptions. One way to state these assumptions is the following:

\begin{enumerate}
\item Before weak gauging, the Lagrangian responsible for EWSB has a $SU\left(
2\right) _{L}\times SU\left( 2\right) ^{N}\times SU\left( 2\right) _{R}$
global symmetry, with $SU\left( 2\right) ^{N}$ gauged, spontaneously broken to
the diagonal $SU\left( 2\right) _{d}$ by a generic non-linear sigma model.
%The theory has a global symmetry $SU\left(2\right)_{L}\times SU\left(2\right)^{N}\times SU\left(2\right)_{R}$ (with the $SU\left(2\right)_{L}\times SU\left(2\right)^{N}\times U\left(1\right)_{Y}$ subgroup gauged) that is broken to the diagonal $SU\left(2\right)_{d}$ ($U\left(1\right)_{\text{em}}$ gauged) by a general non-linear sigma model;

\item Only one vector triplet $V_{\mu}^{a}$ of the $SU\left( 2\right) ^{N}$
gauge group has a mass below the cutoff $\Lambda\approx3\,\text{TeV}$, while
all the other heavy vectors can be integrated out. Furthermore the new vector
triplet $V_{\mu}^{a}$ couples to fermions only through the mixing with the
weak gauge bosons of $SU$(2)$_{L}\times U$(1)$_{Y}$, $Y=T_{3R}+1/2(B-L)$.

\item The spectrum also contains a scalar singlet of $SU\left(  2\right)
_{d}$ with a relatively low mass $m_{h}\lesssim v$.
\end{enumerate}

We believe that these assumptions may represent a physically interesting situation. Under these assumptions, it follows that the interactions among the composite singlet scalar, composite triplet heavy vectors, Goldstone bosons and the SM gauge fields can be described by a model independent $SU(2)_L\times SU(2)_R/SU(2)_{L+R}$ Chiral Lagrangian given by:
%Summarizing, in the framework of a strongly interacting dynamics for the EWSB, the interactions among the composite singlet scalar, composite triplet heavy vectors and the SM gauge bosons and SM fermions can be described by the following model independent $SU(2)_L\times SU(2)_R/SU(2)_{L+R}$ Chiral Lagrangian:
\be\label{ltot}
\La_{\text{eff}}=\La_{\chi}+\La^{V}+\La_{h}+\La_{h-V},
\ee
where the different terms will be explained in the following.\\

The term $\La_{\chi}$ is the usual lowest order chiral Lagrangian for the $SU(2)_L\times SU(2)_R/SU(2)_{L+R}$ Goldstone fields with the addition of the invariant kinetic terms for the $W$ and $B$ bosons and has the following form:
\be\label{lbasic}
\La_{\chi}=\f{v^{2}}{4}\left<D_{\mu}U\(D^{\mu}U\)^{\dag}\right>-\f{1}{2g^{2}}\left<W_{\mu\nu}W^{\mu\nu}\right>-\f{1}{2g^{\prime 2}}\left<B_{\mu\nu}B^{\mu\nu}\right>\,.
\ee\\
The Lagrangian $\La^{V}$ which contains the kinetic and mass terms for the heavy spin-1 fields, the vector self-interactions as well as the interactions of these vectors with the Goldstone bosons and SM gauge fields is given by:
%This Lagrangian is given by: 
\begin{eqnarray}
\La^{V}&=&-\frac{1}{4}\left\langle \hat{V}^{\mu\nu}\hat{V}%
_{\mu\nu}\right\rangle +\frac{M_{V}^{2}}{2}\left\langle {V}^{\mu}{V}%
_{\mu}\right\rangle-\f{ig_{V}}{2\sqrt{2}}\left<\hat{V}_{\mu\nu}[u^{\mu},u^{\nu}]\right>-\f{g_{V}}{\sqrt{2}}\left<\hat{V}_{\mu\nu}\(uW^{\mu\nu}u^{\dag}+u^{\dag}B^{\mu\nu}u\)\right>\notag\\
&&+\frac{i}{2}\left\langle V_{\mu }V_{\nu }\left( uW^{\mu \nu }u^{\dagger }+u^{\dagger}B^{\mu \nu }u\right) \right\rangle+\f{ig_{K}}{4\sqrt{2}}\left<\hat{V}_{\mu\nu}[V^{\mu},V^{\nu}]\right>-\f{1}{8}\left<[V_{\mu},V_{\nu}][u^{\mu},u^{\nu}]\right>\notag\\
&&+\f{g_{V}^{2}}{8}\left<[u_{\mu},u_{\nu}][u^{\mu},u^{\nu}]\right>\,.
\end{eqnarray}\\
The Lagrangian $\La_{h}$ includes the kinetic and mass terms for the scalar as well as the interactions of this scalar with the Goldstone bosons and SM gauge fields and is given by:
%The kinetic and mass terms for the scalar as well as the interactions of this scalar with the Goldstone bosons and SM gauge fields are contained in the Lagrangian:
\be\label{lscalar}
\La_{h}=\f{1}{2}\demub h \demua h +\f{m_{h}^{2}}{2}h^{2}+\f{v^{2}}{4}\left<D_{\mu}U\(D^{\mu}U\)^{\dag}\right>\(2a\f{h}{v}+b\f{h^{2}}{v^{2}}\).
\ee\\
The term $\La_{h-V}$ is the scalar-vector interaction Lagrangian:
%The Lagrangian $\La_{h-V}$ which describes the interaction between the scalar and the heavy vector $V$ is
\be\label{lvhinteraction}
\La_{h-V}=\f{dv}{8g_{V}^{2}}h\left<V_{\mu}V^{\mu}\right>.
\ee\\
The light scalar that we are considering could be a Strongly Interacting Light Higgs (SILH) boson in the sense of \cite{ggpr} or a more complicated object arising from an unknown strong dynamics. The couplings of this particle to the SM particles and to the heavy vector $V$ will be strongly related to the mechanism that generates it. The measurement of the different cross sections that are sensitive to the different couplings, hopefully at the LHC but eventually also at a future Linear Collider, could give information about this mechanism.\\

We show in Appendix that the Lagrangian \eqref{ltot}, for the special values
\be\label{gaugeparameters}
a=\f{1}{2}\,,\quad b=\f{1}{4}\,,\quad d=1\,,\quad g_{K}=\f{1}{g_{V}}\,,\quad g_{V}=\f{v}{2M_{V}}\,,
\ee
is obtained from a gauge theory based on $SU\(2\)_{L}\times SU\(2\)_{C}\times U\(1\)_{Y}$ spontaneously broken by two Higgs doublets (with the same \textsc{vev}) in the limit $m_{H}\gg \Lambda$ for the mass of the $L$-$R$-parity odd scalar $H$\footnote{As we discuss in Appendix the mass of the $L$-$R$-parity odd scalar $H$ can be simply raised above the cut-off without any further hypothesis on the low energy physics.}.%\ref{app1}
\subsection{Two body $W_{L}W_{L}$ scattering amplitudes}\label{sec3}
In this Section we compute the scattering amplitudes:
\be\label{processes}
\begin{array}{lcl}
\displaystyle \Amp\(W^{a}_{L}W^{b}_{L}\to W^{c}_{L}W^{d}_{L}\)\qquad\qquad&&\qquad\qquad \displaystyle \Amp\(\pi^{a}\pi^{b}\to \pi^{c}\pi^{d}\)\\
\displaystyle \Amp\(W^{a}_{L}W^{b}_{L}\to V^{c}_{L}V^{d}_{L}\)&\Longrightarrow&\qquad\qquad \displaystyle -\Amp\(\pi^{a}\pi^{b}\to V^{c}_{L}V^{d}_{L}\)\\
\displaystyle \Amp\(W^{a}_{L}W^{b}_{L}\to hh\) &\sqrt{s} \gg M_{W}&\qquad\qquad \displaystyle -\Amp\(\pi^{a}\pi^{b}\to hh\)\\
\displaystyle \Amp\(W^{a}_{L}W^{b}_{L}\to V^{c}_{L}h\)&&\qquad\qquad \displaystyle -\Amp\(\pi^{a}\pi^{b}\to V^{c}_{L}h\)\,,
\end{array}
\ee
where we make use of the Equivalence Theorem to relate the scattering amplitudes involving the Goldstone bosons with the high energy limit of those ones involving the longitudinal polarization of the weak gauge bosons\footnote{The minus sign in the last three amplitudes in \eqref{processes} is due to the fact that the Equivalence Theorem has a factor $\(-i\)^{N}$ where $N$ is the number of external longitudinal vector bosons.}. As before, to simplify the explicit formulae we take the limit $g'=0$ (that implies $Z\approx W^{3}$) so that the $SU\(2\)_{L+R}$ invariance is preserved by the scattering amplitudes.\\ 

We can study the four processes one by one.
\begin{itemize}
\item $\pi^{a}\pi^{b}\to \pi^{c}\pi^{d}$ scattering amplitude\\
Using the $SU\(2\)_{L+R}$ invariance and the Bose symmetry the amplitude for the four pion scattering can be written in the form
\be
\Amp\(\pi^{a}\pi^{b}\to \pi^{c}\pi^{d}\)=\Amp\(s,t,u\)^{\pi\pi\to \pi\pi}\delta^{ab}\delta^{cd}+\Amp\(t,s,u\)^{\pi\pi\to \pi\pi}\delta^{ab}\delta^{cd}+\Amp\(u,t,s\)^{\pi\pi\to\pi\pi}\delta^{ab}\delta^{cd}\,.
\ee
It receives contributions from the four pion contact interaction $\pi^{4}$ and from the exchange of $W$, $V$ and $h$. The contribution coming from the exchange of a $W$ boson is sub-leading in the sense of the Equivalence Theorem, i.e. is of order $M_{W}/\sqrt{s}$ and therefore we can write
\be
\Amp\(\pi^{a}\pi^{b}\to \pi^{c}\pi^{d}\)=\Amp\(\pi^{a}\pi^{b}\to \pi^{c}\pi^{d}\)_{\pi^{4}}+\Amp\(\pi^{a}\pi^{b}\to \pi^{c}\pi^{d}\)_{V}+\Amp\(\pi^{a}\pi^{b}\to \pi^{c}\pi^{d}\)_{h}\,,
\ee
so that we obtain
\be\label{pipipipi}
\Amp\(s,t,u\)^{\pi\pi\to \pi\pi}=\f{s}{v^{2}}+\f{g_{V}^{2}M_{V}^{2}}{v^{4}}\Big[-3s+M_{V}^{2}\(\f{\(u-s\)}{t-M_{V}^{2}}+\f{\(t-s\)}{u-M_{V}^{2}}\)\Big]-\f{a^{2}}{v^{2}}\(\f{s^{2}}{s-m_{h}^{2}}\)\,.
\ee

\item $\pi^{a}\pi^{b}\to V^{c}_{L}V^{d}_{L}$ scattering amplitude\\
The amplitude can be reduced to
\be
\Amp\(\pi^{a}\pi^{b}\to V^{c}_{L}V^{d}_{L}\)=\Amp\(s,t,u\)^{\pi\pi\to VV}\delta^{ab}\delta^{cd}+\mathcal{B}\(s,t,u\)^{\pi\pi\to VV}\delta^{ab}\delta^{cd}+\mathcal{B}\(s,u,t\)^{\pi\pi\to VV}\delta^{ab}\delta^{cd}\,.
\ee
It receives contributions from the $\pi^{2}V^{2}$ contact interaction and the exchange of $\pi$, $V$ and $h$
\be\label{vvcontributions}
\begin{array}{lll}
\displaystyle \Amp\(\pi^{a}\pi^{b}\to VV\)&=&\displaystyle \Amp\(\pi^{a}\pi^{b}\to VV\)_{\pi^{2}V^{2}}+\Amp\(\pi^{a}\pi^{b}\to VV\)_{\pi}\\
&& \displaystyle+ \Amp\(\pi^{a}\pi^{b}\to VV\)_{V}+\Amp\(\pi^{a}\pi^{b}\to VV\)_{h}\,.
\end{array}
\ee
The explicit forms obtained for $\Amp\(s,t,u\)^{\pi\pi\to VV}$ and $\mathcal{B}\(s,t,u\)^{\pi\pi\to VV}$ are
\begin{align}
& \displaystyle \Amp\(s,t,u\)^{\pi\pi\to VV}=  \f{g_{V}^{2}M_{V}^{2}s}{v^{4}\(s-4M_{V}^{2}\)}\Big[\f{\(t+M_{V}^{2}\)^{2}}{t}+\f{\(u+M_{V}^{2}\)^{2}}{u}\Big]+\f{ad}{2v^{2}}\(\f{s}{s-m_{h}^{2}}\)\(s-2M_{V}^{2}\)\,,\label{pipiVVA}\\
& \displaystyle \mathcal{B}\(s,t,u\)^{\pi\pi\to VV} = \f{t-u}{2v^{2}}-\f{g_{V}^{2}M_{V}^{2}s\(u+M_{V}^{2}\)^{2}}{v^{4}u\(s-4M_{V}^{2}\)}+\f{s\(u-t\)}{4v^{2}M_{V}^{2}}\(g_{V}g_{K}\f{s+2M_{V}^{2}}{s-M_{V}^{2}}-1\)\,.\label{pipiVVB}
\end{align}

\item $\pi^{a}\pi^{b}\to hh$ scattering amplitude\\
The amplitude can be written as
\be
\Amp\(\pi^{a}\pi^{b}\to hh \)=\Amp\(s,t,u\)^{\pi\pi\to hh}\delta^{ab}\,.
\ee
This amplitude receives contributions from the $\pi^{2}h^{2}$ contact interaction and the exchange of $\pi$ and $h$
\be\label{hhcontributions}
\Amp\(\pi^{a}\pi^{b}\to hh \)= \Amp\(\pi^{a}\pi^{b}\to hh \)_{\pi^{2}h^{2}}+\Amp\(\pi^{a}\pi^{b}\to hh \)_{\pi}+ \Amp\(\pi^{a}\pi^{b}\to hh \)_{h}\,.
\ee
In this case $\Amp\(s,t,u\)^{\pi\pi\to hh}$ is given by
\be\label{pipihh}
\Amp\(s,t,u\)^{\pi\pi\to hh}= -\f{1}{v^{2}}\(s\(b-a^{2}\)+\f{3as m_{h}^{2}}{2\(s-m_{h}^{2}\)}-2a^{2}m_{h}^{2}+\f{a^{2}m_{h}^{4}}{t}+\f{a^{2}m_{h}^{4}}{u}\)\,.
\ee

\item $\pi^{a}\pi^{b}\to V^{c}_{L}h$ scattering amplitude\\
The $SU\(2\)_{L+R}$ invariance implies
\be
\Amp\(\pi^{a}\pi^{b}\to V^{c}_{L}h \)=\Amp\(s,t,u\)^{\pi\pi\to Vh}\epsilon^{abc}\,.
\ee
The amplitude receives contributions from the exchange of $\pi$ and $V$
\be
\Amp\(\pi^{a}\pi^{b}\to V^{c}_{L}h \)= \Amp\(\pi^{a}\pi^{b}\to V^{c}_{L}h \)_{\pi}+ \Amp\(\pi^{a}\pi^{b}\to V^{c}_{L}h \)_{V}
\ee
so that the explicit value of $\Amp\(s,t,u\)^{\pi\pi\to Vh}$ is
\be\label{pipiVh}
\begin{array}{lll}
\Amp\(s,t,u\)^{\pi\pi\to Vh}
&=&\displaystyle \f{i\(t-u\)}{2v\sqrt{\(M_{V}^{2}+m_{h}^{2}-s\)^{2}-4m_{h}^{2}M_{V}^{2}}}\Bigg[\f{d}{4g_{V}M_{V}}\f{s}{s-M_{V}^{2}}\(m_{h}^{2}-M_{V}^{2}-s\)\\
&&\displaystyle +\f{2ag_{V}M_{V}}{v^{2}tu}\Big[m_{h}^{2}M_{V}^{2}\(m_{h}^{2}-M_{V}^{2}+s\)+tu\(M_{V}^{2}-m_{h}^{2}+s\)\Big]\Bigg]\,.
\end{array}
\ee
\end{itemize}

\subsection{Asymptotic amplitudes and parameter constraints}\label{sec4}
In the very high energy limit in which $s\gg M_{V}^{2}\gg m_{h}^{2}$ we can summarize the amplitudes \eqref{pipipipi}, \eqref{pipiVVA}, \eqref{pipiVVB}, \eqref{pipihh} and \eqref{pipiVh} as follows:
\begin{subequations}\label{asymptamplitudes}
\begin{align}
& \displaystyle \Amp\(s,t,u\)^{\pi\pi\to\pi\pi} \approx \f{s}{v^{2}}\(1-a^{2}-\f{3g_{V}^{2}M_{V}^{2}}{v^{2}}\)+\f{g_{V}^{2}M_{V}^{4}}{v^{4}}\Big[\(\f{\(u-s\)}{t}+\f{\(t-s\)}{u}\)\Big]\label{asymptamplitudes1}\,,\\
& \displaystyle \Amp\(s,t,u\)^{\pi\pi\to VV} \approx \(\f{ad}{2v^{2}}-\f{1}{4v^{2}}\)\(s-2M_{V}^{2}\)\label{asymptamplitudes2}\,,\\
& \hspace{-1.5mm}\begin{array}{lll} \displaystyle \mathcal{B}\(s,t,u\)^{\pi\pi\to VV} &\approx&\displaystyle \f{u-t}{2v^{2}}\bigg[\f{s}{2M_{V}^{2}}\(g_{V}g_{K}-1\)-1+\f{3g_{V}g_{K}}{2}\(1+\f{M_{V}^{2}}{s}\)\bigg]\vspace{2mm}\\
&&\displaystyle -\f{g_{V}^{2}M_{V}^{2}u}{v^{4}}\(1+\f{4M_{V}^{2}}{s}+\f{2M_{V}^{2}}{u}\)\,,\end{array}\label{asymptamplitudes3}\\
& \displaystyle \Amp\(s,t,u\)^{\pi\pi\to hh} \approx -\f{1}{v^{2}}\Big[\(b-a^{2}\)s+\f{a m_{h}^{2}}{2}\(3-4a\)\Big]\label{asymptamplitudes4}\,,\\
& \hspace{-1.5mm}\begin{array}{lll} \displaystyle \Amp\(s,t,u\)^{\pi\pi\to Vh} &\approx&\displaystyle \f{ig_{V}M_{V}\(t-u\)}{v}\Bigg[\f{a}{v^{2}}-\f{d}{8g_{V}^{2}M_{V}^{2}}\Bigg]\vspace{2mm}\\
&&\displaystyle+\f{ig_{V}M_{V}\(t-u\)}{vs}\Bigg[\f{a}{v^{2}}\(M_{V}^{2}-m_{h}^{2}\)+\f{d}{8g_{V}^{2}M_{V}^{2}}\(m_{h}^{2}-2M_{V}^{2}\)\Bigg]\,.\end{array}\label{asymptamplitudes5}
\end{align}
\end{subequations}
For generic values of the parameters, all these amplitudes grow with the c.o.m. energy like $s$ except $\mathcal{B}\(s,t,u\)^{\pi\pi\to VV}$ that grows like $s^{2}$. On the other hand, with the parameters as in \eqref{gaugeparameters} the amplitudes reduce to
\begin{subequations}\label{asymptamplitudesgauge}
\begin{align}
& \displaystyle \Amp\(s,t,u\)^{\pi\pi\to\pi\pi} \approx \f{M_{V}^{2}}{4v^{2}}\Big[\(\f{\(u-s\)}{t}+\f{\(t-s\)}{u}\)\Big]+O\(\f{m_{h}^{2}}{v^{2}}\)\label{asymptamplitudesgauge1}\,,\\
& \displaystyle \Amp\(s,t,u\)^{\pi\pi\to VV} \approx O\(\f{m_{h}^{2}}{v^{2}}\)\label{asymptamplitudesgauge2}\,,\\
& \displaystyle \mathcal{B}\(s,t,u\)^{\pi\pi\to VV} \approx -\f{t}{4v^{2}}-\f{M_{V}^{2}}{4v^{2}}\(\f{u+3t}{s}+2\)\label{asymptamplitudesgauge3}\,,\\
& \displaystyle \Amp\(s,t,u\)^{\pi\pi\to hh} \approx -\f{m_{h}^{2}}{4v^{2}}\label{asymptamplitudesgauge4}\,,\\
& \displaystyle \Amp\(s,t,u\)^{\pi\pi\to Vh} \approx \f{iM_{V}^{2}\(u-t\)}{4v^{2}s}+O\(\f{m_{h}^{2}}{v^{2}}\)\label{asymptamplitudesgauge5}\,.
\end{align}
\end{subequations}\\
From the last relations we see that with the choice \eqref{gaugeparameters} of the parameters, that corresponds to the choice of the $SU\(2\)_{L}\times SU\(2\)_{C}\times U\(1\)_{Y}$ gauge model spontaneously broken by two Higgs doublets in the limit of very heavy $L$-$R$-parity odd scalar $H$, all the amplitudes except for $\mathcal{B}\(s,t,u\)^{\pi\pi\to VV}$ have a constant asymptotic behavior. As shown in the Appendix \ref{app1} if we also add to the spectrum the $H$ scalar we can also regulate the $\mathcal{B}\(s,t,u\)^{\pi\pi\to VV}$ amplitude making the theory asymptotically well behaved and fully perturbative.\\

The choice of parameters as in \eqref{gaugeparameters} is however too restrictive. Other than $g_{V}g_{K}=1$, so that the $\pi\pi \to VV$ scattering amplitude grows at most like $s$, we only pretend that the exchange of the scalar and of the vector lead together to a good asymptotic behavior of elastic $W_{L}W_{L}$ scattering, i.e. 
\be\label{unitarityrelation}
a=\sqrt{1-\f{3G_{V}^{2}}{v^{2}}}\,,\qquad\qquad G_{V}\equiv g_{V}M_{V}\,.
\ee
The processes \eqref{processes} are all important at the LHC in order to understand the underlying mechanism that can generate the spectrum that we consider. In fact the pair production of new states can be very useful to measure the different couplings and to constrain the parameter space. Both the scalar and vector pair productions have been recently studied in \cite{Contino:2010} and \cite{Barbieri:2010}, respectively. The phenomenology studied in those works changes as follows in the present approach:
\begin{itemize} 
\item Scalar pair production\\
Equations \eqref{hhcontributions} shows that there are no contributions of the heavy vector to the scalar pair production so that the results of \cite{Contino:2010} exactly hold also in this case.
%\footnote{The existence of an $hVV$ vertex can in principle modify the width of the scalar. However it is reasonable to consider negligible this effect.}. 
\item Vector pair production\\
From equation \eqref{vvcontributions} we see that there is a contribution to the heavy vectors pair production coming from the scalar exchange. However, having imposed in this Chapter relation (\ref{unitarityrelation}) so that $\Amp\(W_LW_L\to W_LW_L\)\simeq const$ at high energy, one has $G_V\leq v/\sqrt{3}$, which leads to a $W_LW_L\to VV$ cross section well below the values found in Chapter 2.

%Unfortunately this contribution is not big enough to compensate the decrease of the cross section due to the lowering of $G_{V}$. In other words, we see that the longitudinal $WW\to WW$ scattering unitarity relation \eqref{unitarityrelation} implies $G_{V}\leq v/\sqrt{3}$. This region of values of $G_{V}$ is rather below the value considered in \cite{Barbieri:2010} that is $G_{V}=200$~GeV. This effect leads to a fast decrease of the total cross sections that quickly fall out of the LHC accessible region. 
\end{itemize}

It remains to study the associated $Vh$ production not considered before. The associated production can be generated both by Vector Boson Fusion (VBF) and by Drell-Yan (DY) $q\bar{q}$ annihilation.  In the next section we discuss the total cross sections for the associated production by VBF and by DY.
%For these reasons we leave out of this paper the discussion of the total cross sections at order $\alpha_{S}^{2}$ for which a dedicated paper is probably needed. In the next section we discuss the total cross sections for the associated production by VBF and by DY
%The gluon-gluon fusion associated production at order $\alpha_{S}^{2}$ could be another relevant production channel, i.e. comparable to the VBF or the DY productions. However, in addition to the loop factor suppression, the absence of a direct coupling of the vector to the quarks (or at least to the top) introduces a further suppression coming from the $WV$ mixing with respect to the analogous case for the Higgs pair production by gluon-gluon fusion in the SM. There can be relevant two-loop contributions of order $\alpha^2_S$ to the total cross sections but their estimation requires a large and careful computation which is beyond the scope of this work. For this reason we defer for a future work the study of the top quark effects in the Vh associated production.

\subsection{Associated production of $Vh$ total cross sections}\label{sec5}
In this section we discuss the total cross section for the associated $Vh$ production of the heavy vector and the light scalar. There are three possible final states for the associated production, corresponding to the three charge states of the $V$: $hV^{-}$, $hV^{0}$ and $hV^{+}$. According to the constraints discussed in the previous Section on the parameter space we can compute the total cross sections for some reference values of the independent parameters that we choose to be $G_{V}$ and $d$. Some values of the total cross sections at the LHC for $\sqrt{s}=14$ TeV for different values of the parameters and for a scalar mass $m_{h}=180$ GeV are listed in Tables \ref{table1}, \ref{table2} and \ref{table3} for the production of $hV^{-}$, $hV^{0}$ and $hV^{+}$ respectively. 
\begin{table}[htb!]
\begin{minipage}[b]{8.2cm}
\centering
\resizebox{7cm}{!} {
\begin{tabular}[c]{|c|c|c|c|c|}
		\hline
		$G_{V}$ &  $a$ &  $d$ & VBF (fb) & DY (fb) \\
		\hline
		$\sqrt{5} v/4$  &$1/4$ & $0$ & $0.05$& $0$ \\
		\hline
		$\sqrt{5} v/4$  &$1/4$ & $1$ & $0.09$ & $3.31$ \\
		\hline
		$\sqrt{5} v/4$  &$1/4$ & $2$ & $0.62$& $13.24$ \\
		\hline
		$v/2$  & $1/2$ & $0$ & $0.15$ & $0$ \\
		\hline
		$v/2$  & $1/2$ & $1$ & $0.05$ & $4.14$ \\
		\hline
		$v/2$  & $1/2$ & $2$ & $0.56$ & $16.56$ \\
		\hline
		$v/\sqrt{6}$  & $1/\sqrt{2}$ & $0$ & $0.20$  & $0$ \\
		\hline
		$v/\sqrt{6}$  & $1/\sqrt{2}$ & $1$ & $0.08$  & $6.20$ \\
		\hline
		$v/\sqrt{6}$  & $1/\sqrt{2}$ & $2$ & $0.89$  & $24.80$ \\
		\hline
\end{tabular}}
\\\vspace{4mm} \footnotesize{(\ref*{table1}.a)}
 \end{minipage}
\ \hspace{2mm} \hspace{3mm} \ 
\begin{minipage}[b]{8.2cm}
\centering
\resizebox{7cm}{!} {
\begin{tabular}[c]{|c|c|c|c|c|}
		\hline
		$G_{V}$ &  $a$ &  $d$ & VBF (fb) & DY (fb) \\
		\hline
		$\sqrt{5} v/4$  &$1/4$ & $0$ & $0.02$ & $0$ \\
		\hline
		$\sqrt{5} v/4$  &$1/4$ & $1$ & $0.08$ & $1.23$ \\
		\hline
		$\sqrt{5} v/4$  &$1/4$ & $2$ & $0.49$ & $4.92$ \\
		\hline
		$v/2$  & $1/2$ & $0$ & $0.07$ & $0$ \\
		\hline
		$v/2$  & $1/2$ & $1$ & $0.06$ & $1.54$ \\
		\hline
		$v/2$  & $1/2$ & $2$ & $0.48$ & $6.16$ \\
		\hline
		$v/\sqrt{6}$  & $1/\sqrt{2}$ & $0$ & $0.09$  & $0$ \\
		\hline
		$v/\sqrt{6}$  & $1/\sqrt{2}$ & $1$ & $0.09$  & $2.30$ \\
		\hline
		$v/\sqrt{6}$  & $1/\sqrt{2}$ & $2$ & $0.75$  & $9.20$ \\
		\hline
\end{tabular}}
\\\vspace{4mm} \footnotesize{(\ref*{table1}.b)}
 \end{minipage}
\vspace{-7mm}\caption{Total cross sections for the associated production of $hV^{-}$ final state by VBF and DY at the LHC for $\sqrt{s}=14$ TeV as functions of the different parameters for $M_{V}=700$ GeV (\ref*{table1}.a) and $M_{V}=1$ TeV (\ref*{table1}.b). The parameter $a$ is fixed by the value of $G_{V}$ (and vice versa) according to equation \eqref{unitarityrelation}.}\label{table1}
\end{table}
We have chosen $m_{h}=180$ GeV to maximize both the total cross sections and the branching ratio for $h\to W^{+}W^{-}$. In this case signals of the associated productions can appear in the multi-lepton channels. In particular if the final state contains at least a pair of equal sign $W$'s there can be signals in the same-sign di-lepton and tri-lepton final states from W decays that are much simpler to be separated from the background than those corresponding to the hadronic final states.
%In particular if the final state contains at least a pair of equal sign $W$'s there can be signals in the same-sign di-lepton and tri-lepton final states from W decays that are much more simpler to be separated from the background than to the hadronic final states. 
\begin{table}[htb!]
\begin{minipage}[b]{8.2cm}
\centering
\resizebox{7cm}{!} {
\begin{tabular}[c]{|c|c|c|c|c|}
		\hline
		$G_{V}$  &$a$ &  $d$ & VBF(fb) & DY(fb) \\
		\hline
		$\sqrt{5} v/4$ & $1/4$ & $0$ & $0.08$& $0$ \\
		\hline
		$\sqrt{5} v/4$ & $1/4$ & $1$ & $0.14$ & $6.14$ \\
		\hline
		$\sqrt{5} v/4$ & $1/4$ & $2$ & $0.99$& $24.56$ \\
		\hline
		$v/2$  & $1/2$ & $0$ & $0.24$  & $0$ \\
		\hline
		$v/2$ & $1/2$ & $1$ & $0.08$ & $7.67$ \\
		\hline
		$v/2$ & $1/2$ & $2$ & $0.90$  & $30.68$ \\
		\hline
		$v/\sqrt{6}$ & $1/\sqrt{2}$ & $0$ & $0.32$  & $0$ \\
		\hline
		$v/\sqrt{6}$ & $1/\sqrt{2}$ & $1$ & $0.13$ & $11.51$ \\
		\hline
		$v/\sqrt{6}$ & $1/\sqrt{2}$ & $2$ & $1.42$ & $46.04$ \\
		\hline
\end{tabular}}
\\\vspace{4mm} \footnotesize{(\ref*{table2}.a)}
 \end{minipage}
\ \hspace{2mm} \hspace{3mm} \ 
\begin{minipage}[b]{8.2cm}
\centering
\resizebox{7cm}{!} {
\begin{tabular}[c]{|c|c|c|c|c|}
		\hline
		$G_{V}$  &$a$ &  $d$ & VBF(fb) & DY(fb) \\
		\hline
		$\sqrt{5} v/4$ & $1/4$ & $0$ & $0.04$ & $0$ \\
		\hline
		$\sqrt{5} v/4$ & $1/4$ & $1$ & $0.13$ & $2.43$ \\
		\hline
		$\sqrt{5} v/4$ & $1/4$ & $2$ & $0.79$ & $9.74$ \\
		\hline
		$v/2$  & $1/2$ & $0$ & $0.11$  & $0$ \\
		\hline
		$v/2$ & $1/2$ & $1$ & $0.09$ & $3.04$ \\
		\hline
		$v/2$ & $1/2$ & $2$ & $0.78$ & $12.16$ \\
		\hline
		$v/\sqrt{6}$ & $1/\sqrt{2}$ & $0$ & $0.15$  & $0$ \\
		\hline
		$v/\sqrt{6}$ & $1/\sqrt{2}$ & $1$ & $0.15$ & $4.57$ \\
		\hline
		$v/\sqrt{6}$ & $1/\sqrt{2}$ & $2$ & $1.22$ & $18.28$ \\
		\hline
\end{tabular}}
\\\vspace{4mm} \footnotesize{(\ref*{table2}.b)}
 \end{minipage}
\vspace{-8mm}\caption{Total cross sections for the associated production of $hV^{0}$ final state by VBF and DY at the LHC for $\sqrt{s}=14$ TeV as functions of the different constants for $M_{V}=700$ GeV (\ref*{table2}.a) and $M_{V}=1$ TeV (\ref*{table2}.b). The parameter $a$ is fixed by the value of $G_{V}$ (and vice versa) according to equation \eqref{unitarityrelation}.}\label{table2}
\end{table}
Obviously different values of $m_{h}$ are possible: in that case the detection of a signal can be disfavored by the large branching ratio for $h\to b\bar{b}$ for $m_{h}<2M_{W}$, by the large branching ratio for $h\to ZZ$ for $m_{h}>2M_{Z}$ and by the small cross sections for $m_{h}\apprge 250$ GeV (see Fig. \ref{fig2}).\\

\begin{table}[htb!]
\begin{minipage}[b]{8.2cm}
\centering
\resizebox{7cm}{!} {
\begin{tabular}[c]{|c|c|c|c|c|}
		\hline
		$G_{V}$  & $a$ &  $d$ & VBF(fb) & DY(fb) \\
		\hline
		$\sqrt{5} v/4$ & $1/4$ & $0$ & $0.10$ & $0$ \\
		\hline
		$\sqrt{5} v/4$ & $1/4$ & $1$ & $0.18$  & $7.30$ \\
		\hline
		$\sqrt{5} v/4$ & $1/4$ & $2$ & $1.28$ & $29.20$ \\
		\hline
		$v/2$ & $1/2$ & $0$ & $0.33$  & $0$ \\
		\hline
		$v/2$ & $1/2$ & $1$ & $0.10$  & $9.12$ \\
		\hline
		$v/2$ & $1/2$ & $2$ & $1.15$  & $36.48$ \\
		\hline
		$v/\sqrt{6}$ & $1/\sqrt{2}$ & $0$ & $0.43$  & $0$ \\
		\hline
		$v/\sqrt{6}$ & $1/\sqrt{2}$ & $1$ & $0.17$  & $13.68$ \\
		\hline
		$v/\sqrt{6}$ & $1/\sqrt{2}$ & $2$ & $1.82$  & $54.72$ \\
		\hline
\end{tabular}}
\\\vspace{4mm} \footnotesize{(\ref*{table3}.a)}
 \end{minipage}
\ \hspace{2mm} \hspace{3mm} \ 
\begin{minipage}[b]{8.2cm}
\centering
\resizebox{7cm}{!} {
\begin{tabular}[c]{|c|c|c|c|c|}
		\hline
		$G_{V}$  & $a$ &  $d$ & VBF(fb) & DY(fb) \\
		\hline
		$\sqrt{5} v/4$ & $1/4$ & $0$ & $0.05$ & $0$ \\
		\hline
		$\sqrt{5} v/4$ & $1/4$ & $1$ & $0.18$  & $3.03$ \\
		\hline
		$\sqrt{5} v/4$ & $1/4$ & $2$ & $1.10$ & $12.12$ \\
		\hline
		$v/2$ & $1/2$ & $0$ & $0.16$  & $0$ \\
		\hline
		$v/2$ & $1/2$ & $1$ & $0.12$  & $3.79$ \\
		\hline
		$v/2$ & $1/2$ & $2$ & $1.07$  & $15.16$ \\
		\hline
		$v/\sqrt{6}$ & $1/\sqrt{2}$ & $0$ & $0.22$  & $0$ \\
		\hline
		$v/\sqrt{6}$ & $1/\sqrt{2}$ & $1$ & $0.20$  & $5.69$ \\
		\hline
		$v/\sqrt{6}$ & $1/\sqrt{2}$ & $2$ & $1.66$  & $22.76$ \\
		\hline
\end{tabular}}
\\\vspace{4mm} \footnotesize{(\ref*{table3}.b)}
 \end{minipage}
\vspace{-8mm}\caption{Total cross sections for the associated production of $hV^{+}$ final state by VBF and DY at the LHC for $\sqrt{s}=14$ TeV as functions of the different constants for $M_{V}=700$ GeV (\ref*{table3}.a) and $M_{V}=1$ TeV (\ref*{table3}.b). The parameter $a$ is fixed by the value of $G_{V}$ (and vice versa) according to equation \eqref{unitarityrelation}.}\label{table3}
\end{table}
The total cross sections have been computed using the Matrix Element Generator CalcHEP \cite{calchep} with the CTEQ5M NLO parton distribution functions, the model was implemented in it using the FeynRules Mathematica package \cite{feynrules}. For the calculation of the VBF total cross sections the acceptance cuts $p_{T\,j}>30$ GeV and $|\eta|<5$ for the forward quark jets have been imposed. From the tables we immediately see that the DY total cross sections are much greater than the corresponding VBF ones.
This is due in part to the structure of the phase space, which for the DY is a $2\to2$ and for the VBF is a $2\to4$ and in part to the structure of the squared amplitude which for the DY is proportional to
\be
|\Amp\(q\bar{q}\to Vh\)|^{2}\propto g_{V}^{2}\f{d^{2}}{g_{V}^{4}}=\f{d^{2}}{g_{V}^{2}}\,,
\ee
%\begin{eqnarray}
%\sum_{polarizations}\left\vert A\left( u\overline{d}\rightarrow
%V^{+}h\right) \right\vert ^{2} &=&2\sum_{polarizations}\left\vert A\left( u%
%\overline{u}\rightarrow V^{0}h\right) \right\vert
%^{2}=2\sum_{polarizations}\left\vert A\left( d\overline{d}\rightarrow
%V^{0}h\right) \right\vert ^{2}  \notag \\
%&=&\frac{g^{4}d^{2}v^{2}}{128g_{V}^{2}M_{V}^{2}s^{2}\left(
%s-M_{V}^{2}\right) {}^{2}}\left\{ 6M_{V}^{2}s^{3}+4M_{h}^{6}\left(
%M_{V}^{2}+s\right) \right.  \notag \\
%&&-\left. M_{h}^{4}\left[ 3s^{2}+2s\left( u-t+M_{V}^{2}+M_{h}^{2}\right)
%+8M_{V}^{2}s+6M_{V}^{4}\right] \right.  \notag \\
%&&+\left. 2M_{h}^{2}\left( M_{V}^{2}+s\right) \left[ 2M_{V}^{4}-s^{2}+2s%
%\left( u-t+M_{V}^{2}+M_{h}^{2}\right) \right] \right.  \notag \\
%&&+\left. M_{V}^{4}\left[ s^{2}-2s\left( u-t+M_{V}^{2}+M_{h}^{2}\right) %
%\right] -M_{h}^{8}-M_{V}^{8}\right.  \notag \\
%&&+\left. 4s^{2}\left[ ut+\frac{1}{2}\left( M_{V}^{2}+M_{h}^{2}\right)
%\left( s+t-u-M_{V}^{2}-M_{h}^{2}\right) \right] \right\}  \notag \\
%&\approx &\frac{g^{4}d^{2}v^{2}ut}{32g_{V}^{2}M_{V}^{2}s^{2}}  \label{ady}
%\end{eqnarray}
while the VBF squared amplitude includes a strong dependance on $d-a$ and has a more complicated structure than the DY squared amplitude.\\
%One important result that emerges from the tables is that if the VBF total cross sections are too small to expect a signal at the LHC, the DY ones can give rise to a signal for a large region of the parameter space. In the next section we give the expected rates of multi-lepton events coming from the total cross sections listed in Tables \ref{table1}-\ref{table3}.

\begin{figure}[!htb]
\centering
\vspace{5mm}\includegraphics[width=13cm]{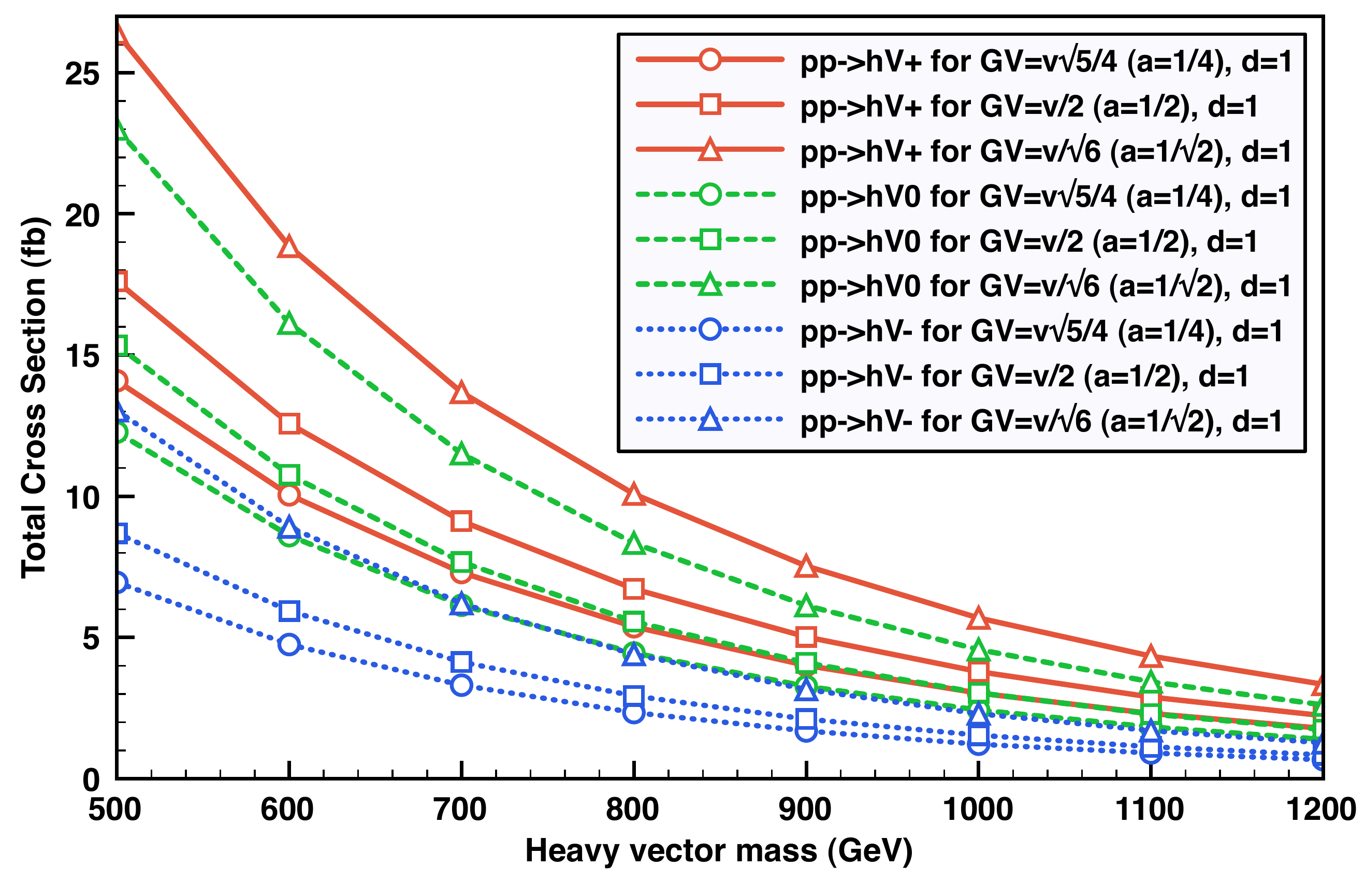}
\vspace{-4mm}\caption{Total cross sections for the $Vh$ associated productions via Drell--Yan $q\bar{q}$ annihilation as functions of the heavy vector mass at the LHC for $\sqrt{s}=14$ TeV, for $m_{h}=180$ GeV, for different values of $G_{V}$ (corresponding to different values of $a$ according to \eqref{unitarityrelation}) and for $d=1$. Since the DY total cross sections are proportional to $d^{2}$ the results can be simply generalized to different values of $d$.}
\label{fig1}
\end{figure}
The Figure \ref{fig1} shows the total cross sections for the DY associated production at the LHC for $\sqrt{s}=14$ TeV as functions of the heavy vector mass for different values of the parameter $G_{V}$ (and therefore of $a$ according to \eqref{unitarityrelation}). We see that even for $d=1$ that corresponds to the choice of the gauge model coupling (see App. \ref{app1}), the total cross sections are of order of $10$ fb for a vector mass between $500$ GeV and $800$ GeV. Furthermore, since the DY total cross sections grow with $d^{2}$, deviations from $d=1$ could result in a strong increase of the values given in Figure \ref{fig1}.\\

Finally, to give an idea of the dependence of the total cross sections on the scalar mass $m_{h}$, we plot in Fig. \ref{fig2} the total cross sections for the $Vh$ associated production as functions of the scalar mass for $150GeV<m_{h}<300GeV$. From Fig. \ref{fig2} we immediately see that the total cross sections have almost halved, going from $m_h=180$ GeV to $m_h=270$ GeV. Taking also into account the relevant branching ratio of $h$ we can conclude that a scalar with a mass between $2M_{W}$ and $2M_{Z}$ is the most favorable situation to find a signal of the associated production, while it can be much more difficult to access a signal for $m_{h}<2M_{W} $ or $m_{h}>2M_{Z}$ than for $2M_W<m_{h}<2M_{Z}$.

\begin{figure}[!htb]
\centering
\vspace{5mm}\includegraphics[width=13cm]{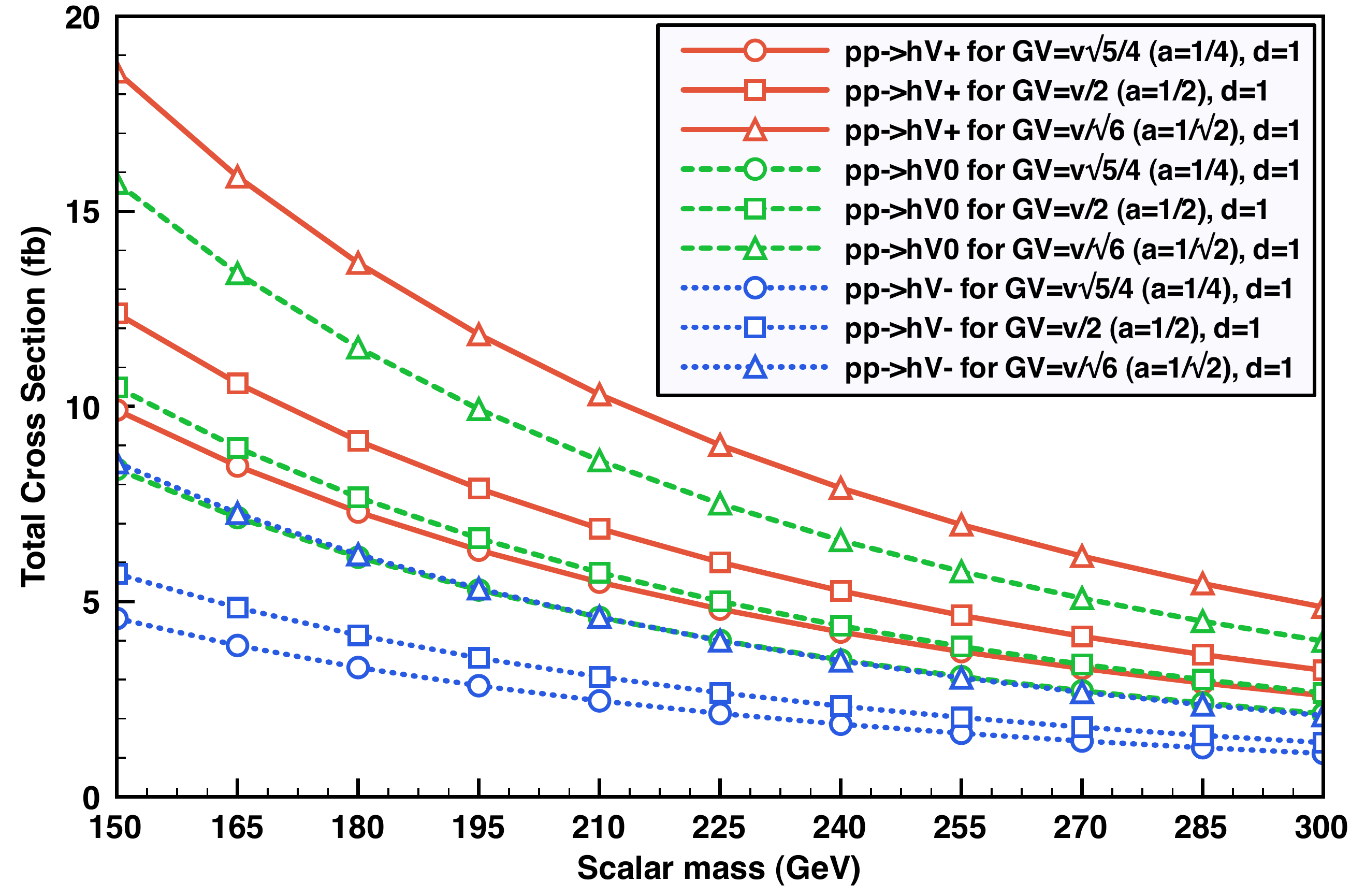}
\vspace{-4mm}\caption{Total cross sections for the $Vh$ associated productions via Drell--Yan $q\bar{q}$ annihilation as functions of the scalar mass at the LHC for $\sqrt{s}=14$ TeV, for $M_{V}=700$ GeV, for different values of $G_{V}$ (corresponding to different values of $a$ according to \eqref{unitarityrelation}) and for $d=1$. Since the DY total cross sections are proportional to $d^{2}$ the results can be simply generalized to different values of $d$.}
\label{fig2}
\end{figure}

\subsection{Same-sign di-lepton and tri-lepton events}\label{sec6}
The number of multi-lepton events is strongly dependent on the decay modes of the light scalar and the heavy vector. As the vector couples to the fermions only via the mixing with the weak gauge bosons, the decay width of $V$ into fermions is strongly suppressed with respect to the decay width into gauge bosons. In the limit $g'\approx 0$ we can write \cite{Barbieri:2008}
\be
\f{\Gamma\(V^{0}\to \bar{\psi}\psi\)}{\Gamma\(V^{0}\to W_{L}^{+}W_{L}^{-}\)}\approx \f{4M_{W}^{4}}{M_{V}^{4}}\,,
\ee
so that we can take the branching ratios
\be
\text{BR}\(V^{+}\to W_{L}^{+}Z_{L}\)\approx \text{BR}\(V^{0}\to W_{L}^{+}W_{L}^{-}\)\approx 1\,.
\ee\\
For what concerns the scalar $h$ we neglect $\Gamma\(h\to \bar{\psi}\psi\)$ with respect to $\Gamma\(h\to W^{+}W^{-}\)$.

\begin{table}[htb!]
\centering
\begin{tabular}[c]{|c|c|c|c|}
		\hline
		Decay Mode & di-leptons ($\%$)& tri-leptons ($\%$)\\
		\hline  
		$V^{0}h\to W^{+}W^{-}W^{+}W^{-}$ & $8.9$ & $3.2$\\
		\hline  
		$V^{\pm}h\to W^{\pm}ZW^{+}W^{-}$ & $4.5$ & $1.0$\\
		\hline  
%		$V^{0}h\to W^{+}W^{-}W^{+}W^{-}$ & $1$ & $98.7$ & $8.8$ & $3.2$\\
%		\hline  
%		$V^{\pm}h\to W^{\pm}ZW^{+}W^{-}$ & $1$ & $98.7$ & $4.4$ & $1.0$\\
%		\hline
\end{tabular}\caption{Decay modes and cumulative branching ratios for the different charge configurations of the $hV$ system assuming $BR\(h\to W^{+}W^{-}\)\approx 1$. For the same sign di-lepton and tri-lepton branching ratios we consider only the $e$ and $\mu$ leptons coming from the $W$ decays.}\label{BR}
\end{table}
%Using the values of the BRs given in Table \ref{BR} and a reference integrated luminosity of $\int\mathcal{L}dt=100~\text{fb}^{-1}$ we obtain the total number of same sign di-lepton and tri-lepton events given in Table \ref{events}.
Using the values of the branching fractions given in Table \ref{BR} and a reference integrated luminosity of $\int\mathcal{L}dt=100~\text{fb}^{-1}$ we obtain the total number of same sign di-lepton and tri-lepton events given in Table \ref{events}.
\begin{table}[htb!]
\begin{minipage}[b]{8.2cm}
\centering
\begin{tabular}[c]{|c|c|c|c|}
		\hline
		$G_{V}$ &  $a$ & di-leptons & tri-leptons \\
		\hline
		$\sqrt{5}v/4$  &$1/4$ & $102.4$ & $30.3$ \\
		\hline
		$v/2$  & $1/2$ & $128.0$ & $37.8$ \\
		\hline
		$v/\sqrt{6}$  & $1/\sqrt{2}$ & $192.0$  & $56.7$ \\
		\hline
\end{tabular}
\\\vspace{3mm} \footnotesize{(\ref*{events}.a)}
 \end{minipage}
\ \hspace{2mm} \hspace{3mm} \ 
\begin{minipage}[b]{8.2cm}
\centering
\begin{tabular}[c]{|c|c|c|c|c|}
		\hline
		$G_{V}$ &  $a$ & di-leptons & tri-leptons \\
		\hline
		$\sqrt{5}v/4$  &$1/4$ & $41.0$ & $12.0$ \\
		\hline
		$v/2$  & $1/2$ & $51.0$ & $15.1$ \\
		\hline
		$v/\sqrt{6}$  & $1/\sqrt{2}$ & $76.6$  & $22.6$ \\
		\hline
\end{tabular}
\\\vspace{3mm} \footnotesize{(\ref*{events}.b)}
 \end{minipage}
\caption{Total number of same sign di-lepton and tri-lepton events ($e$ or $\mu$ from $W$ decays) for the DY $Vh$ associated production at the LHC for $\sqrt{s}=14$ TeV and $\int\mathcal{L}dt=100$ fb$^{-1}$ for $M_{V}=700$ GeV (\ref*{events}.a) and $M_{V}=1$ TeV (\ref*{events}.b) for different values of the parameter $G_{V}$ (or $a$ according to equation \eqref{unitarityrelation}) and for $d=1$. Since the DY total cross sections are proportional to $d^{2}$ the results can simply be generalized to different values of $d$.}\label{events}
\end{table}

\newpage
\hspace{2cm}\vspace{40cm}
%\hspace{2cm}
\section{Conclusions}
While being disfavored relative to a weak coupling picture, the possibility that EWSB be due to a new strong interaction at about $4\pi v$ remains open. As a matter of fact the difficulties that different models of this kind encounter in reproducing the experimental data may have something to do with the lack of reliable computational tools in strong coupling theories. In turn this may obscure the emergence of the right dynamics or even of the right explicit model for EWSB. The lack so far of a thorough experimental exploration of the energy range at or well above the Fermi scale should also not be forgotten. A way to provisionally overcome this difficult situation may be offered by the EWCL with the inclusion of some ``composite'' particles. EWCL are a minimal way to describe massive vectors consistently with gauge invariance. Their problems and their limitations are well known. Yet they offer a conceptual framework to describe the phenomenology of a strong dynamics maybe responsible of EWSB in a way that may help unravelling its structure.\\   

In the framework of a strongly interacting dynamics for EWSB, composite heavy vector and scalar states may exist. The interactions among themselves and with the Standard Model gauge bosons can be described by a $SU(2)_L\times SU(2)_R/SU(2)_{L+R}$ Effective Chiral Lagrangian. These composite heavy vector and composite scalar resonances play a special role in preserving unitarity in longitudinal $WW$ scattering. In the first part of the thesis we have consider the case in which a $SU(2)_{L+R}$-triplet of composite vectors with a mass lower than $\Lambda \approx 4\pi v$ is relevant. The pair production of such composite vectors at the LHC by Vector Boson Fusion and Drell-Yan annihilation has been studied in this framework. The effective Lagrangian description of the interactions of these vectors,
among themselves or with the Standard Model gauge bosons, has several free parameters and gives rise in general to scattering amplitudes with bad asymptotic behaviour. In order to avoid the saturation of perturbative unitarity, relations among the different parameters should exist. These relations have been used to constrain the parameter space. The connection between a composite vector and a gauge vector of a spontaneously broken gauge symmetry has been discussed. For a reasonable effective theory approach one can only accept relatively small deviations of the parameters from those corresponding to a good asymptotic behavior of the various physical amplitudes, since large deviations quickly lower the cutoff to unacceptably small values. The total cross sections at the LHC for the vector pair production by Vector Boson Fusion and Drell-Yan annihilation are of order of few fb. The numbers of same sign di-lepton and trilepton events at the LHC with an integrated luminosity of $100$ $\text{fb}^{-1}$ are of the order of $10$.\\
%The parameters of the model should not deviate significantly from the gauge model scenario in order to avoid an unnaceptable low cutoff of the theory. 

In the second part of the thesis a Higgs-like scalar $h$ and a vector $V^{a}$, triplet under the custodial $SU(2)_{L+R}$, have been considered in the framework of a $SU(2)_{L}\times SU(2)_{R}/SU(2)_{L+R}$ Effective Lagrangian which describes the interactions of these states. 
%These composite triplet of heavy vectors and composite scalar singlet play a special role in preserving unitarity in longitudinal $WW$ scattering. 
In order to have a reasonable Effective Lagrangian description of these particles, the interactions of the $V^{a}$ among themselves and with the electroweak gauge bosons have been restricted to those resulting from a $SU(2)_{L}\times SU(2)^{N}\times SU(2)_{R}$ gauge theory spontaneously broken to the diagonal $SU(2)_{L+R}$ subgroup by a generic non-linear sigma model. In this framework, the two body amplitudes for the scattering of the $W_{L}W_{L}$ initial state into the $W_L W_L, hh, V_{L}V_{L}, V_{L}h$ final states have been computed in terms of five couplings ($a$, $b$, $d$, $g_{V}$ and $g_{K}$) and two masses ($m_{h}$ and $M_{V}$). The relation of these amplitudes with those arising from an explicit $SU(2)_{L}\times SU(2)_{C}\times U(1)_{Y}$ gauge model spontaneously broken by Higgs multiplets has been clarified. The parameter space has been restricted by requiring a good high energy behaviour of the elastic $W_{L}W_{L}\to W_{L}W_{L}$ scattering amplitude. From a phenomenological point of view the associated production of a scalar and a heavy vector by Vector Boson Fusion and Drell-Yan annihilation has been studied. It has been found that for a vector with a mass between $500$ GeV and $1$ TeV and for $m_{h}=180$ GeV, the main production mechanism at the LHC of a composite vector together with a composite scalar is by Drell-Yan annihilation. The order of magnitude of the cross sections is about $10$ fb for a reasonable choice of the parameters. This value can also be strongly increased since it depends quadratically on the scalar-vector coupling $d$. The expected same sign di-lepton and tri-lepton events are of the order of $10-100$ for an integrated luminosity of $100$ fb$^{-1}$. 
Further detailed studies, which are beyond the scope of this work will have to be made to assess the detectability at the LHC of composite vector pairs and composite vector-composite scalar final states above the Standard Model backgrounds.\\

The experimental investigation of all the processes that we have studied will only be possible at the LHC with its maximum energy and intensity. Before that, the single direct production of any composite state, if they exist at all, will have been discovered. Nevertheless, to unravel the structure of the underlying dynamics, the study of the processes considered in this thesis will probably be necessary. To this end the tools and the considerations developped here will hopefully prove useful.

\newpage
\addcontentsline{toc}{section}{Acknowledgments}
\section*{Acknowledgments}
I am very grateful to my advisor Professor Riccardo Barbieri for accepting me as a student, for his mentoring, for introducing me to this field and for his big contribution in my becoming a theoretical physicist. I woe much to Gennaro Corcella, Enrico Trincherini and Riccardo Torre for their contributions and collaboration in the papers related to this thesis. I am especially thankful to Riccardo Torre for many interesting and stimulating 
discussions. I also thank the organizers of the XVIII International Workshop on Deep-Inelastic 
Scattering and Related Subjects, Gennaro Corcella in particular, for inviting me to present 
a talk related to this Thesis. I would like to thank Gennaro Corcella and Rakibur Rahman 
for the parallel projects and for fruitful discussions, from which I have learnt a lot. 
I am indebted to Henry Dupont, Rakibur Rahman, Jayne Thompson, Santosh Gopal and Ashwanth 
Francis for proof-reading. I express deepest gratitude to my parents Antonio Nicol\'as and 
Yadira Esther, to my sister Eliana Marcela, to my brother Juan David, to the rest of my family, 
and to all my friends for their continual support, love and encouragement. My  deepest gratitude is extended to Emeline for her love, care and support. My thanks go for Emeline's Grandmother Mrs Couge for her kind hospitality, and to Loic Bahier for lending me an office. Finally, above all, I deeply thank God for loving me, protecting me, blessing me and helping me to overcome all the difficulties.

\clearpage

\addcontentsline{toc}{section}{References}
%\section{Bibliography}

\newpage

\addcontentsline{toc}{section}{Appendix: A well behaved theory at all energies}
\section*{Appendix: A well behaved theory at all energies}\label{app1}
Let us consider the following $SU(2)_{L}\times SU(2)_{C}\times U(1)_{Y}$ invariant non-linear sigma model Lagrangian: 
\begin{equation}\label{la0}
\mathcal{L}^{\text{gauge}}=\mathcal{L}_{\chi}^{\text{gauge}}-\frac{1}{2g_{C}^{2}}\left< v_{\mu\nu}v^{\mu\nu}\right> -\frac{1}{2g^{2}}\left<  W_{\mu\nu}W^{\mu\nu}\right>  -\frac{1}{2g^{\prime 2}}\left<  B_{\mu\nu}B^{\mu\nu}\right>  -V\left(  \Sigma_{YC},\Sigma_{CL}\right)\,,
\end{equation} 
where 
\begin{equation}\label{l2a}
v_{\mu}=\frac{g_{C}}{2}v_{\mu}^{a}\tau^{a}
\end{equation} 
is the $SU(2)_{C}$-gauge vector and 
\begin{equation}\label{l3}
\mathcal{L}_{\chi}^{\text{gauge}}=\frac{v^{2}}{2}\left\langle D_{\mu}\Sigma_{YC}\left(  D^{\mu}\Sigma_{YC}\right)  ^{\dag}\right\rangle +\frac{v^{2}}{2}\left\langle D_{\mu}\Sigma_{CL}\left(  D^{\mu}\Sigma_{CL}\right)  ^{\dag}\right\rangle
\end{equation} 
is the symmetry breaking Lagrangian and $V\left(  \Sigma_{YC},\Sigma_{CL}\right)  $ is the scalar potential, which has the form
\begin{align} \label{l4}
V\left(  \Sigma_{YC},\Sigma_{CL}\right)   &  =\f{\mu^{2}v^{2}}{2}\left\langle \Sigma_{YC}\Sigma_{YC}^{\dag}\right\rangle +\f{\mu^{2}v^{2}}{2}\left\langle \Sigma_{CL}\Sigma_{CL}^{\dag}\right\rangle -\f{\lambda v^{4}}{4}\left(  \left\langle\Sigma_{YC}\Sigma_{YC}^{\dag}\right\rangle \right)^{2}\nonumber\\ 
&  -\f{\lambda v^{4}}{4}\left(  \left\langle \Sigma_{CL}\Sigma_{CL}^{\dag}\right\rangle \right)  ^{2}-\kappa v^{4}\left\langle \Sigma_{YC}\Sigma_{CL}^{\dag}\Sigma_{CL}\Sigma_{YC}^{\dag}\right\rangle\,.
\end{align}\\
To ensure the correct normalization for the Goldstone bosons kinetic terms, $\Sigma_{YC}$ and $\Sigma_{CL}$ are defined as: 
\begin{equation}\label{s17}
\Sigma_{YC}=\left(  1+\frac{h+H}{2v}\right)  U_{YC}\,,\hspace{2cm}\hspace{2cm}U_{YC}=\exp\left[  \frac{i}{2v}\left(  \pi+\sigma \right)\right] \,,
\end{equation}
\begin{equation}\label{s18}
\Sigma_{CL}=\left(  1+\frac{h-H}{2v}\right)  U_{CL}\,,\hspace{2cm}\hspace{2cm}U_{CL}=\exp\left[  \frac{i}{2v}\left(  \pi-\sigma\right)\right]  \,,
\end{equation} 
where $\pi=\pi^{a}\tau^{a}$ and $\sigma=\sigma^{a}\tau^{a}$, being $\pi^{a}$ and $\sigma^{a}$ the Goldstone bosons respectively associated with the EW gauge bosons $W_{\mu}^{a}$ and with the heavy vectors $v_{\mu}^{a}$ and $\tau^{a}$ the usual Pauli matrices. Furthermore $h$ and $H$ are the physical $L$-$R$-parity even and odd scalars respectively and are assumed to have the same \textsc{vev} $v$ and to have the following masses
\be\label{s17b}
m_{h}^{2}=4v^{2}\(\lambda+\kappa\)\,,\qquad\qquad m_{H}^{2}=4v^{2}\(\lambda-\kappa\)\,.
\ee\\
The two Higgs doublets realize the spontaneous breaking of the $SU(2)_{L}\times SU(2)_{C}\times U(1)_{Y}$ local symmetry to $U(1)_{\text{em}}$, while the global group $G=SU(2)_{L}\times SU(2)_{C}\times SU(2)_{R}$ is broken to the diagonal subgroup $H=SU(2)_{L+C+R}$. The covariant derivatives appearing in \eqref{l3} are given by
\begin{equation}\label{l5a}
D_{\mu}U_{YC}=\partial_{\mu}U_{YC}-iB_{\mu}U_{YC}+iU_{YC}v_{\mu}\,,\hspace{2cm}D_{\mu}U_{CL}=\partial_{\mu}U_{CL}-iv_{\mu}U_{CL}+iU_{CL}W_{\mu}\,.
\end{equation}\\ 
The $U$ fields can be written as $U_{YC}=\sigma_{Y}\sigma_{C}^{\dag}$ and $U_{CL}=\sigma_{C}\sigma_{L}^{\dag}$ where the $\sigma_{L,C,Y}$ are elements of $SU\left(  2\right)_{L,C,R}/H$ respectively\footnote{Remember that only the generator $T^{3}$ of $SU\(2\)_{R}$ is gauged.}. These $\sigma_{I}$ with $I=L,C,Y$ transform under the full $SU\left(2\right)  _{L}\times SU\left(2\right)  _{C}\times U\left(  1\right)  _{Y}$ as $\sigma_{I}\rightarrow g_{I}\sigma_{I}h^{\dag}$. By applying the gauge transformation
\begin{equation}\label{x6}
v_{\mu}^{I}\rightarrow\sigma_{I}^{\dag}v_{\mu}^{I}\sigma_{I}+i\sigma_{I}^{\dag}\partial_{\mu}\sigma_{I}=\Omega_{\mu}^{I},\hspace{2cm}\hspace{2cm}U_{IJ}\rightarrow\sigma_{I}^{\dag}U_{IJ}\sigma_{J}=1\,,
\end{equation} 
the symmetry breaking Lagrangian takes the form
\begin{equation}\label{l6a}
\mathcal{L}_{\chi}^{\text{gauge}}=\frac{v^{2}}{2}\left(  1+\frac{h+H}{2v}\right)^{2}\left\langle \left(  \Omega_{\mu}^{Y}-\Omega_{\mu}^{C}\right)  ^{2}\right\rangle +\frac{v^{2}}{2}\left(  1+\frac{h-H}{2v}\right)  ^{2}\left\langle \left(  \Omega_{\mu}^{L}-\Omega_{\mu}^{C}\right)^{2}\right\rangle \,.
\end{equation}\\ 
After the gauge fixing $\sigma_{Y}=\sigma_{L}^{\dag}=u^{2}=U=e^{\f{i\hat{\pi}}{v}}$ and $\sigma_{C}=1$, which implies that $U_{YC}=U_{CL}$ (i.e. $\hat{\sigma}=0$) corresponding to the unitary gauge in which we get rid of the Goldstone bosons associated with the heavy vectors $v_{\mu}^{a}$, the Lagrangian of the previous expression becomes
\begin{equation}\label{lt}
\mathcal{L}_{\chi}^{\text{gauge}}=v^{2}\left(  1+\frac{h^{2}+H^{2}}{4v^{2}}+\frac{h}{v}\right)\left(  \left\langle \left(  v_{\mu}-i\Gamma_{\mu}\right)  ^{2}\right\rangle+\frac{1}{4}\left\langle u_{\mu}u^{\mu}\right\rangle \right)  -\frac{1}{2}\left(  2vH+hH\right)  \left\langle u^{\mu}\left(  v_{\mu}-i\Gamma_{\mu}\right)  \right\rangle\,,
\end{equation} 
where
\begin{equation}\label{x10}
u_{\mu}=\Omega_{\mu}^{Y}-\Omega_{\mu}^{L}=iu^{\dag}D_{\mu}U u^{\dag},\hspace{2cm}\Gamma_{\mu}=\frac{1}{2i}\left(  \Omega_{\mu}^{Y}+\Omega_{\mu}^{L}\right)=\f{1}{2}\Big[u^{\dag}\(\demub -iB_{\mu}\)u+u\(\demub-iW_{\mu}\)u^{\dag}\Big]\,.
\end{equation}\\ 
Now by setting
\begin{equation}\label{x11}
v_{\mu}=V_{\mu}+i\Gamma_{\mu}\,,
\end{equation} 
by using the identity \cite{Ecker:1989yg}
\begin{equation}\label{l2}
v_{\mu\nu}=V_{\mu\nu}-i\left[  V_{\mu},V_{\nu}\right]  +\frac{i}{4}\left[u_{\mu},u_{\nu}\right]  +\frac{1}{2}f_{\mu\nu}^{+}\,,
\end{equation} 
where $f_{\mu\nu}^{+}=uW_{\mu\nu}u^{\dagger}+u^{\dagger}B_{\mu\nu}u$, 
and by redefining $V_{\mu}\rightarrow\frac{g_{C}}{\sqrt{2}}V_{\mu}$, we obtain the following effective Lagrangian
\begin{equation}\label{s20}
\mathcal{L}^{\text{gauge}}=\mathcal{L}_{h=H=0}+\mathcal{L}_{h,H}\,,
\end{equation} 
where $\mathcal{L}_{h=H=0}$ and $\mathcal{L}_{h,H}$ are given by:
\begin{align}\label{l9}
\mathcal{L}_{h=H=0} &  = -\frac{1}{2g^{2}}\left\langle W_{\mu\nu}W^{\mu\nu}\right\rangle -\frac{1}{2g^{\prime2}}\left\langle B_{\mu\nu}B^{\mu\nu}\right\rangle -\frac{1}{4}\left\langle V_{\mu\nu}V^{\mu\nu}\right\rangle+\frac{v^{2}}{4}\left\langle D_{\mu}U\(D^{\mu}U\)^{\dag}\right\rangle  +\frac{M_{V}^{2}}{2}\left\langle V_{\mu}V^{\mu}\right\rangle \nonumber\\ 
&  +\frac{ig_{C}}{2\sqrt{2}}\left\langle V_{\mu\nu}\left[  V^{\mu},V^{\nu}\right]  \right\rangle  -\frac{g_{C}^{2}}{8}\left<  \left[  V_{\mu},V_{\nu}\right]  \left[V^{\mu},V^{\nu}\right]  \right>-\frac{i}{4\sqrt{2}g_{C}}\left\langle V_{\mu\nu}\left[  u^{\mu},u^{\nu}\right]  \right\rangle  \nonumber\\ 
&  -\frac{1}{8}\left\langle \left[  V_{\mu},V_{\nu}\right]  \left[  u^{\mu},u^{\nu}\right]  \right\rangle -\frac{1}{2\sqrt{2}g_{C}}\left\langle V_{\mu\nu}f^{+\mu\nu}  \right\rangle  +\frac{i}{4}\left\langle \left[  V^{\mu},V^{\nu}\right]  f^{+\mu\nu}  \right\rangle  \nonumber\\ 
&  +\frac{1}{32g_{C}^{2}}\left\langle \left[  u_{\mu},u_{\nu}\right]  \left[  u^{\mu},u^{\nu}\right]  \right\rangle-\frac{1}{8g_{C}^{2}}\left\langle f_{\mu\nu}^{+} f^{+\mu\nu}  \right\rangle -\frac{i}{8g_{C}^{2}}\left\langle \left[  u^{\mu},u^{\nu}\right]  f^{+\mu\nu}  \right\rangle\,,
\end{align}
\begin{align}\label{s22}
\mathcal{L}_{h,H} &  =v^{2}\left(\frac{h^{2}+H^{2}}{4v^{2}}+\frac{h}{v}\right)\left(  \frac{g_{C}^{2}}{2}\left\langle V_{\mu}V^{\mu}\right\rangle +\frac{1}{4}\left\langle D_{\mu}U\(D^{\mu}U\)^{\dag}\right\rangle \right)   \nonumber\\ 
&  -\frac{g_{C}}{2\sqrt{2}}\left(  2vH+hH\right)  \left\langle u^{\mu}V_{\mu}\right\rangle+\frac{1}{2}\left[  \left(  \partial_{\mu}h\right)  ^{2}+\left(\partial_{\mu}H\right)  ^{2}\right]  -V\left(  h,H\right)\,,
\end{align} 
and with the potential $V\left(h,H\right)$ given by
\begin{align}\label{s23}
V\left(  h,H\right)   &  =-\mu^{2}v^{2}\left(  1+\frac{h+H}{2v}\right)^{2}-\mu^{2}v^{2}\left(  1+\frac{h-H}{2v}\right)  ^{2}+2\kappa v^{4}\left(1+\frac{h+H}{2v}\right)  ^{2}\left(  1+\frac{h-H}{2v}\right)  ^{2}\nonumber\\ 
&  +\lambda v^{4}\left(  1+\frac{h+H}{2v}\right)  ^{4}+\lambda v^{4}\left(1+\frac{h-H}{2v}\right)^{4}\,.
\end{align}\\ 
By taking the mass of the $L$-$R$-parity odd $H$ given in \eqref{s17b} infinitely large (so that it is decoupled from the theory), $\mathcal{L}^{\text{gauge}}$ coincides with $\mathcal{L}_{\text{eff}}$ in \eqref{ltot} up to operators irrelevant for the processes \eqref{processes}, only for the values of the parameters: 
\begin{equation}\label{l5}
\begin{array}{l}
\displaystyle g_{V}=\frac{1}{2g_{C}}=\frac{1}{g_{K}}=\frac{v}{2M_{V}},\hspace{2cm} f_{V}=2g_{V},\hspace{2cm}M_{V}=g_{C}v=\frac{1}{2}g_{K}v=\frac{v}{2g_{V}}\,,\vspace{1mm}\\
\displaystyle a=\frac{1}{2},\hspace{2cm}b=\frac{1}{4},\hspace{2cm}d=1,\hspace{2cm} G_{V}=\frac{v}{2}\,.
\end{array}
\end{equation} 
This implies that when the relations \eqref{l5} are satisfied, $\mathcal{L}_{\text{eff}}$ in \eqref{ltot} reduces to $\mathcal{L}^{\text{gauge}}$ in \eqref{s20} in the limit $m_{H}\gg \Lambda$. Since the theory described by $\mathcal{L}^{\text{gauge}}$ is well behaved at all energies, the relations \eqref{l5} allow to take under control the unitarity of the model under consideration.
%\end{appendix}
%\subsection*{Acknowledgements}


\begin{thebibliography}{99}
\bibitem{Quigg} C. Quigg, [\href{http://arxiv.org/abs/hep-ph/0704.2232v2}{arXiv:hep-ph/0704.2232v2}].

\bibitem{Djouadi} A. Djouadi, [\href{http://arxiv.org/abs/hep-ph/0503172v2}{arXiv:hep-ph/0503172v2}].

\bibitem{Herrero} M. J. Herrero, [\href{http://arxiv.org/abs/hep-ph/9601286v1}{arXiv:hep-ph/9601286v1}].

\bibitem{Manohar} A. V. Manohar, [\href{http://arxiv.org/abs/hep-ph/9606222v1}{arXiv:hep-ph/9606222v1}].

\bibitem{Pich} A. Pich, [\href{http://arxiv.org/abs/hep-ph/9806303v1}{arXiv:hep-ph/9806303v1}].

%\bibitem{Dawson} S. Dawson, {Nucl. Phys.} \textbf{B 249 } (1985) 42.

\bibitem{Dawson:1985} S. Dawson, [\href{http://arxiv.org/abs/hep-ph/9901280v1}{arXiv:hep-ph/9901280v1}].
                                          
\bibitem{Isidori}G.~Isidori,
%``Effective theories of electroweak symmetry breaking,''
[\href{http://arxiv.org/abs/0911.3219v1}{arXiv:0911.3219v1 [hep-ph]}].

\bibitem {Kaplan:1983} D.~B.~Kaplan and H.~Georgi,
%``SU(2) X U(1) Breaking By Vacuum Misalignment,''
Phys.\ Lett.\ B \textbf{136} (1984) 183.

\bibitem {Chivukula:1993} R.~S.~Chivukula and V.~Koulovassilopoulos,
%``Phenomenology of a nonstandard Higgs,''
Phys.\ Lett.\ B \textbf{309}, 371 (1993)
\href{http://arxiv.org/abs/hep-ph/9304293}{arXiv:hep-ph/9304293}].

\bibitem {ggpr}G. F. Giudice, C. Grojean, A. Pomarol and R. Rattazzi, JHEP
0706 (2007) 045
[\href{http://arxiv.org/abs/hep-ph/0703164}{arXiv:hep-ph/0703164}].

\bibitem{Zerwekh:2010}A. R. Zerwekh, Mod. Phys. Lett. A A25 (2010), 423 [\href{http://arxiv.org/abs/hep-ph/0907.4690}{arXiv:hep-ph/0907.4690}].
%"Two Composite Higgs Doublets: Is it the Low Energy Limit of a Natural Strong Electroweak Symmetry Breaking Sector?", a model of two composite higgs doublets is proposed motivated by the MInimal Walking Technicolor. Two composite higgs doublet are proposed in order to account for the top quark mass. One composite higgs boson is assumed to come from the MInimal Walking Technicolor-Extended Technicolor sector while the other one is assumed to be originated from the Top color sector. The main production mechanisms at the LHC of the lighter scalar are found to be the associated production with a SM gauge boson and the Vector Boson Fusion.

\bibitem {Low:2009}I.~Low, R.~Rattazzi and A.~Vichi,
%``Theoretical Constraints on the Higgs Effective Couplings,''
[\href{http://arxiv.org/abs/hep-ph/0907.5413}{arXiv:hep-ph/0907.5413}]

\bibitem {Contino:2009}R.~Contino,
%``New Physics at the LHC: Strong vs Weak symmetry breaking,''
[\href{http://arxiv.org/abs/0908.3578}{arXiv:0908.3578 [hep-ph]}].

\bibitem {Bagger}J.~Bagger~\emph{et al.,} Phys.\ Rev.\ D \textbf{49} (1994) 1246.

%\cite{Kaplan:1983fs}

\bibitem{Pelaez:1996} J.~R.~Pel\'aez, Phys.\ Rev.\ D \textbf{55} (1997) 4193 [\href{http://arxiv.org/abs/hep-ph/9609427}{arXiv:hep-ph/9609427}].

\bibitem {SekharChivukula:2001} R.~S.~Chivukula, D.~A.~Dicus and H.~J.~He,
%``Unitarity of compactified five dimensional Yang-Mills theory,''
Phys.\ Lett.\ B \textbf{525} (2002) 175
[\href{http://arxiv.org/abs/hep-ph/0111016}{arXiv:hep-ph/0111016}].

\bibitem {Csaki:2003}C.~Csaki, C.~Grojean, H.~Murayama, L.~Pilo and
J.~Terning,
%``Gauge theories on an interval: Unitarity without a Higgs,''
Phys.\ Rev.\ D \textbf{69}, 055006 (2004)
[\href{http://arxiv.org/abs/hep-ph/0305237}{arXiv:hep-ph/0305237}].


\bibitem {Barbieri:2008}R. Barbieri, G. Isidori, V. S. Rychkov and E. Trincherini,
Phys. Rev. D \textbf{78} (2008) 036012
[\href{http://arxiv.org/abs/0806.1624}{arXiv:0806.1624 [hep-ph]}].

%``Drell--Yan production of Heavy Vectors in Higgsless models,''
\bibitem{Cata:2009iy} O.~Cata, G.~Isidori and J.~F.~Kamenik, 
Nucl.\ Phys.\ B \textbf{822} (2009) 230 [\href{http://arXiv.org/pdf/0905.0490}{arXiv:0905.0490[hep-ph]}].

\bibitem{Barbieri:2010}R.~Barbieri, A.~E.~C\'arcamo Hern\'andez, G.~Corcella, R.~Torre and
 E.~Trincherini, JHEP \textbf{03} (2010) 068
[\href{http://arXiv.org/pdf/0911.1942}{arXiv:0911.1942[hep-ph]}]
%http://www.springerlink.com/content/w2181848228pkj71/

\bibitem{Zerwekh:2006}A. R. Zerwekh, Eur. Phys. J. C 46 (2006) 791 [\href{http://arxiv.org/abs/hep-ph/0512261}{arXiv:hep-ph/0512261}].
%Associated Higgs and Gauge boson production at Hadron colliders in a Model with vector resonances" studiano la produzione associata di un Higgs composto leggero e un SM gauge boson in the framework of a Effective Lagrangian with light SM Higgs and vector resonances. In that article the vector resonances have interactions with the SM quarks and the vector resonances have masses between 200GeV-500GeV. Interactions type pho_{mu} W+{mu}H are considered so that the exchange of the heavy resonance pho in the s channel gives a contribution to the HW+ associated production via Drell Yan process. For some values of the parameters an enhancement of the total cross section at the LHC for the process pp to HW+ is obtained with respect to the predicted by the SM.

%\bibitem{Carcamo:2010p} A.~E.~C\'arcamo Hern\'andez [\href{http://arXiv.org/pdf/1006.1065}{arXiv:1006.1065[hep-ph]}]

\bibitem{Carcamo:2010} A.~E.~C\'arcamo Hern\'andez and R.~Torre [\href{http://arXiv.org/pdf/1005.3809}{arXiv:1005.3809[hep-ph]}], Nuclear Physics B 2010, [\href{http://dx.doi.org/10.1016/j.nuclphysb.2010.08.004}{dx.doi.org/10.1016/j.nuclphysb.2010.08.004}].
%Nucl.\ Phys.\  B {\bf 365} (2010) 259.


%Composite fermions in strong EWSB
\bibitem{Barbieri:2008b} R. Barbieri, G. Isidori and D.
Pappadopulo, JHEP {\bf 02} (2009) 029
  [\href{http://arxiv.org/abs/0811.2888}{arXiv:0811.2888 [hep-ph]}].

\bibitem {He:2007ge}H.~J.~He \textit{et al.},
%``LHC Signatures of New Gauge Bosons in Minimal Higgsless Model,''
Phys.\ Rev.\ D \textbf{78} (2008) 031701
[\href{http://arXiv.org/pdf/0708.2588}{arXiv:0708.2588[hep-ph]}].

%\cite{Accomando:2008jh}

\bibitem {Accomando:2008jh}E.~Accomando, S.~De Curtis, D.~Dominici and
L.~Fedeli,
%``Drell--Yan production at the LHC in a four site Higgsless model,''
Phys. Rev. D \textbf{79} (2009) 055020 \href{http://arXiv.org/abs/hep-ph/0807.5051}{arXiv:hep-ph/0807.5051}]; Nuovo
Cim.\ \textbf{123B} (2008) 809 \href{http://arXiv.org/abs/hep-ph/0807.2951}{arXiv:hep-ph/0807.2951}].


\bibitem {Belyaev:2008yj}A.~Belyaev, R.~Foadi, M.~T.~Frandsen, M.~Jarvinen,
F.~Sannino and A.~Pukhov,
%``Technicolor Walks at the LHC,''
Phys.\ Rev.\ D \textbf{79} (2009) 035006, \href{http://arXiv.org/abs/hep-ph/0809.0793}{arXiv:hep-ph/0809.0793}].

\bibitem{Hirn:2007}
  J.~Hirn, A.~Martin and V.~Sanz,
  %``Benchmarks for new strong interactions at the LHC,''
  JHEP {\bf 0805} (2008) 084
 \href{http://arXiv.org/abs/hep-ph/0712.3783}{arXiv:hep-ph/0712.3783}].
  %%CITATION = JHEPA,0805,084;%%
  %\bibitem{Hirn:2008tc}
  %J.~Hirn, A.~Martin and V.~Sanz,
  %``Describing viable technicolor scenarios,''
  Phys.\ Rev.\  D {\bf 78} (2008) 075026
  \href{http://arXiv.org/abs/hep-ph/0807.2465}{arXiv:hep-ph/0807.2465 }].
  %%CITATION = PHRVA,D78,075026;%%\

  %%CITATION = EPHJA,C34,447;%%


\bibitem {Ecker:1988te}G.~Ecker, J.~Gasser, A.~Pich and E.~de Rafael,
%``The Role Of Resonances In Chiral Perturbation Theory,''
Nucl.\ Phys.\ B \textbf{321} (1989) 311.

\bibitem {Ecker:1989yg}G.~Ecker, J.~Gasser, A.~Pich and E.~de Rafael,
%%J.~Gasser, H.~Leutwyler, A.~Pich and E.~de Rafael,
%%``Chiral Lagrangians for Massive Spin 1 Fields,''
Phys.\ Lett.\ B \textbf{223} (1989) 425;
%%CITATION = PHLTA,B223,425;%%

\bibitem {coleman}S.~R.~Coleman \textit{et. al.} Phys.\ Rev.\ \textbf{177},
2239, 2247 (1969);
%%CITATION = PHRVA,177,2247;%%
C.G.~Callan, \textit{et. al.}
%``Structure of phenomenological lagrangians. 2,''
Phys.\ Rev.\ \textbf{177} (1969) 2247.
%%CITATION = PHRVA,177,2247;%%
%For recent reviews, see e.g.:~G.~Ecker,
%%``Chiral Perturbation Theory,''
%Prog.\ Part.\ Nucl.\ Phys.\ \textbf{35}, 1 (1995) [arXiv:hep-ph/9501357];
%%%CITATION = PPNPD,35,1;%%
%G.~Colangelo and G.~Isidori,
%%``An introduction to CHPT,''
%arXiv:hep-ph/0101264.
%%%CITATION = HEP-PH/0101264;%%

\bibitem{Contino:2010t}C.~Contino,
%Tasi lectures: Higgs as a Composite Nambu-Goldstone Boson.
[\href{http://arxiv.org/abs/1005.4269v1}{arXiv:1005.4269v1 [hep-ph]}].

\bibitem {Contino:2010}R.~Contino, C.~Grojean, M.~Moretti, F.~Piccinini and
R.~Rattazzi, [\href{http://arXiv.org/pdf/1002.1011}{arXiv:1002.1011[hep-ph]}]


\bibitem {Casalbuoni:1985}R.~Casalbuoni, S.~De Curtis, D.~Dominici and
R.~Gatto,
%``Effective Weak Interaction Theory With Possible New Vector Resonance From A
%Strong Higgs Sector,''
Phys.\ Lett.\ B \textbf{155} (1985) 95; Nucl.\ Phys.\ B \textbf{282} (1987) 235.


\bibitem {Nomura}Y.~Nomura,
%``Higgsless theory of electroweak symmetry breaking from warped space,''
JHEP \textbf{0311} (2003) 050 \href{http://arXiv.org/abs/hep-ph/0309189}{arXiv:hep-ph/0309189}].

%\cite{Barbieri:2003pr}


\bibitem {Barbieri:2003pr}R.~Barbieri, A.~Pomarol and R.~Rattazzi,
%``Weakly coupled Higgsless theories and precision electroweak tests,''
Phys.\ Lett.\ B \textbf{591} (2004) 141  [\href{http://arxiv.org/abs/hep-ph/0310285}{arXiv:hep-ph/0310285}].

\bibitem {Foadi:2003xa}R.~Foadi, S.~Gopalakrishna and C.~Schmidt,
%``Higgsless electroweak symmetry breaking from theory space,''
JHEP \textbf{0403} (2004) 042 [\href{http://arxiv.org/abs/hep-ph/0312324}{arXiv:hep-ph/0312324}].
%%CITATION = JHEPA,0403,042;%%


\bibitem {Georgi:2004iy}H.~Georgi,
%``Fun with Higgsless theories,''
Phys.\ Rev.\ D \textbf{71} (2005) 015016 [\href{http://arxiv.org/abs/hep-ph/0408067}{arXiv:hep-ph/0408067}].
%%CITATION = PHRVA,D71,015016;%%

\bibitem {SekharChivukula:2008mj} R.~S.~Chivukula, H.~J.~He, M.~Kurachi,
E.~H.~Simmons and M.~Tanabashi,
%``General Sum Rules for WW Scattering in Higgsless Models: Equivalence
%Theorem and Deconstruction Identities,''
Phys.\ Rev.\ D \textbf{78}, 095003 (2008), [\href{http://arxiv.org/abs/hep-ph/0808.1682}{arXiv:hep-ph/0808.1682}].

\bibitem {calchep}A. Pukhov, A. Belyaev and N. Christensen,
\href{http://theory.sinp.msu.ru/~pukhov/calchep.html}{http://theory.sinp.msu.ru/$\sim
$pukhov/calchep.html}.


\bibitem{Foadi:2008} R.~Foadi, M.~J\"arvinen and F.~Sannino, Phys.\ Rev.\ D \textbf{79} (2008) 035010 [\href{http://arXiv.org/abs/0811.3719}{arXiv:0811.3719 [hep-ph]}].

\bibitem {feynrules}N. Christensen, C. Duhr and B. Fucks,
\href{http://feynrules.phys.ucl.ac.be/}{http://feynrules.phys.ucl.ac.be/}


\bibitem {Chivukula:2003}R.~S.~Chivukula, D.~A.~Dicus, H.~J.~He and
S.~Nandi,
%``Unitarity of the higher dimensional standard model,''
Phys.\ Lett.\ B \textbf{562} (2003) 109 \href{http://arXiv.org/abs/hep-ph/030226}{arXiv:hep-ph/030226}].


\bibitem {Birkedal:2005yg}A.~Birkedal, K.~T.~Matchev and M.~Perelstein,
%``Phenomenology of Higgsless models at the LHC and the ILC,''
\textit{In the Proceedings of 2005 International Linear Collider Workshop
(LCWS 2005), Stanford, California, 18-22 Mar 2005, pp 0314}
[\href{http://arXiv.org/pdf/0508185}{arXiv:0508185[hep-ph]}]


\bibitem{Kaplan:1991dc}
  D.~B.~Kaplan,
  %``Flavor at SSC energies: A New mechanism for dynamically generated fermion
  %masses,''
  Nucl.\ Phys.\  B {\bf 365} (1991) 259.

%\bibitem{Carcamo:2010ckm}A. E. C\'{a}rcamo Hern\'{a}ndez and Rakibur Rahman [\href{http://arXiv.org/pdf/1007.0447}{arXiv:1007.0447[hep-ph]}].
%
%\bibitem{Carcamo:2010back}A. E. C\'{a}rcamo Hern\'{a}ndez, G.~Corcella and R.~Torre, work in progress.

\bibitem{Grojean}C.~Grojean,
%``New Theories for the Fermi scale,''
[\href{http://arxiv.org/abs/0910.4976v1}{arXiv:0910.4976v1 [hep-ph]}].


%\cite{He:2007ge}

\bibitem{Carcamo:2010ggtoVV}A. E. C\'{a}rcamo Hern\'{a}ndez, [\href{http://arXiv.org/pdf/1008.1039}{arXiv:1008.1039[hep-ph]}].

\bibitem{Hirn:2004}
  J.~Hirn and J.~Stern,
  %``The role of spurions in Higgs-less electroweak effective theories,''
  Eur.\ Phys.\ J.\  C {\bf 34} (2004) 447 \href{http://arXiv.org/abs/hep-ph/0401032}{arXiv:hep-ph/0401032}];

%%\cite{Accomando:2008dm}
%
\bibitem{Accomando:2008dm}
E.~Accomando, S.~De Curtis, D.~Dominici and L.~Fedeli, \href{http://arXiv.org/abs/hep-ph/0807.2951}{arXiv:hep-ph/0807.2951}].
%%``The four site Higgsless model at the LHC,''
%Nuovo Cim.\  {\bf 123B} (2008) 809
%[arXiv:0807.2951 [hep-ph]].


%\cite{Belyaev:2008yj}







%\cite{SekharChivukula:2008mj}



%References Professor Robert Shrock
\bibitem {Appelquist:2003}T.~Appelquist and R.~Shrock, Phys. Rev. Lett. 90, 201801 (2003), [\href{http://arxiv.org/abs/hep-ph/0301108}{arXiv:hep-ph/0301108}].

\bibitem{Quigg:2009}C.~Quigg and R.~Shrock, Phys. \ Rev. \ D \textbf{79} :096002 (2009) [\href{http://arxiv.org/abs/hep-ph/0901.3958}{arXiv:hep-ph/0901.3958}].
%Gedanken Worlds without Higgs: QCD-Induced Electroweak Symmetry Breaking

%Dynamical Symmetry Breaking of Extended Gauge Symmetries, Phys. Rev. Lett. 90, 201801 (2003) (hep-ph/0301108). 
%reasonably ultraviolet complete theory with dynamical breaking of two gauge groups with left-right electroweak structure are constructed and studied, namely
%SU(3)_c \times SU(2)_L \times SU(2)_R \times U(1)_{B-L}
%and
%SU(4) \times SU(2)_L \times SU(2)_R 

\bibitem{Han}T. Han, D. L. Rainwater and G. Valencia, Phys. \ Rev. \ D \textbf{68} 015003 (2003) [\href{http://arxiv.org/abs/hep-ph/0301039}{arXiv:hep-ph/0301039}].
%TeV resonances in top physics at the LHC" si estudiano resonanzie scalare S pesante con massa in torna al TeV e vettoriale V_{mu} associated with the strongly interacting EWSB sector. Si considerano interazioni tipo VVS, WWV, Sttbar e interazioni di la resonancia vettoriale V con fermioni. Partial wave unitarity conditions constrain the couplings of these resonances. Se considera la posibilita di observare la segnale al LHC in il canale WW \to ttbar over the large QCD background.

\bibitem{Sanino:2008}R. Foadi, M. Jarvinen and F. Sannino, Phys. \ Rev. \ D \textbf{79} (2008) 035010 [\href{http://arxiv.org/abs/hep-ph/0811.3719}{arXiv:hep-ph/0811.3719}].
%"Unitarity in Technicolor", the unitariy in the elastic WW scattering is studied up to a cuttoff of 3TeV in the framework of a Technicolor model with one composite light singlet scalar (mH=200GeV), a parity odd spin one resonance and a spin two resonance (which has the same properties of a massive graviton). The heavy parity odd spin one resonance is assumed to have a mass of the order TeV. In the study of the unitarity in the elastic WW scattering they concentrate in the I=J=0 partial wave amplitude. They also study the constrains of the axial vector mass Mv and g_{V\pi\pi} coupling coming from a small S parameter.
            
%\bibitem {calchep}A. Pukhov, A. Belyaev and N. Christensen, http://theory.sinp.msu.ru/~pukhov/calchep.html.

%\cite{Pallante:1992qe}


\bibitem {Pallante:1992qe}E.~Pallante and R.~Petronzio,
%``Anomalous effective Lagrangians and vector resonance models,''
Nucl.\ Phys.\ B \textbf{396} (1993) 205.

%\cite{Borasoy:1995ds}


\bibitem {Borasoy:1995ds}B.~Borasoy and U.~G.~Meissner,
%``Chiral Lagrangians for Baryons coupled to massive Spin-1 Fields,''
Int.\ J.\ Mod.\ Phys.\ A \textbf{11} (1996) 5183  [\href{http://arxiv.org/abs/hep-ph/9511320}{arXiv:hep-ph/9511320}].

%\cite{Harada:2003jx}


\bibitem {Harada:2003jx}M.~Harada and K.~Yamawaki,
%``Hidden local symmetry at loop: A new perspective of composite gauge boson
%and chiral phase transition,''
Phys.\ Rept.\ \textbf{381} (2003) 1 [arXiv:hep-ph/0302103].

%\cite{Bijnens:1995ii}


\bibitem {Bijnens:1995ii}J.~Bijnens and E.~Pallante,
%``On the tensor formulation of effective vector Lagrangians and duality
%transformations,''
Mod.\ Phys.\ Lett.\ A \textbf{11} (1996) 1069 [\href{http://arxiv.org/abs/hep-ph/9510338}{arXiv:hep-ph/9510338}].

%\cite{Cirigliano:2006hb}


\bibitem {Cirigliano:2006hb}V.~Cirigliano, G.~Ecker, M.~Eidemuller, R.~Kaiser,
A.~Pich and J.~Portoles,
%``Towards a consistent estimate of the chiral low-energy constants,''
Nucl.\ Phys.\ B \textbf{753} (2006) 139 [\href{http://arxiv.org/abs/hep-ph/0603205}{arXiv:hep-ph/0603205}].

%\cite{Kampf:2006yf}


\bibitem {Kampf:2006yf}K.~Kampf, J.~Novotny and J.~Trnka,
%``On different lagrangian formalisms for vector resonances within chiral
%perturbation theory,''
Eur.\ Phys.\ J.\ C \textbf{50} (2007) 385 [\href{http://arXiv.org/pdf/0608051}{arXiv:[hep-ph/0608051]}]



%\bibitem{Aaltonen:2007je}
  %T.~Aaltonen {\it et al.}  [CDF Collaboration], To be cited only if composite fermions are included
  %``Search for New Particles Leading to $Z +$ jets Final States in $p \bar{p}$
  %Collisions at $\sqrt{s}$ = 1.96-TeV,''
  %Phys.\ Rev.\  D {\bf 76} (2007) 072006 [\href{http://arXiv.org/pdf/0706.3264}{arXiv:[hep-ex/0706.3264]}].
  %[arXiv:0706.3264 [hep-ex]].
  %%CITATION = PHRVA,D76,072006;%%
\end{thebibliography}
\end{document}